\documentclass[a4paper,11pt]{article}

\usepackage{jheppub}

\usepackage{amsmath, amssymb, xcolor, physics, indentfirst, enumitem, comment, tikz, array}
\usepackage{orcidlink}
\usepackage[export]{adjustbox}
\usepackage[dvipsnames]{xcolor}
\usepackage{bbm}

\usetikzlibrary{
    arrows.meta,             
    decorations.pathmorphing,  
    decorations.pathreplacing, 
    positioning,             
    calc,                     
    patterns,
    matrix, 
    angles,                 
    quotes                  
}

\usepackage[numbers]{natbib}

\usepackage{hyperref}
\hypersetup{
	colorlinks=true,
	urlcolor=RoyalBlue,
	linkcolor=orange!80!black,
	citecolor=RoyalBlue,
	pdftitle={The Unitary Architecture of Renormalization},
	pdfauthor={},
	pdfdisplaydoctitle=true,
	pdfstartview=FitH
}

\usepackage[mathscr]{eucal}

\usepackage[normalem]{ulem}

\usepackage{empheq}
\usepackage[most]{tcolorbox}
\newtcbox{\sentencebox}[1][]{%
	nobeforeafter, math upper, tcbox raise base,
	enhanced, colframe=gray!10,
	colback=gray!10, boxrule=0pt, arc=0pt,
	#1}

\arxivnumber{2511.10613} 

\numberwithin{equation}{section}

\def\equationautorefname~#1\null{(#1)\null}

\renewcommand{\Im}{\mathrm{Im}}
\def \be {\begin{equation}}
\def \ee {\end{equation}}

\def \leq {\leqslant}
\def \geq {\geqslant}

\def \M {\mathcal{M}}

\title{\boldmath The Unitary Architecture of Renormalization}

\author{Ameya Chavda \orcidlink{0000-0002-1173-1605},}
\author{Daniel McLoughlin \orcidlink{0009-0005-3535-2334},}
\author{Sebastian Mizera \orcidlink{0000-0002-8066-5891},}
\author{John Staunton \orcidlink{0009-0004-1661-9577}}

\affiliation{Center for Theoretical Physics, Department of Physics,\\
Columbia University, Pupin Hall,
538 West 120th Street, New York, NY 10027, USA}

\emailAdd{ameya.chavda@columbia.edu, dcm2183@columbia.edu, sebastian.mizera@columbia.edu, j.staunton@columbia.edu}

\abstract{We set up a bootstrap problem for renormalization. Working in the massless four-dimensional $O(N)$ model and the $\lambda \phi^4$ theory, we prove that unitarity leads to all-loop recursion relations between coefficients of scattering amplitudes with different multiplicities. These turn out to be equivalent to the identities imposed by renormalization of the coupling and the wavefunction through subleading logarithmic order, except with different initial conditions. Matching the initial conditions thus fixes the beta function and wavefunction anomalous dimension to these orders. We explain how to connect this new on-shell renormalization picture with the standard renormalized perturbation theory, highlighting a rich interplay between finiteness, dimensional regularization, and unitarity cuts.}

\begin{document}

\setcounter{tocdepth}{2}
\maketitle
\setcounter{page}{2}

\section{Introduction}
\label{Introduction}

\textbf{Background.}
Perhaps one of the most important missing pieces in $S$-matrix theory is an on-shell understanding of renormalization. The renormalization group (RG) describes how physics changes across different energy scales. 
There are a couple of complementary ways of interpreting RG and its physical consequences, including the Wilsonian and the continuum approaches, see, e.g., \cite{Georgi:1993mps, Schwartz:2014, Cohen:2019wxr}. In the Wilsonian approach, one concedes lack of knowledge about physics above a cutoff energy scale, $\Lambda$ (e.g. the scale of interatomic separation in hydrodynamics), that is large compared to the energy scale of the experiment. Then, the detailed physics at energies above the cutoff $\Lambda$ is swept into a set of constants (e.g. viscosity) that are \textit{renormalized} as $\Lambda$ changes. An example implementation of this approach is through Polchinski's RG equation, which uses a smooth cutoff function applied to the path integral \cite{Polchinski:1984}. On the other hand, in the continuum approach, one defines these same couplings at a given scale $\mu$, which does not itself need to be large, in terms of a physical observable. Then, as $\mu$ changes, so do the couplings in such a way that leaves physical observables independent of the arbitrary energy scale $\mu$. The property of the parameters describing the physics of a system changing with the scale $\mu$, or the cutoff $\Lambda$, is referred to as \textit{running} or \textit{RG flow}. It is this phenomenon that currently lacks a satisfactory explanation in terms of on-shell quantities. 

\textbf{Motivation.}
The goal of this paper is to derive aspects of running and RG flow in terms of the $S$-matrix. This objective stands in contrast with the conventional way of talking about RG, which is based on renormalized perturbation theory involving Feynman diagrams and counterterms, see, e.g., \cite{Collins:1984}. There are many motivations for pursuing the on-shell program. The most important one is the efficiency of computations. As illustrated by recent advances in scattering amplitudes \cite{Elvang:2015,Travaglini:2022uwo}, thinking directly in terms of on-shell quantities provides a way of avoiding redundancies in the usual Feynman-diagrammatic approach. A famous example is given by gluon scattering, where millions of diagrams turn out to simplify to a single-line expression or even cancel out entirely \cite{Parke:1986gb}. Such cancellations were later understood in terms of on-shell recursion relations \cite{Britto:2005fq} and related structures \cite{Arkani-Hamed:2012zlh}.

The case of renormalization is not too dissimilar. It is known that RG allows us to ``recycle'' computations between different loop orders. The intuitive picture is that the same divergence that appears at one loop order is going to appear nested within every higher loop order amplitude. Likewise every two loop divergence is going to reappear, etc. Renormalization therefore makes it obvious that perturbative computations are highly redundant. There is a good reason for this: perturbation theory organizes physics order-by-order in the coupling, while the modern understanding of the RG and effective field theories is that we are really supposed to organize it scale-by-scale. Feynman diagrams obscure this elegant picture. This motivates a search for another approach that does not make any reference to Feynman diagrams or counterterms.

\textbf{Summary of the results.}
In this paper, we set up a version of a bootstrap problem in which we write a parametrization of the S-matrix elements consistent with all symmetries and analytic properties and then impose physical constraints. The most important constraint will be that of unitarity, which comes from the fact that the $S$-matrix operator, $\hat{S}$, satisfies $\hat{S} \hat{S}^\dag = \mathbbm{1}$. It embodies the physical principle of probability conservation. Working within the realm of massless quartic theories, the goal of the paper is to demonstrate that the underlying architecture of renormalization emerges as a result of imposing these constraints. 

\begin{figure}[t]
\label{f:RecursionTowers}
\hspace{-1em}
\begin{tikzpicture}
\matrix (m) [
        matrix of math nodes, 
        nodes={text width=1cm, minimum height=1cm, anchor=center, align=center}, 
        row sep=-\pgflinewidth, 
        column sep=-\pgflinewidth,  
        label={[font=\bfseries]above: {\qquad\qquad Unitarity}}
    ]
    {
        |[text width=2.2cm]| \text{\bf\small tree} & m_{0,0} &  &  &  & \\
        |[text width=2.2cm]|\text{\bf\small one loop} & |[fill=orange!20]| m_{1,1} &  m_{1,0} &  &  & \\
        |[text width=2.2cm]|\text{\bf\small two loops} & |[fill=orange!20]| m_{2,2} & |[fill=orange!20]| m_{2,1} &  m_{2,0} & \\
        |[text width=2.2cm]|\text{\bf\small three loops} & |[fill=orange!20]| m_{3,3} & |[fill=orange!20]| m_{3,2}  & |[fill=gray!10]| m_{3,1} & m_{3,0} &  \\
        |[text width=2.2cm]|{\bf \vdots} & |[fill=orange!20]| \vdots & |[fill=orange!20]| \vdots & |[fill=gray!10]| \vdots & |[fill=gray!10]| \vdots & \ddots  \\
        |[text width=2.2cm]| & |[text width=2cm, rotate=90]|\text{\bf\small leading} & |[text width=2cm, rotate=90]|\text{\bf\small subleading} & \dots  &  &  \\
    };
    \draw [<-, orange!80!black, shorten <=-4pt, shorten >=-4pt] (m-1-2) -- (m-2-2);
    \draw [<-, orange!80!black, shorten <=-4pt, shorten >=-4pt] (m-2-2) -- (m-3-2);
    \draw [<-, orange!80!black, shorten <=-4pt, shorten >=-4pt] (m-3-2) -- (m-4-2);
    \draw [<-, orange!80!black, shorten <=-4pt, shorten >=-4pt] (m-4-2) -- (m-5-2);
    \draw [<-, orange!80!black, shorten <=-4pt, shorten >=-4pt] (m-2-3) -- (m-3-3);
    \draw [<-, orange!80!black, shorten <=-4pt, shorten >=-4pt] (m-3-3) -- (m-4-3);
    \draw [<-, orange!80!black, shorten <=-4pt, shorten >=-4pt] (m-4-3) -- (m-5-3);
    \draw [<-, orange!80!black, shorten <=-4pt, shorten >=-4pt] (m-2-2) -- (m-3-3);
    \draw [<-, gray, shorten <=-4pt, shorten >=-4pt] (m-3-4) -- (m-4-4);
    \draw [<-, gray, shorten <=-4pt, shorten >=-4pt] (m-4-4) -- (m-5-4);
    \draw [<-, gray, shorten <=-4pt, shorten >=-4pt] (m-3-3) -- (m-4-4);
    \draw [<-, gray, shorten <=-4pt, shorten >=-4pt] (m-4-5) -- (m-5-5);
    \draw [<-, gray, shorten <=-4pt, shorten >=-4pt] (m-4-4) -- (m-5-5);
\end{tikzpicture}
\begin{tikzpicture}
\matrix (m) [
        matrix of math nodes, 
        nodes={text width=1cm, minimum height=1cm, anchor=center, align=center}, 
        row sep=-\pgflinewidth, 
        column sep=-\pgflinewidth,  
        label={[font=\bfseries]above:Renormalization}
    ]
    {
        m_{0,0} & (\beta_1, \gamma_1) &  &  & \\
        |[fill=orange!20]| m_{1,1} &  m_{1,0} & (\beta_2, \gamma_2) &  & \\
        |[fill=orange!20]| m_{2,2} & |[fill=orange!20]| m_{2,1} &  m_{2,0} & (\beta_3, \gamma_3) \\
        |[fill=orange!20]| m_{3,3} & |[fill=orange!20]| m_{3,2} & |[fill=orange!20]| m_{3,1} & m_{3,0} & (\beta_4, \gamma_4) \\
        |[fill=orange!20]| \vdots & |[fill=orange!20]| \vdots & |[fill=orange!20]| \vdots & |[fill=orange!20]| \vdots &  \ddots  \\
        |[text width=2cm, rotate=90]|\text{\bf\small leading} & |[text width=2cm, rotate=90]|\text{\bf\small subleading} & \dots & &  \\
    };
    \draw [<-, orange!80!black, shorten <=-4pt, shorten >=-4pt] (m-1-1) -- (m-2-1);
    \draw [<-, orange!80!black, shorten <=-4pt, shorten >=-4pt] (m-2-1) -- (m-3-1);
    \draw [<-, orange!80!black, shorten <=-4pt, shorten >=-4pt] (m-3-1) -- (m-4-1);
    \draw [<-, orange!80!black, shorten <=-4pt, shorten >=-4pt] (m-4-1) -- (m-5-1);
    \draw [<-, orange!80!black, shorten <=-4pt, shorten >=-4pt] (m-1-2) -- (m-2-1);
    \draw [<-, orange!80!black, shorten <=-4pt, shorten >=-4pt] (m-2-2) -- (m-3-2);
    \draw [<-, orange!80!black, shorten <=-4pt, shorten >=-4pt] (m-3-2) -- (m-4-2);
    \draw [<-, orange!80!black, shorten <=-4pt, shorten >=-4pt] (m-4-2) -- (m-5-2);
    \draw [<-, orange!80!black, shorten <=-4pt, shorten >=-4pt] (m-2-3) -- (m-3-2);
    \draw [<-, orange!80!black, shorten <=-4pt, shorten >=-4pt] (m-2-1) -- (m-3-2);
    \draw [<-, orange!80!black, shorten <=-4pt, shorten >=-4pt] (m-3-3) -- (m-4-3);
    \draw [<-, orange!80!black, shorten <=-4pt, shorten >=-4pt] (m-4-3) -- (m-5-3);
    \draw [<-, orange!80!black, shorten <=-4pt, shorten >=-4pt] (m-3-4) -- (m-4-3);
    \draw [<-, orange!80!black, shorten <=-4pt, shorten >=-4pt] (m-3-2) -- (m-4-3);
    \draw [<-, orange!80!black, shorten <=-4pt, shorten >=-4pt] (m-4-4) -- (m-5-4);
    \draw [<-, orange!80!black, shorten <=-4pt, shorten >=-4pt] (m-4-5) -- (m-5-4);
    \draw [<-, orange!80!black, shorten <=-4pt, shorten >=-4pt] (m-4-3) -- (m-5-4);
\end{tikzpicture}
    \caption{\label{fig:intro-table}The unitary architecture of renormalization. Each row represents a given order in perturbation theory, while each column indicates the logarithmic order. The directions of the arrows denote proven (orange) and conjectured (gray) relationships between amplitude coefficients $m_{L,k}$. Note that $m_{0,0}=1$ and all other $m_{L,0} = 0$ in the on-shell scheme. Unitarity and renormalization give similar recursion relations, except for the initial conditions. \textbf{Left:} Unitarity fixes each column solely in terms of the $m_{L,0}$ as initial conditions, see \autoref{fig:results} for the concrete realization. \textbf{Right:} Renormalization fixes each column in terms of $\beta_L$, $\gamma_L$, and $m_{L,0}$ as initial conditions, see \autoref{fig:resultsRG2} and \autoref{fig:resultsRG} for the concrete realization.}
\end{figure}
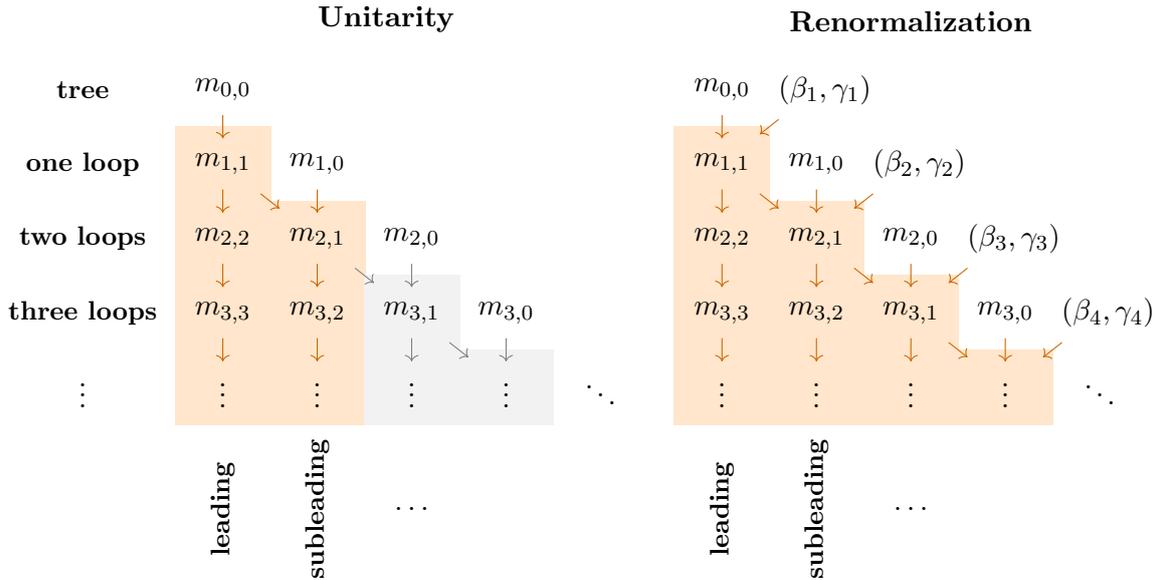

Let us illustrate a much simplified version of the argument. Consider the 4-particle S-matrix element in a quartic theory with $\mathbb{Z}_2$ symmetry, $\M_4$, which can be parametrized as
\begin{equation}
\begin{aligned}
\label{eq:M4-intro}
\M_4 = &\quad\; -\lambda\;\;\, m_{0,0} \\
&+ \left(-\lambda\right)^2 \left[ m_{1,1} \left(\log_{-s} + \log_{-t} + \log_{-u}\right) + m_{1,0 }\right] \\
&+ \left(-\lambda\right)^3 \left[ m_{2,2} \left(\log_{-s}^2 + \log_{-t}^2 + \log_{-u}^2\right) + m_{2,1} \left(\log_{-s} + \log_{-t} + \log_{-u}\right) + m_{2,0} \right] \\
& + \ldots
\end{aligned}
\end{equation}
The amplitude depends on the Mandelstam invariants $s$, $t$, and $u$ satisfying $s + t + u=0$. It is symmetric under permuting $(s,t,u)$ since we scatter identical bosons. Each line represents a given order in perturbation theory, starting from tree level, one loop order, etc. The dependence on the invariants enters through the logarithms, which we notate as
\begin{equation}
\log_x \equiv \log\left(\frac{\Lambda^2}{x}\right) \qquad \text{or} \qquad \log\left(\frac{\mu^2}{x}\right),
\end{equation}
depending on whether we use the Wilsonian or continuum approach. At the $L^{\text{th}}$ order in perturbation theory, we can have at most $L$ such logarithms coming from the UV divergences (later on, we will also show that mixed combinations of the form $\log_s^{n} \log_t^{m}$ do not appear as a consequence of unitarity). Moreover, we work in the hard scattering regime in which $|s| \sim |t| \sim |u| \ll \Lambda^2$ or $\gg \mu^2$ so that all the logarithms are large. From this perspective, the only objects not fixed by symmetries are the real constants $m_{L,k}$, which are the ``on-shell data'' describing the scattering process. Terms in the first column, proportional to $m_{L,L}$, are called leading logarithms, those in the second column, proportional to $m_{L,L-1}$, are called subleading logarithms, and so on. In the conventional picture, all these coefficients would be computed by summing over all Feynman diagrams at a given order and extracting the coefficients of the relevant logarithm. This, however, would be a lot of redundant work.

The coefficients $m_{L,k}$ are actually not uniquely determined because the amplitude itself $\M_4$ is not. Only the difference of the amplitude evaluated at different cutoffs/scales is physically measurable. Hence we need to agree on some reference scale relative to which we measure the couplings. One convenient choice is to demand that $\M_4 = -\lambda$ when the amplitude is continued to the symmetric point $s = t = u = -\Lambda^2$ or $-\mu^2$. In this case, all logarithms vanish and we have $\M_4 = -\lambda m_{0,0} + \left(-\lambda\right)^2 m_{1,0} + \left(-\lambda\right)^3 m_{2,0}  + \ldots$,  leading to the solution $m_{0,0} = 1$ and $m_{1,0} = m_{2,0} = \ldots = 0$. This choice is known as the \textit{on-shell scheme}.

Instead of the diagrammatic approach, we proceed by imposing unitarity on \autoref{eq:M4-intro}. Unitarity gives a quadratic constraint on the $S$-matrix elements, which translates to consistency conditions between the amplitude coefficients $m_{L,k}$. Explaining these constraints will be the subject of \autoref{s:UnitarityconstraintsPhi4}. They turn out to mix together scattering amplitudes $\M_n$ at all multiplicities $n$. However, if we focus our attention on the leading and subleading coefficients only, it is sufficient to consider two-, four-, and six-particle amplitudes. Unitarity then gives recursion relations illustrated in \autoref{fig:intro-table} (left): all the leading coefficients $m_{L,L}$ are fixed recursively in terms of $m_{0,0}$, while all the subleading coefficients $m_{L,L-1}$ are fixed recursively in term of $m_{0,0}$ and $m_{1,0}$. Conjecturally, each column is fixed solely in terms of $m_{L,0}$, which act as initial conditions for the recursions. But recall that these coefficients are scheme dependent and we are free to pick them at will. Hence the whole architecture of the on-shell data is fixed, at least conjecturally, by unitarity!

At first sight, it might seem like we gained a lot of information with little to no effort. Where has all the difficulty of computing renormalized Feynman integrals gone to? Indeed, there is no free lunch. The most computationally-intensive part of the above story is calculating the on-shell phase space integrals associated with unitarity cuts. To push the computation to higher subleading logarithmic orders, one needs to be able to evaluate more and more complicated unitarity cuts. However, once this step is done, a recursion is obtained to all loop orders.

We can finally circle back to renormalization. It is known that renormalization also gives a web of identities between logarithmic orders implied by the renormalization group equations, see, e.g., \cite{Cohen:2019wxr} for a pedagogical introduction. We illustrate it schematically in \autoref{fig:intro-table} (right). It takes a form reminiscent of the recursion relations derived from unitarity. The main difference is in the initial conditions, which are expressed in terms of the beta function, $\beta = \beta_1 \left(-\lambda\right)^2 + \beta_2 \left(-\lambda\right)^3 + \ldots$ (running of the coupling) and the anomalous dimension, $\gamma = \gamma_1 \left(-\lambda\right) + \gamma_2 \left(-\lambda\right)^2 + \ldots$ (running of the wavefunction). For example, leading logarithms are fixed in terms of $m_{0,0}$, as well as the one-loop beta function $\beta_1$ and the one-loop gamma function $\gamma_1$. Likewise, the subleading logarithms are determined in terms of $m_{0,0}$, $m_{1,0}$, $\beta_1$, $\beta_2$, $\gamma_1$, and $\gamma_2$, now including two-loop corrections. A detailed derivation will be given in \autoref{s:RGconstraintsPhi4}. The similarity of the unitarity and renormalization constraints inspired us to formulate the Unitarity Flow Conjecture \cite{Chavda:2025}, now proven through the subleading logarithmic order, which states that identities obtained from unitarity are the same as those obtained from renormalization. We can put it to practical use by matching the initial conditions between the two sides of \autoref{fig:intro-table}. This procedure achieves the on-shell derivation of the first few beta function coefficients:
\be\label{eq:intro-beta}
\beta_1 = \frac{3}{16 \pi^2}\, , \qquad \beta_2 = \frac{17}{3\left(16\pi^2\right)^2} + 0 \cdot m_{1,0}\, .
\ee
One observation is that $m_{1,0}$ entirely drops out after the matching. This is an on-shell derivation of the fact that the one- and two-loop beta functions are scheme-independent \cite{Arnone:2003pa}. Likewise, for the anomalous dimension we get
\be\label{eq:intro-gamma}
\gamma_1 = 0\, , \qquad \gamma_2 = \frac{1}{12 \left(16\pi^2\right)^2} + 0 \cdot m_{1,0}\, .
\ee
Indeed, these match the known results computed in the $\overline{\mathrm{MS}}$ scheme \cite{Kleinert:2001}. We will explain the connection between the on-shell framework and the traditional renormalized perturbation theory in \autoref{s:ConenctionToRPT}.

\textbf{Relations to previous work.}
Aspects of the relationship between unitarity and renormalization have already been studied in the literature, though not in the same way we argue for in this paper. Let us summarize the connections to our work.

Koschinski, Polyakov, and Vladimirov \cite{Koschinski:2010} showed that the leading logarithms of the 4-particle amplitude, expressed in the partial-wave basis, can be computed using unitarity, analyticity, and permutation symmetry in a general quartic scalar theory. The main technical difference to our work is in the way analyticity is implemented: we use an ansatz of the form \autoref{eq:M4-intro}, while \cite{Koschinski:2010} uses dispersion relations and an analyticity assumption. Both results are compatible at leading logarithmic order. Additionally, we extend the results to the subleading logarithmic order, which means we have to implement inelastic unitarity that mixes four-point amplitudes with other multiplicities.

Caron-Huot and Wilhelm \cite{Caron-Huot:2016} related the phase of the S-matrix to RG, by noting the relationship $\frac{2}{\pi}\, \Im \log_{p^2} = \mu \dv{\mu} \log_{p^2},$ for timelike momenta, $p^2 < 0$. The imaginary part is then related to the long-range propagation of on-shell states through the optical theorem. This intuition, dating back to the Regge-physics literature \cite{Fadin:1995xg}, allows the authors to extract the one- and two-loop beta function for gauge theories. The idea was then extended to the forward limit where rapidity renormalization is important by \cite{Rothstein:2023}. The procedures used in these papers have also been applied to processes in the Standard Model Effective Field Theory \cite{EliasMiro:2020tdv,Baratella:2020lzz,Jiang:2020mhe,Bern:2020}. Our work differs from \cite{Caron-Huot:2016} in the chain of logic: we do not assume the renormalization group equation, but instead we derive its consequences from unitarity.

A recent pathway has been to explore positivity constraints of unitarity to put bounds on beta function and anomalous dimension coefficients \cite{Chala:2023jyx,Chala:2023xjy,Liao:2025} in certain effective field theories. In this sense, we provide a stronger result since we not only fix the sign, but also the precise value of the coefficients, as in \autoref{eq:intro-beta} and \autoref{eq:intro-gamma}, though we only consider a specific field theory without higher-derivative operators.

Another approach based on generalized cuts has been taken in \cite{Huang:2012,Arkani-Hamed:2017jhn} at one-loop order. It is known that the one-loop beta function is fixed by the UV divergence of the bubble and these papers reconstruct the coefficients using generalized cuts \cite{Bern:1994cg}. This strategy is further from a purely on-shell approach we argue for here, which does not mention Feynman diagrams at all. These results proved useful in giving an on-shell symmetry explanation for the vanishing of various Standard Model one-loop anomalous dimensions \cite{Cheung:2015}.

Finally, let us mention that RG in the massive $\lambda \phi^4$ theory has been studied extensively.
A textbook treatment up to five-loop order is given by \cite{Kleinert:2001}, with extensions to the six-loop order in \cite{Kompaniets:2016, Kompaniets:2017}, and seven-loop order in \cite{Schnetz:2022nsc}. While our proof-of-principle results are nowhere near the state-of-the-art of beta function computations in the $\lambda \phi^4$ theory, we hope a new computational strategy could contribute to these precision calculations.

\textbf{Outline.}
As a warm-up, we provide a discussion of our results in the simplified case of the $O(N)$ model in the large $N$ limit in \autoref{s:ONModel}, which illustrates the main ideas without too many technical details. We then discuss the constraints of unitarity in the massless $\lambda \phi^4$ theory, including multi-particle cuts in \autoref{s:UnitarityconstraintsPhi4}. Comparison with the RG constraints is given in \autoref{s:RGconstraintsPhi4}, where we show how to match the initial conditions of the recursion relations. In \autoref{s:ConenctionToRPT}, we explain how the results of our work fit into the usual discussion of renormalized perturbation theory, including aspects of finiteness and dimensional regularization. We conclude in \autoref{s:Discussion} with a discussion of possible generalizations and future directions. A number of technical computations are relegated to appendices: \autoref{a:onShellLargeON} provides an on-shell argument for restrictions we put on the O($N$) model amplitude, \autoref{a:ExactResults} contains explicit computations in perturbation theory for $\lambda \phi^4$ that confirm our results, and \autoref{a:PhaseSpaceIntegrals} has details of the phase-space integration.

\textbf{Notation and conventions.} Throughout these notes, we will use the QFT conventions of \cite{Srednicki:2007}. In particular, we will use the mostly plus metric signature $\left(-, +, +, +\right)$ in four dimensions. When we do calculations using dimensional regularization, the number of dimensions will be $d = 4 - 2\epsilon$. A generic number of particles is given by $n$. The $S$-matrix operator $\hat{S}$ is decomposed as
\begin{equation}
\hat{S} = \mathbbm{1} + i \hat{T},
\end{equation}
with the $n$-particle amplitude related to the transfer matrix operator $\hat{T}$ as 
\begin{equation}
\mathcal{M}_n \left(2\pi\right)^d\delta^{(d)}\left(p_1 + \dots + p_{n_\text{in}} - p_{n_\text{in}+1} - \dots - p_{n}\right) = \mel{p_{n_\text{in} + 1}, \dots, p_{n}}{\hat{T}}{p_1, \dots, p_{n_\text{in}}} ,
\end{equation}
where $n_\text{in}$ is the number of incoming particles.
Since the majority of our computations involve a $2\rightarrow 2$ scattering process, it is helpful to define the Mandelstam variables
\begin{equation}
s = -\left(p_1 + p_2\right)^2, \qquad t = -\left(p_2 - p_3\right)^2, \qquad \text{and} \qquad u = -\left(p_2 - p_4\right)^2.
\end{equation}
Next, the propagator is given by
\begin{equation}
\begin{gathered}
\vspace{-0.75cm}
\begin{tikzpicture}[scale=0.5]
\draw[line width=0.3mm] (-1, 0) -- (1, 0);
\draw[line width=0.2mm, -Stealth] (-0.5, -0.3) -- node[anchor=north] {$\ell$} (0.5, -0.3);
\end{tikzpicture}
\end{gathered} = \frac{-i}{\ell^2 - i\varepsilon}.
\end{equation}
Finally, cut propagator, on the other hand, is given by:
\begin{equation}
\label{eq:CutPropagator}
\begin{gathered}
\vspace{-0.5cm}
\begin{tikzpicture}[scale=0.5]
\draw[line width=0.3mm] (-1, 0) -- (1, 0);
\draw[line width=0.2mm, -Stealth] (-0.5, -0.3) -- node[anchor=north, xshift=-0.25cm] {$\ell$} (0.5, -0.3);
\node at (0.25, -1.25) {$\hphantom{}$};
\draw[line width=0.5mm, orange, dashed] (0, 0.75) -- (0, -0.75);
\end{tikzpicture}
\end{gathered} = 2\pi \delta\left(\ell^2\right) \Theta\left(\ell^{\:0}\right).
\end{equation}
We refer to the procedure of putting $k$ particles on shell as a $k$-cut. 

\section{Warm-Up: The Large \texorpdfstring{$N$}{N} Limit of the O(\texorpdfstring{$N$}{N}) Model}
\label{s:ONModel}

As a warm-up we begin by considering the large–$N$ limit of the O$\left(N\right)$ model, where $N$ indicates the number of particle flavors. In \autoref{a:onShellLargeON}, we derive the most general 4-particle amplitude ansatz for such a theory, consistent with our assumptions in \autoref{Introduction} of masslessness, marginal coupling, IR finiteness, and hard scattering ($|s| \sim |t| \sim |u| \ll \Lambda^2$ or $\gg \mu^2$). For any model of $N$ flavors, the amplitude can be decomposed into flavor-ordered constituents based on the tensor structure. A key simplification brought about by the large–$N$ limit is that all dependence on different kinematic channels completely decouples among the flavor-ordered pieces: 
\begin{equation}
\M_{ij;kl}\left(s,t,u\right)=\frac{1}{N}\left[\M_{s}\left(s\right)\delta_{ij}\delta_{kl}+\M_{t}\left(t\right)\delta_{jk}\delta_{il}+\M_{u}\left(u\right)\delta_{ik}\delta_{jl}\right].
\end{equation}
We demonstrate this in \autoref{a:onShellLargeON}. We can then consider each flavor-ordered amplitude individually, and choose to focus on 
\begin{equation}
\M_s = \sum_{L = 0}^{\infty} \frac{\left(-\lambda\right)^{L + 1}}{\left(16\pi^2\right)^L} \M_s^{\left(L\right)},
\end{equation}
where
\begin{equation}
\label{eq:O(N)ansatz4pt}
\M_s^{\left(L\right)} = \sum_{k = 0}^{L} m_{L, k} \log^k_{-s}. 
\end{equation}
We will refer to the $m_{L,k}$ as \textit{amplitude coefficients}, where the first subscript $L$ denotes the $\left(L+1\right)^{\mathrm{th}}$ power of the coupling and the second subscript $k$ denotes the logarithmic order. We call $m_{L,L}$ the \textit{leading coefficients} and $m_{L,L-1}$ the \textit{subleading coefficients} etc. Note that, since we demand that the tree-level amplitude is $-\lambda$, we will impose that $m_{0,0} = 1$ as an input.

As outlined in \autoref{Introduction}, we will demonstrate that the recursion relations among amplitude coefficients derived by applying unitarity to \autoref{eq:O(N)ansatz4pt} are equivalent to those that follow from RG through subleading logarithmic order. As we show in \autoref{a:onShellLargeON}, this theory serves as an apt warm-up for our arguments for two main reasons, both following from the large–$N$ limit. On the unitarity side, only 2-particle cuts (2-cuts) contribute to the imaginary part of the amplitude, with all higher-particle cuts suppressed by powers of $N$. On the RG side, the large $N$ limit suppresses wavefunction renormalization, meaning that the wavefunction anomalous dimension is zero to leading $N$ order. This theory therefore serves as a pedagogical toy model that highlights the essence of our conjecture while avoiding many technical complications. 

\subsection{Unitarity Constraints}
To relate the leading and subleading coefficients we use the unitarity equation $\hat{S}^\dagger \hat{S} = \mathbbm{1}$. At the level of the $n$-particle amplitudes, $\M_n$, the unitarity equation relates its imaginary part to a product of amplitudes at lower order in perturbation theory. In this section, we derive the non-linear constraints from unitarity on the amplitude coefficients. For the $n_1 \rightarrow n_2$ particle amplitude, the unitarity equation states
\begin{equation}
\label{eq:UnitarityEquation}
\Im\, \M_{n_1 + n_2} = \frac{1}{2} \sum_{k\text{-cuts}} \int \M_{n_1 + k}\, \M^\ast_{n_2 + k} \dd{\Phi}_{k},
\end{equation}
where $\dd{\Phi}_{\text{k}}$ denotes the $k$-particle Lorentz-invariant phase space, see, e.g., \cite{Britto:2024} for a review. As argued in \autoref{a:onShellLargeON}, the leading contribution to the right-hand side of the unitarity equation is the 2-cut contribution, with all higher-particle cuts suppressed. Diagrammatically this reads, 
\begin{equation}
\label{eq:cut-formula}
2\: \text{Im}\left(\hspace{0.1cm}
\begin{gathered}
\begin{tikzpicture}[scale=0.6]
\draw[line width=0.3mm] (-1, 1) -- (-0.5*0.707, 0.5*0.707);
\draw[line width=0.3mm] (-1, -1) -- (-0.5*0.707, -0.5*0.707);
\filldraw[pattern=north west lines] (0, 0) circle (0.5cm);
\draw[line width=0.3mm] (0, 0) circle (0.5cm);
\draw[line width=0.3mm] (0.5*0.707, 0.5*0.707) -- (1, 1);
\draw[line width=0.3mm] (0.5*0.707, -0.5*0.707) -- (1, -1);
\end{tikzpicture}
\end{gathered}\hspace{0.1cm}\right) =  \: \:
\begin{gathered}
\begin{tikzpicture}[scale=0.6]
\draw[line width=0.3mm] (-1, 1) -- (-0.5*0.707, 0.5*0.707);
\draw[line width=0.3mm] (-1, -1) -- (-0.5*0.707, -0.5*0.707);
\draw[line width=0.3mm] (0, 0) circle (0.5cm);
\filldraw[pattern=north west lines] (0, 0) circle (0.5cm);
\draw[line width=0.3mm] (0.5*0.866, 0.5*0.5) -- (2 - 0.5*0.866, 0.5*0.5);
\draw[line width=0.3mm] (0.5*0.866, -0.5*0.5) -- (2 - 0.5*0.866, -0.5*0.5);
\draw[line width=0.3mm] (2, 0) circle (0.5cm);
\filldraw[pattern=north west lines] (2, 0) circle (0.5cm);
\draw[line width=0.3mm] (2 + 0.5*0.707, 0.5*0.707) -- (3, 1);
\draw[line width=0.3mm] (2 + 0.5*0.707, -0.5*0.707) -- (3, -1);
\filldraw[orange!10, opacity=0.5] (1, 1.1) rectangle (3.2, -1.1);
\draw[line width=0.5mm, dashed, orange] (1, 1) -- (1, -1);
\end{tikzpicture}
\end{gathered} \:\: ,
\end{equation}
where shading indicates complex conjugation. 

By plugging in the ansatz, we are able to write the unitarity equation in terms of $\M_s^{\left(L\right)}$ in \autoref{eq:O(N)ansatz4pt}, 
\begin{equation}
\text{Im}\left(\M_s^{\left(L\right)}\right) = \frac{16\pi^2}{2} \sum_{L^\prime = 0}^{L - 1} \int \M_s^{\left(L^\prime\right)} \M_s^{\left(L - L^\prime - 1\right)} \dd{\Phi}_2.
\end{equation}
One way to see why the unitarity equation takes this form is to examine the powers of $\lambda$ on each side of the equation. On the left-hand side, $\M_s^{\left(L\right)}$ is associated to order $\lambda^{L + 1}$. As such, the right-hand side is associated to order $\lambda^{L^\prime + 1 + L - L^\prime} = \lambda^{L + 1}$. The equation as stated simply follows from matching powers of $\lambda$ in the unitarity equation.

We will plug the ansatz, \autoref{eq:O(N)ansatz4pt}, into the left-hand side of the unitarity equation. The imaginary part comes from the logarithms in the ansatz since for physical values of momenta, $s > 0$, so $\log^k_{-s}$ has a nonzero imaginary part. The imaginary part of the logarithm can be found by expanding to first order
\begin{equation}
\label{eq:ExpansionInLogs}
\log^k_{-s} = \left(\log_{s} + i\pi\right)^k = \log^k_{s} + i\pi k \log^{k-1}_{s} + \dots
\end{equation}
In particular, the imaginary part of the amplitude up to subleading logarithmic order is 
\begin{equation}
\label{eq:ImaginaryPartON}
\Im\left(\M_s^{\left(L\right)}\right) = L \pi m_{L, L} \log^{L - 1}_{s} + \left(L - 1\right)\pi m_{L, L - 1} \log^{L - 2}_{s} + \dots
\end{equation}
The ellipsis contains further subleading logarithmic powers.

On the right-hand side of the unitarity equation, the phase space integral is simplified by the fact that the integrand does not depend on the cut momenta, as determined by the ansatz \autoref{eq:O(N)ansatz4pt}. As such, using the result of computing the 2-particle phase space, \autoref{eq:2particlePhaseSpace}, we find
\begin{equation}
\label{eq:UnitarityEqMsL}
\Im\left(\M_s^{\left(L\right)}\right) = \frac{\pi}{2}\sum_{L^\prime = 0}^{L - 1} \M_s^{\left(L^\prime\right)} \left(\M_s^{\left(L - L^\prime - 1\right)}\right)^{\ast}. 
\end{equation}
The leading order in logarithm on the right-hand side of this equation will arise from products of leading logarithms of the constituent amplitudes. Meanwhile, the subleading logarithmic order is determined by products of leading logarithms of the amplitude on the left of the cut and subleading logarithms on the right of the cut and vice versa. Explicitly, the leading contribution on the right-hand side is
\begin{equation}
\eval{\Im\left(\M_s^{\left(L, L\right)}\right)}_{2-\text{cuts}} = \frac{\pi}{2} \log^{L - 1}_{s} \sum_{L^\prime = 0}^{L - 1} m_{L^\prime, L^\prime} m_{L - L^\prime - 1, L - L^\prime - 1},
\end{equation}
where the second superscript on the left-hand side specifies that this is the leading logarithm contribution. Matching logarithmic order to \autoref{eq:ImaginaryPartON} yields a recursion relation for the leading coefficients 
\begin{equation}
m_{L, L} = \frac{1}{2L} \sum_{L^\prime = 0}^{L - 1} m_{L^\prime, L^\prime} m_{L - L^\prime - 1, L - L^\prime - 1},
\end{equation}
with the initial data $m_{0, 0} = 1$ set by the tree-level value of $\M_s$. The solution to this equation is
\begin{equation}
\label{eq:O(N)UnitarityLeadingM}
m_{L, L} = \frac{1}{2^L} .
\end{equation}

Pushing the expansion of the right-hand side of \autoref{eq:UnitarityEqMsL} to subleading logarithmic order as described above, we find 
\begin{equation}
\eval{\Im\left(\M_s^{\left(L, L - 1\right)}\right)}_{2-\text{cuts}} \!\! = \frac{\pi}{2} \log^{L - 2}_{s} \sum_{L^\prime = 0}^{L - 1} \left[m_{L^\prime, L^\prime - 1} m_{L - L^\prime - 1, L - L^\prime - 1} \! + \!  m_{L^\prime, L^\prime} m_{L - L^\prime - 1, L - L^\prime - 2}\right].
\end{equation}
Here, the superscript $L - 1$ indicates subleading logarithmic order. If we change the dummy variable in the second term in the square brackets from $L^\prime$ to $L - L^\prime - 1$, we find that it is equivalent to the first term. Then, plugging in the result we obtained for the leading coefficient, we find that unitarity constrains the subleading coefficient as
\begin{equation}
m_{L, L - 1} = \frac{1}{2^{L - 1}\left(L - 1\right)} \sum_{L^\prime = 1}^{L - 1} 2^{L^\prime} m_{L^\prime, L^\prime - 1}.
\end{equation}
Notice that the sum starts at $L^\prime = 1$ instead of $L^\prime = 0$ since $m_{0, -1} = 0$. The initial data for this recursion relation is the scheme-dependent, and therefore arbitrary, $m_{1, 0}$, yielding
\begin{equation}
\label{eq:O(N)UnitaritySubLeadingM}
m_{L, L - 1} = \frac{L}{2^{L - 1}} m_{1, 0}.
\end{equation}
The interpretation of these results is that the whole tower of leading and subleading amplitude coefficients is fixed by unitarity and the input data $m_{0, 0}$ (for leading) and $m_{1, 0}$ (for subleading). We will rediscover these constraints from the RG equation in the next subsection, with the key difference that the required input data is different. 

\subsection{Reconstruction from the RG Flow}
Now we turn to the RG perspective. Our starting point here is the \textit{Callan--Symanzik} (CS) equation (also referred to as the \textit{renormalization group equation} or RGE) \cite{Callan:1970, Symanzik:1970, Symanzik:1971}, which follows from the fact that S-matrix elements run with the scale $\mu$: 
\begin{equation}
\label{eq:CSonM}
\mu\partial_\mu \M_n = - \beta \partial_\lambda \M_n + n \gamma \M_n\, ,
\end{equation}
where we use the shorthand notation $\partial_x = \frac{\partial}{\partial x}$.
The running of the coupling is encoded in the beta function, which follows from the chain rule,
\begin{equation}
\beta\left(\lambda\right) = \mu\partial_\mu \lambda.
\end{equation}
There is also generically a wavefunction anomalous dimension $\gamma$, which follows from the fact that the one-particle states $\ket{p_i}$ also get renormalized, 
\begin{equation}
\gamma\left(\lambda\right) = \mu \partial_\mu \ket{p_i}.
\end{equation}
The factor of $n$ in \autoref{eq:CSonM} comes from the amplitude constructed out of $n$-particle states.

As discussed in \autoref{a:onShellLargeON}, at leading $N$ order in the large $N$ limit there is no wavefunction renormalization. Consequently, the $4$-particle flavor-ordered amplitude ansatz $\M_s\left(\mu, \lambda\left(\mu\right)\right)$, defined in \autoref{eq:O(N)ansatz4pt}, obeys a simplified version of the CS equation:
\begin{equation}
\label{eq:CSequationO(N)}
\mu\partial_\mu \M_s + \beta \partial_\lambda \M_s = 0.
\end{equation}

Just as the amplitude ansatz \autoref{eq:O(N)ansatz4pt} admits a Taylor expansion in the coupling $\lambda$, so too does the $\beta$ function: 
\begin{equation}
\label{eq:betaO(N)model}
\beta\left(\lambda\right)=\sum_{L=1}^{\infty}\frac{\beta_{L}\left(-\lambda\right)^{L+1}}{\left(16\pi^{2}\right)^{L}},
\end{equation}
where we have neglected the possibility of constant and linear-in-$\lambda$ terms for $\beta$ since these are immediately set to zero upon substitution into the Callan--Symanzik equation \autoref{eq:CSequationO(N)}. We choose the summation index and normalization to match
common conventions.

Our strategy will be as follows. We will plug the Taylor expansions of the amplitude and the $\beta$ function, given in \autoref{eq:O(N)ansatz4pt} and \autoref{eq:betaO(N)model} respectively, into the Callan--Symanzik equation \autoref{eq:CSequationO(N)} and then match orders in the coupling $\lambda$ and the logarithm order. In doing so, we will find the emergence of recursion relations among the amplitude coefficients. For clarity, we will proceed term-by-term. The first term in \autoref{eq:CSequationO(N)} is given by 
\begin{equation}
\label{eq:O(N)CSFirstTerm}
\mu\partial_{\mu}\mathcal{M}_{s}=\sum_{L=1}^{\infty}\sum_{k=0}^{L-1}\frac{\left(-\lambda\right)^{L+1}}{\left(16\pi^{2}\right)^{L}}2\left(k+1\right)m_{L,k+1}\log_{-s}^{k}.
\end{equation}
The second term in \autoref{eq:CSequationO(N)} is more involved. We first compute the derivative, 
\begin{equation}
\beta\left(\lambda\right)\partial_{\lambda}\mathcal{M}_{s}=-\sum_{L=0}^{\infty}\sum_{L^{\prime}=1}^{\infty}\sum_{k=0}^{L}\frac{\left(-\lambda\right)^{L^{\prime}+1}\left(-\lambda\right)^{L}}{\left(16\pi^{2}\right)^{L+L^{\prime}}}\left(L+1\right)\beta_{L^{\prime}}m_{L,k}\log^{k}_{-s},
\end{equation}
then shift the starting points of the sums so that we can apply the double sum identity
\begin{equation}
\label{eq:DoubleSum}
\sum_{n=0}^{\infty}\sum_{m=0}^{\infty}a_{n,m}=\sum_{n=0}^{\infty}\sum_{m=0}^{n}a_{n-m,m}
\end{equation}
to give 
\begin{equation}
\beta\left(\lambda\right)\partial_{\lambda}\mathcal{M}_{s}=-\sum_{L=0}^{\infty}\sum_{L^{\prime}=0}^{L}\sum_{k=0}^{L-L^{\prime}}\frac{\left(-\lambda\right)^{L+2}}{\left(16\pi^{2}\right)^{L+1}}\left(L-L^{\prime}+1\right)\beta_{L^{\prime}+1}m_{L-L^{\prime},k}\log_{-s}^{k},
\end{equation}
before finally switching the order of the sums over $L^{\prime}$ and $k$:
\begin{equation}
\label{eq:O(N)CSSecondTerm}
\begin{aligned}
\beta\left(\lambda\right)\partial_{\lambda}\mathcal{M}_{s} & =-\sum_{L=1}^{\infty}\sum_{L^{\prime}=0}^{L-1}\frac{\left(-\lambda\right)^{L+1}}{\left(16\pi^{2}\right)^{L}}\left(L-L^{\prime}\right)\beta_{L^{\prime}+1}m_{L-L^{\prime}-1,0}\\
 & \hphantom{=}-\sum_{L=1}^{\infty}\sum_{k=1}^{L-1}\sum_{L^{\prime}=0}^{L-k-1}\frac{\left(-\lambda\right)^{L+1}}{\left(16\pi^{2}\right)^{L}}\left(L-L^{\prime}\right)\beta_{L^{\prime}+1}m_{L-L^{\prime}-1,k}\log_{-s}^{k}.
\end{aligned}
\end{equation}

We now recombine these terms as in \autoref{eq:CSequationO(N)} and find new constraints for each term in the sum over $k$, isolating the coefficients of different powers of logarithm. We will choose three specific values of $k$ that allow us to extract constraints for the leading coefficients, the subleading coefficients, and the $\beta$ function. For each of these values of $k$ we will compare \autoref{eq:O(N)CSFirstTerm} and \autoref{eq:O(N)CSSecondTerm}. First, we consider the leading logarithmic terms which come from setting $k=L-1$ for $L\geq2$:
\begin{equation}
m_{L,L}=\frac{1}{2}\beta_{1}m_{L-1,L-1}.
\end{equation}
This recursion can be solved to give
\begin{equation}
\label{eq:O(N)RGLeadingM}
m_{L,L}=m_{1,1}\left(\frac{\beta_{1}}{2}\right)^{L-1},
\end{equation}
again valid for $L\geq2$.

A recursion for the subleading amplitude coefficients can be found by considering the $k=L-2$ terms for $L\geq3$:
\begin{equation}
m_{L,L-1}=\frac{L}{2\left(L-1\right)}\beta_{1}m_{L-1,L-2}+\frac{1}{2}\beta_{2}m_{L-2,L-2}.
\end{equation}
The right-hand side has dependence on the leading and subleading coefficients. This recursion can be resummed to
\begin{equation}
\label{eq:O(N)RGSubLeadingM}
m_{L,L-1}=\frac{L}{2}\left(\frac{\beta_{1}}{2}\right)^{L-2}m_{2,1}+\left(\frac{\beta_{1}}{2}\right)^{L-3}\frac{L\beta_{2}}{2}m_{1,1}\left(H_{L}-\frac{3}{2}\right),
\end{equation}
where $H_{L}$ denotes the $L^{\mathrm{th}}$ harmonic number, e.g., $H_1 = 1$, $H_2 = 3/2$, $H_3 = 11/6$.

Although the large $N$ O$\left(N\right)$ model can be done to all orders, we will truncate our analysis at the subleading logarithmic order, since this is as far as we will go in the remaining sections of the paper. Finally, we extract expressions for the $\beta$ function coefficients by setting $k=0$ for $L\geq1$:
\begin{equation}
\beta_{L}=\frac{1}{m_{0,0}}\left[2m_{L,1}-\sum_{L^{\prime}=0}^{L-2}\left(L-L^{\prime}\right)\beta_{L^{\prime}+1}m_{L-L^{\prime}-1,0}\right].
\end{equation}
Through subleading order, we are interested in the first two terms:
\begin{subequations}
\begin{align}
\label{eq:O(N)beta1BC}
\beta_{1}&=2\frac{m_{1,1}}{m_{0,0}},\\
\label{eq:O(N)beta2BC}
\beta_{2}&=\frac{1}{m_{0,0}}\left(2m_{2,1}-2\beta_{1}m_{1,0}\right).
\end{align}
\end{subequations}
The RG recursion relations in \autoref{eq:O(N)RGLeadingM} and \autoref{eq:O(N)RGSubLeadingM} start at one higher order in the coupling $\lambda$ than those that we found from applying unitarity in \autoref{eq:O(N)UnitarityLeadingM} and \autoref{eq:O(N)UnitaritySubLeadingM}. Studying \autoref{eq:O(N)beta1BC} and \autoref{eq:O(N)beta2BC} reveals that the missing link between $m_{1,1}$ and $m_{0,0}$ in the leading case and between $m_{2,1}$ and $m_{1,0}$ in the subleading case is provided by $\beta_{1}$ and $\beta_{2}$. Said another way, as a result of starting the recursion relations at one higher order in the coupling than in unitarity, the RG recursions require additional input data that is given by the $\beta$ function coefficients. We can make this more obvious by substituting \autoref{eq:O(N)beta1BC} into \autoref{eq:O(N)RGLeadingM} to eliminate all dependence on $m_{1,1}$:
\begin{equation}
\label{eq:O(N)RGLeadingMFinal}
m_{L,L}=\frac{1}{2}m_{0,0}\beta_{1}\left(\frac{\beta_{1}}{2}\right)^{L-1},
\end{equation}
and substituting \autoref{eq:O(N)beta1BC} and \autoref{eq:O(N)beta2BC} into \autoref{eq:O(N)RGSubLeadingM} to eliminate all dependence on $m_{1,1}$ and $m_{2,1}$:
\begin{equation}
\label{eq:O(N)RGSubLeadingMFinal}
m_{L,L-1}=\frac{L}{2}\left(\frac{\beta_{1}}{2}\right)^{L-2}\left(\frac{1}{2}m_{0,0}\beta_{2}+\beta_{1}m_{1,0}\right)+\left(\frac{\beta_{1}}{2}\right)^{L-2}\frac{L\beta_{2}}{2}m_{0,0}\left(H_{L}-\frac{3}{2}\right).
\end{equation}
We now see that the recursive towers extend down to the same ground level and require the same input data, $m_{0, 0}$ (for leading) and $m_{1, 0}$ (for subleading), as in the unitarity case, but with the need for additional input data provided by $\beta_{1}$ for the leading case and $\beta_{1}$ and $\beta_{2}$ for the subleading case. This concept, which we will rediscover through subleading logarithmic order in the full $\lambda\phi^{4}$ theory in subsequent sections, is summarized in \autoref{fig:intro-table}.

We conclude this warm-up section by demonstrating that we can insert the recursion relations obtained from the unitarity analysis into the Callan--Symanzik equation and extract the numerical values of the $\beta$ function coefficients. The recursion tower that follows from unitarity for the leading amplitude coefficients was derived in \autoref{eq:O(N)UnitarityLeadingM}. In particular, this recursion gives that $m_{1,1}=1/2$. Substituting this value, along with the input datum $m_{0,0}$ which follows from demanding that the tree-level term is $-\lambda$, into \autoref{eq:O(N)beta1BC} gives
\begin{equation}
\beta_{1}=1.
\end{equation}
We then turn to the subleading logarithm recursion derived from unitarity in \autoref{eq:O(N)UnitaritySubLeadingM}. This constraint gives $m_{2,1}=m_{1,0}$, which we keep as an unknown, scheme-dependent input. Substituting this relation, along with the now known value of $\beta_{1}=1$ into \autoref{eq:O(N)beta2BC} returns
\begin{equation}
\beta_{2}=0.
\end{equation}
This matches with our knowledge of the O$\left(N\right)$ model in the large $N$ limit: it is leading logarithm-exact.

\section{Unitarity Constraints on \texorpdfstring{$\lambda \phi^{4}$}{lambda phi 4}}
\label{s:UnitarityconstraintsPhi4}
In the previous section, we demonstrated, in the large $N$ limit of the O($N$) model, that the tower of constraints on the amplitude coefficients $m_{L, k}$, where $L + 1$ is the order in coupling and $k$ is the logarithmic order, are equivalent to those obtained from the Callan--Symanzik equation through subleading log order, with the difference that the recursion from unitarity starts at one lower order in $L$. We now show that this equivalence extends to full $\lambda \phi^4$ theory. This introduces two technical complications not present in the large $N$ limit: 
\begin{enumerate}[label=(\arabic*),leftmargin=*]
\item We must include all three Mandelstam channels and higher-particle cuts. We will organize our unitarity computation by the number of cut internal particles, which efficiently encodes the hierarchy of logarithm powers. 
\item The propagator corrections to the tree-level 2-particle amplitude are no longer vanishing. In the CS equation, this means that we need to account for wavefunction renormalization. 
\end{enumerate}

Accounting for these changes, we write down new ans\"{a}tze for the 4-particle amplitude of a massless marginal theory with no IR divergences in the hard scattering limit: 
\begin{equation}
\M_4 = \sum_{L = 0}^{\infty} \frac{\left(-\lambda\right)^{L + 1}}{\left(16\pi^2\right)^L} \M_4^{\left(L\right)}.
\end{equation}
where
\begin{equation}
\label{eq:3ch4PtAnsatz}
\M_4^{\left(L\right)} = \sum_{0 \leq k_1 + k_2 + k_3 \leq L} m_{L, \left\{k_1, k_2, k_3\right\}} \log^{k_1}_{-s} \log^{k_2}_{-t} \log^{k_3}_{-u}.
\end{equation}
Here, $m_{L, \left\{k_1, k_2, k_3\right\}}$ indicates order $L + 1$ in the coupling and $k_i$ in logarithm. Since we demand that the tree level 4-particle amplitude is $-\lambda$, we have $m_{0, \left\{0, 0, 0\right\}} = 1$. By permutation symmetry, $m_{L, \left\{k_1, k_2, k_3\right\}}$ is completely symmetric in $\left\{k_1, k_2, k_3\right\}$. As before, we refer to the case where the combined powers of the logarithms is equal to $L$ as leading logarithms, with the associated coefficients called leading coefficients. Similarly, we refer to the case where the total powers of the logarithms is equal to $L - 1$ as subleading logarithms, with the associated coefficients called subleading coefficients. We also introduce the notation for the 6-particle amplitude, which will be needed to compute the 4-cuts,
\begin{equation}
\M_6 = \sum_{L = 0}^{\infty} \frac{\left(-\lambda\right)^{L + 2}}{\left(16\pi^2\right)^L} \M_6^{\left(L\right)},
\end{equation}
where we will specify $\M_6^{\left(L\right)}$ later. Lastly, the 2-particle amplitude is 
\begin{equation}
\M_2 = p^2 - \sum_{L = 1}^{\infty} \sum_{k = 0}^{L} \frac{\left(-\lambda\right)^L}{\left(16\pi^2\right)^L} \M_2^{\left(L\right)},
\end{equation}
where 
\begin{equation}
\label{eq:3ch2PtAnsatz}
\M_2^{\left(L\right)} = p^2 \sum_{k = 0}^{L} n_{L, k} \log^k_{p^2}, 
\end{equation} 
and $p^2 < 0$ is off shell. Here, $n_{L, k}$ is the coefficient at $L$ order in $\lambda$ and $k$ logarithmic order. The additional minus sign in the second term of $\M_2$ is conventional. We choose it to match to the typical definition $\M_2 \equiv p^2 - \Sigma$, where $\Sigma$ is the sum of all 1-particle irreducible diagrams. Hence $\M_2^{\left(L\right)}$ is equivalent to the $\Sigma^{\left(L\right)}$.

In this section, we will focus exclusively on the constraints obtained from unitarity, leaving the relationship to RG for the next section. Before delving into detailed calculations, let us sketch out how we will obtain the main results of this section: the recursion relations for the leading and subleading coefficients. First, for the 4-particle amplitude, we now include 4-cuts:
\begin{equation}
\label{eq:4ParticlePicture}
\vphantom{\rule{0pt}{1.75cm}} 
2\: \Im\left(\hspace{0.1cm}
\begin{gathered}
\begin{tikzpicture}[scale=0.6]
\draw[line width=0.3mm] (-1, 1) -- (-0.5*0.707, 0.5*0.707);
\draw[line width=0.3mm] (-1, -1) -- (-0.5*0.707, -0.5*0.707);
\filldraw[pattern=north west lines] (0, 0) circle (0.5cm);
\draw[line width=0.3mm] (0, 0) circle (0.5cm);
\draw[line width=0.3mm] (0.5*0.707, 0.5*0.707) -- (1, 1);
\draw[line width=0.3mm] (0.5*0.707, -0.5*0.707) -- (1, -1);
\end{tikzpicture}
\end{gathered}\hspace{0.1cm}\right) =
\begin{gathered}
\begin{tikzpicture}[scale=0.6, remember picture]
\draw[line width=0.3mm] (-1, 1) -- (-0.5*0.707, 0.5*0.707);
\draw[line width=0.3mm] (-1, -1) -- (-0.5*0.707, -0.5*0.707);
\filldraw[pattern = north west lines] (0, 0) circle (0.5cm);
\draw[line width=0.3mm] (0, 0) circle (0.5cm);
\draw[line width=0.3mm] (0.5*0.866, 0.5*0.5) -- (2 - 0.5*0.866, 0.5*0.5);
\draw[line width=0.3mm] (0.5*0.866, -0.5*0.5) -- (2 - 0.5*0.866, -0.5*0.5);
\filldraw[pattern=north west lines] (2, 0) circle (0.5cm);
\draw[line width=0.3mm] (2, 0) circle (0.5cm);
\draw[line width=0.3mm] (2 + 0.5*0.707, 0.5*0.707) -- (3, 1);
\draw[line width=0.3mm] (2 + 0.5*0.707, -0.5*0.707) -- (3, -1);
\filldraw[orange!10, opacity=0.5] (1, 1.1) rectangle (3.2, -1.1);
\draw[line width=0.5mm, dashed, orange] (1, 1.1) -- (1, -1.1);
\coordinate (brace1_top) at (2.75, -1.25);
\coordinate (brace1_left) at (-0.75, 1.55);
\coordinate (brace1_bottom) at (-0.75, -1.25);
\end{tikzpicture}
\end{gathered} \: \: + \: \:
\begin{gathered}
\begin{tikzpicture}[scale=0.6, remember picture]
\draw[line width=0.3mm] (-1, 1) -- (-0.5*0.707, 0.5*0.707);
\draw[line width=0.3mm] (-1, -1) -- (-0.5*0.707, -0.5*0.707);
\draw[pattern=north west lines] (0, 0) circle (0.5cm);
\draw[line width=0.3mm] (0, 0) circle (0.5cm);
\draw[line width=0.3mm] (0.5*0.866, 0.5*0.5) -- (2 - 0.5*0.866, 0.5*0.5);
\draw[line width=0.3mm] (0.5*0.985, 0.5*0.174) -- (2 - 0.5*0.985, 0.5*0.174);
\draw[line width=0.3mm] (0.5*0.985, -0.5*0.174) -- (2 - 0.5*0.985, -0.5*0.174);
\draw[line width=0.3mm] (0.5*0.866, -0.5*0.5) -- (2 - 0.5*0.866, -0.5*0.5);
\draw[pattern = north west lines] (2, 0) circle (0.5cm);
\draw[line width=0.3mm] (2, 0) circle (0.5cm);
\draw[line width=0.3mm] (2 + 0.5*0.707, 0.5*0.707) -- (3, 1);
\draw[line width=0.3mm] (2 + 0.5*0.707, -0.5*0.707) -- (3, -1);
\filldraw[orange!10, opacity=0.5] (1, 1.1) rectangle (3.2, -1.1);
\draw[line width=0.5mm, dashed, orange] (1, 1.1) -- (1, -1.1);
\coordinate (brace2_top) at (-0.75, 1.55);
\coordinate (brace2_bottom) at (2.75, 1.55);
\end{tikzpicture}
\end{gathered} \: \:+ \dots
\begin{tikzpicture}[overlay, remember picture]
\draw [decorate, decoration={brace, amplitude=6pt}] 
(brace1_top) -- (brace1_bottom) node [midway, yshift=-0.5cm] {Leading};
\draw [decorate, decoration={brace, amplitude=6pt}] (brace1_left) -- (brace2_bottom) node [midway, yshift=0.5cm] {Subleading};
\end{tikzpicture}
\end{equation}
\\
\\
\noindent Recall that by $\mathbb{Z}_2$ symmetry, the amplitudes on the left and right of the cut must be even point amplitudes, hence there are no three-particle cuts. As we will see, the 2-cuts alone are sufficient to constrain the leading coefficients. Then, both the 2- and 4-cuts will be sufficient to constrain the subleading coefficients. 

The inclusion of 4-cuts, however, means that we also need knowledge of the 6-particle amplitude. The imaginary part of the 6-particle amplitude can also be organized in terms of the number of internal lines being cut and so the leading logarithms of the 6-particle amplitude will come from 1-cuts: 
\begin{equation}
\label{eq:6ParticleFactorization}
\begin{aligned}
2\: \text{Im}\left(\hspace{0.1cm}
\begin{gathered}
\begin{tikzpicture}[scale=0.6]
\draw[line width=0.3mm] (-1, 1) -- (-0.5*0.707, 0.5*0.707);
\draw[line width=0.3mm] (-1, -1) -- (-0.5*0.707, -0.5*0.707);
\filldraw[pattern=north west lines] (0, 0) circle (0.5cm);
\draw[line width=0.3mm] (0, 0) circle (0.5cm);
\draw[line width=0.3mm] (0.5*0.707, 0.5*0.707) -- (1, 1);
\draw[line width=0.3mm] (0.5*0.707, -0.5*0.707) -- (1, -1);
\draw[line width=0.3mm] (0.5*0.97, 0.5*0.26) -- (0.92*0.75 + 0.5, 0.38*0.75);
\draw[line width=0.3mm] (0.5*0.97, -0.5*0.26) -- (0.92*0.75 + 0.5, -0.38*0.75);
\end{tikzpicture}
\end{gathered}
\hspace{0.1cm}\right)
&= \: \:
\begin{gathered}
\begin{tikzpicture}[scale=0.6]
\draw[line width=0.3mm] (-1, 1) -- (-0.5*0.707, 0.5*0.707);
\draw[line width=0.3mm] (-1, -1) -- (-0.5*0.707, -0.5*0.707);
\filldraw[pattern=north west lines] (0, 0) circle (0.5cm);
\draw[line width=0.3mm] (0, 0) circle (0.5cm);
\draw[line width=0.3mm] (0.5*0.707, 0.5*0.707) -- (3.25, 1.25);
\draw[line width=0.3mm] (0.5, 0) -- (1.5, 0);
\filldraw[pattern=north west lines] (2, 0) circle (0.5cm);
\draw[line width=0.3mm] (2, 0) circle (0.5cm);
\draw[line width=0.3mm] (2 + 0.5*0.707, 0.5*0.707) -- (3.25, 0.75);
\draw[line width=0.3mm] (2 + 0.5, 0) -- (3.25, 0);
\draw[line width=0.3mm] (2 + 0.5*0.707, -0.5*0.707) -- (3.25, -0.75);
\filldraw[orange!10, opacity=0.5] (1, 1.5) rectangle (3.5, -1.5);
\draw[line width=0.5mm, dashed, orange] (1, 1.5) -- (1, -1.5);
\end{tikzpicture}
\end{gathered} \: \: + \: \: 
\begin{gathered}
\begin{tikzpicture}[scale=0.6]
\draw[line width=0.3mm] (-1, 1) -- (-0.5*0.707, 0.5*0.707);
\draw[line width=0.3mm] (-1, -1) -- (-0.5*0.707, -0.5*0.707);
\filldraw[pattern=north west lines] (0, 0) circle (0.5cm);
\draw[line width=0.3mm] (0, 0) circle (0.5cm);
\draw[line width=0.3mm] (0.5*0.707, 0.5*0.707) -- (3.25, 0.75);
\draw[line width=0.3mm] (0.5, 0) -- (1.5, 0);
\filldraw[pattern=north west lines] (2, 0) circle (0.5cm);
\draw[line width=0.3mm] (2, 0) circle (0.5cm);
\draw[line width=1.3mm, white] (2 + 0.5*0.707, 0.5*0.707) -- (3.25, 1.25);
\draw[line width=0.3mm] (2 + 0.5*0.707, 0.5*0.707) -- (3.25, 1.25);
\draw[line width=0.3mm] (2 + 0.5, 0) -- (3.25, 0);
\draw[line width=0.3mm] (2 + 0.5*0.707, -0.5*0.707) -- (3.25, -0.75);
\filldraw[orange!10, opacity=0.5] (1, 1.5) rectangle (3.5, -1.5);
\draw[line width=0.5mm, dashed, orange] (1, 1.5) -- (1, -1.5);
\end{tikzpicture}
\end{gathered} \\
&\hspace{0.5cm} + \: \: 
\begin{gathered}
\begin{tikzpicture}[scale=0.6, yscale=-1]
\draw[line width=0.3mm] (-1, 1) -- (-0.5*0.707, 0.5*0.707);
\draw[line width=0.3mm] (-1, -1) -- (-0.5*0.707, -0.5*0.707);
\filldraw[pattern=north west lines] (0, 0) circle (0.5cm);
\draw[line width=0.3mm] (0, 0) circle (0.5cm);
\draw[line width=0.3mm] (0.5*0.707, 0.5*0.707) -- (3.25, 0.75);
\draw[line width=0.3mm] (0.5, 0) -- (1.5, 0);
\filldraw[pattern=north west lines] (2, 0) circle (0.5cm);
\draw[line width=0.3mm] (2, 0) circle (0.5cm);
\draw[line width=1.3mm, white] (2 + 0.5*0.707, 0.5*0.707) -- (3.25, 1.25);
\draw[line width=0.3mm] (2 + 0.5*0.707, 0.5*0.707) -- (3.25, 1.25);
\draw[line width=0.3mm] (2 + 0.5, 0) -- (3.25, 0);
\draw[line width=0.3mm] (2 + 0.5*0.707, -0.5*0.707) -- (3.25, -0.75);
\filldraw[orange!10, opacity=0.5] (1, 1.5) rectangle (3.5, -1.5);
\draw[line width=0.5mm, dashed, orange] (1, 1.5) -- (1, -1.5);
\end{tikzpicture}
\end{gathered} \: \: + \: \: 
\begin{gathered}
\begin{tikzpicture}[scale=0.6, yscale=-1]
\draw[line width=0.3mm] (-1, 1) -- (-0.5*0.707, 0.5*0.707);
\draw[line width=0.3mm] (-1, -1) -- (-0.5*0.707, -0.5*0.707);
\filldraw[pattern=north west lines] (0, 0) circle (0.5cm);
\draw[line width=0.3mm] (0, 0) circle (0.5cm);
\draw[line width=0.3mm] (0.5*0.707, 0.5*0.707) -- (3.25, 1.25);
\draw[line width=0.3mm] (0.5, 0) -- (1.5, 0);
\filldraw[pattern=north west lines] (2, 0) circle (0.5cm);
\draw[line width=0.3mm] (2, 0) circle (0.5cm);
\draw[line width=0.3mm] (2 + 0.5*0.707, 0.5*0.707) -- (3.25, 0.75);
\draw[line width=0.3mm] (2 + 0.5, 0) -- (3.25, 0);
\draw[line width=0.3mm] (2 + 0.5*0.707, -0.5*0.707) -- (3.25, -0.75);
\filldraw[orange!10, opacity=0.5] (1, 1.5) rectangle (3.5, -1.5);
\draw[line width=0.5mm, dashed, orange] (1, 1.5) -- (1, -1.5);
\end{tikzpicture}
\end{gathered} \: \:  + \text{perms} + \dots
\end{aligned}
\end{equation}
In the above, we show the first four permutations of external legs that contribute to the 6-particle amplitude out of the ten that exist. As we can see, the 6-particle amplitude factorizes into products of 4-particle amplitudes at leading logarithmic order and hence will close the recursion for the subleading coefficients in the 4-particle amplitude. Since each 6-particle amplitude contains ten terms from ten permutations of external legs, the 4-cut computation would very quickly become unwieldy. However, we will see in this section and \autoref{a:4ParticlePhaseSpace} that only \textit{four} terms will contribute to constraining the subleading coefficient, namely the product of each of the displayed diagrams with themselves, since these will be the only 4-cuts for whom the phase space integral itself contributes an additional power of log. We can therefore diagrammatically represent the relevant contributions to the 4-cut as: 
\begin{equation}
\label{eq:PropCorrectionCuts}
\begin{gathered}
\begin{tikzpicture}[scale=0.6]
\draw[line width=0.3mm] (-1, 1) -- (-0.5*0.707, 0.5*0.707);
\draw[line width=0.3mm] (-1, -1) -- (-0.5*0.707, -0.5*0.707);
\draw[pattern=north west lines] (0, 0) circle (0.5cm);
\draw[line width=0.3mm] (0, 0) circle (0.5cm);
\draw[line width=0.3mm] (0.5*0.866, 0.5*0.5) -- (2 - 0.5*0.866, 0.5*0.5);
\draw[line width=0.3mm] (0.5*0.985, 0.5*0.174) -- (2 - 0.5*0.985, 0.5*0.174);
\draw[line width=0.3mm] (0.5*0.985, -0.5*0.174) -- (2 - 0.5*0.985, -0.5*0.174);
\draw[line width=0.3mm] (0.5*0.866, -0.5*0.5) -- (2 - 0.5*0.866, -0.5*0.5);
\draw[pattern = north west lines] (2, 0) circle (0.5cm);
\draw[line width=0.3mm] (2, 0) circle (0.5cm);
\draw[line width=0.3mm] (2 + 0.5*0.707, 0.5*0.707) -- (3, 1);
\draw[line width=0.3mm] (2 + 0.5*0.707, -0.5*0.707) -- (3, -1);
\filldraw[orange!10, opacity=0.5] (1, 1.25) rectangle (3.2, -1.25);
\draw[line width=0.5mm, dashed, orange] (1, 1.25) -- (1, -1.25);
\end{tikzpicture}
\end{gathered}  \: \: = \: \: 4 \: \:
\begin{gathered}
\begin{tikzpicture}[scale=0.6]
\draw[line width=0.3mm] (-1, 1) -- (-0.5*0.707, 0.5*0.707);
\draw[line width=0.3mm] (-1, -1) -- (-0.5*0.707, -0.5*0.707);
\draw[pattern=north west lines] (0, 0) circle (0.5cm);
\draw[line width=0.3mm] (0, 0) circle (0.5cm);
\draw[line width=0.3mm] (0.5*0.707, 0.5*0.707) arc(135:45:3cm and 1.75cm);
\draw[line width=0.3mm] (0.5*0.707, -0.5*0.707) -- (1.25, -0.75);
\draw[pattern=north west lines] (1.75, -0.75) circle (0.5cm);
\draw[line width=0.3mm] (1.75, -0.75) circle (0.5cm);
\draw[line width=0.3mm] (1.75 + 0.5*0.707, -0.75 + 0.5*0.707) arc(135:45:0.52cm and 0.3cm); 
\draw[line width=0.3mm] (2.25, -0.75) -- (2.75, -0.75);
\draw[line width=0.3mm] (1.75 + 0.5*0.707, -0.75 - 0.5*0.707) arc(225:315:0.52cm and 0.3cm); 
\draw[pattern=north west lines] (3.25, -0.75) circle (0.5cm);
\draw[line width=0.3mm] (3.25, -0.75) circle (0.5cm);
\draw[line width=0.3mm] (3.75, -0.75) -- (5 - 0.5*0.707, -0.5*0.707);
\draw[pattern=north west lines] (5, 0) circle (0.5cm);
\draw[line width=0.3mm] (5, 0) circle (0.5cm);
\draw[line width=0.3mm] (5 + 0.5*0.707, 0.5*0.707) -- (6, 1);
\draw[line width=0.3mm] (5 + 0.5*0.707, -0.5*0.707) -- (6, -1);
\filldraw[orange!10, opacity=0.5] (2.5, 1.25) rectangle (6.2, -1.75);
\draw[line width=0.5mm, dashed, orange] (2.5, 1.25) -- (2.5, -1.75);
\end{tikzpicture}
\end{gathered} \: \: + \dots \: ,
\end{equation} 
where the ellipsis indicates other products of six-particle amplitudes that ultimately will not contribute to constraining the subleading coefficients of the 4-particle amplitude. This visually demonstrates that diagrams with contributing 4-cuts are propagator corrections in the traditional Feynman diagram approach. From the RG point of view, as we will see, these are included through wavefunction renormalization and the 2-particle amplitude whereas from the on-shell point of view, they arise from a subset of 6-particle amplitudes contributing to 4-cuts.

To complete the story, we will also study the 2-particle amplitude. Since this, by definition, only contains 1-particle irreducible diagrams, there are no 1-cuts. This immediately constrains the leading coefficients of the 2-particle amplitude to be zero. In addition, by $\mathbb{Z}_2$ symmetry, there are no 2-cuts. Therefore, the leading non-trivial contribution comes from 3-cuts, which, as indicated below, will constrain the subleading coefficients of the 2-particle amplitude:
\begin{equation}
\label{eq:2ParticlePicture}
2\: \text{Im}\left(
\begin{gathered}
\begin{tikzpicture}[scale=0.6]
\draw[line width=0.3mm] (-1.5, 0) -- (-0.5, 0);
\draw[pattern= north west lines] (0, 0) circle (0.5cm);
\draw[line width=0.3mm] (0, 0) circle (0.5cm); 
\draw[line width=0.3mm] (0.5, 0) -- (1.5, 0);
\end{tikzpicture}
\end{gathered}
\right) = 
\begin{gathered}
\vspace{-0.85cm}
\begin{tikzpicture}[scale=0.6]
\draw[line width=0.3mm] (-1.5, 0) -- (-0.5, 0);
\draw[pattern= north west lines] (0, 0) circle (0.5cm);
\draw[line width=0.3mm] (0, 0) circle (0.5cm);
\draw[line width=0.3mm] (0.5*0.866, 0.5*0.5) -- (2 - 0.5*0.866, 0.5*0.5);
\draw[line width=0.3mm] (0.5*0.866, -0.5*0.5) -- (2 - 0.5*0.866, -0.5*0.5);
\draw[pattern=north west lines] (2, 0) circle (0.5cm);
\draw[line width=0.3mm] (2, 0) circle (0.5cm);
\draw[line width=0.3mm] (0.5, 0) -- (1.5, 0);
\draw[line width=0.3mm] (2.5, 0) -- (3.5, 0);
\filldraw[orange!10, opacity=0.5] (1, 1) rectangle (3.7, -1);
\draw[line width=0.5mm, dashed, orange] (1, 1) -- (1, -1);
\draw[decorate, decoration={brace, amplitude=6pt}] (3, -1.15) -- node[anchor=east, xshift=1cm, yshift= -0.5cm] {Subleading} (-1, -1.15);
\end{tikzpicture}
\end{gathered} \: \: + \dots
\end{equation}
\\
\\
\noindent The right-hand side of this equation contains products of 4-particle amplitudes. Since we will have already computed the leading coefficients of the 4-particle amplitude, they can be used here to constrain the subleading coefficients of the 2-particle amplitude. The remainder of this section will focus on carrying through these computations to explicitly state the constraints on the leading and subleading coefficients of the 2- and 4-particle amplitudes. 

\subsection{Steinmann-Like Constraints}
\label{ss:Steinmann}
In gapped theories with stable asymptotic states, the double discontinuity in partially overlapping channels is zero when evaluated in physical kinematics. This result is known as the \textit{Steinmann relations} \cite{Steinmann1960a,Steinmann1960b}. If this were to hold for the case at hand, it would imply that for the 4-particle amplitude 
\begin{equation}
\label{eq:Steinmann}
\eval{\text{Disc}_{t} \, \text{Disc}_{s} \: \M_4}_{\text{physical}} = 0,
\end{equation}
and similarly so for double discontinuities in $s$ and $u$ as well as in $t$ and $u$. Here, physical kinematics corresponds to, say, $s > -t >  0$ in the $s$-channel and $t > -s > 0$ in the $t$-channel. Clearly, both cannot be satisfied at the same time, so the Steinmann relations cannot be applied directly to 4-particle amplitudes, see \cite{Britto:2024} for examples.

\begin{figure}
\begin{center}
\begin{tikzpicture}
\draw[line width=0.3mm] (-1, 1) -- (-0.5*0.707, 0.5*0.707);
\draw[line width=0.3mm, -Latex] (-1, 0.8) -- node[anchor=east, xshift=-0.05cm] {$p_2$} (-0.5, 0.3);
\draw[line width=0.3mm] (-1, -1) -- (-0.5*0.707, -0.5*0.707);
\draw[line width=0.3mm] (0, 0) circle (0.5cm);
\filldraw[pattern=north west lines] (0, 0) circle (0.5cm);
\draw[line width=0.3mm] (0.5*0.866, 0.5*0.5) -- (2 - 0.5*0.866, 0.5*0.5);
\draw[line width=0.3mm] (0.5*0.866, -0.5*0.5) -- (2 - 0.5*0.866, -0.5*0.5);
\draw[line width=0.3mm] (2, 0) circle (0.5cm);
\filldraw[pattern=north west lines] (2, 0) circle (0.5cm);
\draw[line width=0.3mm] (2 + 0.5*0.707, 0.5*0.707) -- (3, 1);
\draw[line width=0.3mm, -Latex] (2.5, 0.3) -- node[anchor=west, xshift=0.15cm] {$p_3$} (3, 0.8);
\draw[line width=0.3mm] (2 + 0.5*0.707, -0.5*0.707) -- (3, -1);
\draw[line width=0.3mm, -Latex] (0.6, 0.1) -- (1.4, 0.1);
\node at (0.75, 0.5) {$\ell_1$};
\draw[line width=0.5mm, dashed, orange] (1, 1) -- (1, -1);
\coordinate (L1) at (0, 0.2);
\coordinate (L2) at (0.85, 0.25);
\coordinate (L3) at (-0.6, 0.6);
\coordinate (R1) at (2, 0.2);
\coordinate (R2) at (1.15, 0.25);
\coordinate (R3) at (2.6, 0.6);
\pic[draw, "$\theta_L$", angle eccentricity =1.5, line width=0.2mm] {angle = L2--L1--L3};
\pic[draw, "$\theta_R$", angle eccentricity =1.5, line width=0.2mm] {angle = R3--R1--R2};
\end{tikzpicture}
\end{center}
\caption{Left and right angles. This is the 2-cut for a 4-particle amplitude. The kinematic invariants for the left and right amplitude can be expressed in terms of two angles, $\theta_L$ and $\theta_R$, which have the geometric interpretation of being the angle between $\vb{p}_2$ and $\vb*{{\ell}}_1$ and $\vb{p}_3$ and $\vb*{{\ell}}_1$ respectively.}
\label{fig:AngleFig}
\end{figure}
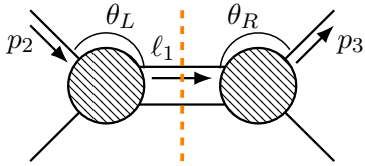

Regardless, we will show that a version of Steinmann relations holds up to subleading logarithmic order as a result of unitarity. More specifically, we will show that mixed channel logarithm terms, such as $\log_{-s} \log_{-t}$ are not present. To demonstrate this, note that the imaginary part of the 4-particle amplitude at order $L$, shown in \autoref{eq:3ch4PtAnsatz}, is
\begin{equation}
\Im\left(\M_4^{\left(L\right)}\right) = \sum_{k_1 + k_2 + k_3 \leq L} \pi k_1 m_{L, \left\{k_1, k_2, k_3\right\}} \log^{k_1}_{-s} \log^{k_2}_{-t} \log^{k_3}_{-u}. 
\end{equation}
This must match onto the contributions from 2- and 4-cuts on the right-hand side of the unitarity equation. The $L$ order contribution to the 2-cuts is
\begin{equation}
\eval{\Im\left(\M_4^{\left(L\right)}\left(s, t\right)\right)}_{2-\text{cuts}} = \frac{16\pi^2}{2} \sum_{L' = 0}^{L - 1} \M_4^{\left(L'\right)}\left(s, t_L\right) \left(\M_4^{\left(L - L' - 1\right)}\left(s, t_R\right)\right)^\ast \dd{\Phi}_2. 
\end{equation}
Here, we have explicitly included the kinematic dependence on the right-hand side. The Mandelstam variables for the amplitude on the left side of the cut are 
\begin{equation}
\label{eq:LeftMandelstam}
t_L = - s \sin^2\left(\frac{\theta_L}{2}\right) \qquad \text{and} \qquad u_L = - s \cos^2\left(\frac{\theta_L}{2}\right),
\end{equation}
where $\theta_L$ is the angle between $p_2$ and $\ell_1$, as shown in \autoref{fig:AngleFig}. This is similarly true for the Mandelstam variables of the amplitude on the right side of the cut, 
\begin{equation}
\label{eq:RightMandelstam}
t_R = - s\sin^2\left(\frac{\theta_R}{2}\right) \qquad \text{and} \qquad u_R = -s \cos^2\left(\frac{\theta_R}{2}\right),
\end{equation}
except the angle, $\theta_R$, is between $\ell_1$ and $p_3$, also shown in \autoref{fig:AngleFig}. Since the 2-particle phase space itself only depends on $p_1$ and $p_2$, there will be no dependence on $t$ and $u$ from 2-cuts even after integration.

While this will become explicit in the computation of the 4-cut, the logic from above will hold for cuts of the form in \autoref{eq:PropCorrectionCuts} since the amplitude on the left and right of the cut, as well as the 4-particle phase space integration measure itself, will only depend on $s$ and not on $t$ nor $u$.

As a result, the coefficients of logarithms involving $t$ and $u$ on the left-hand side of the unitarity equation must be zero. In particular, this implies that 
\begin{equation}
m_{L, \left\{k, 0, 0\right\}} = m_{L, \left\{0, k, 0\right\}} = m_{L, \left\{0, 0, k\right\}}
\end{equation}
are the only non-zero leading ($k = L$) and subleading ($k = L - 1$) coefficients. Since the remaining coefficients are zero, this implies that the double discontinuity in partially overlapping channels is zero, which is reminiscent of the Steinmann relations for gapped theories, \autoref{eq:Steinmann}. We verified that this was true through three-loop order in \autoref{a:ExactResults}, even for sub-subleading logarithms. We moreover demonstrated that this holds for leading and subleading logarithms at four-loop order.

The practical implication of these Steinmann-like relations is that the ansatz for the 4-particle amplitude simplifies to
\begin{equation}
\label{eq:3ch4PtAnsatzSimplified}
\M_4^{\left(L\right)} = \sum_{k = 0}^{L} m_{L, k} \left[\log^k_{-s} + \log^k_{-t} + \log^k_{-u}\right].
\end{equation}
We will use this ansatz instead of \autoref{eq:3ch4PtAnsatz} in the below computations.

\subsection{4-Particle Leading Log Constraints}
As described at the start of this section, the 2-cuts will be sufficient to constrain the leading coefficients of the 4-particle amplitudes. Hence, to leading logarithmic order, we have
\begin{equation}
2\: \Im\left(\hspace{0.1cm}
\begin{gathered}
\begin{tikzpicture}[scale=0.6]
\draw[line width=0.3mm] (-1, 1) -- (-0.5*0.707, 0.5*0.707);
\draw[line width=0.3mm] (-1, -1) -- (-0.5*0.707, -0.5*0.707);
\filldraw[pattern=north west lines] (0, 0) circle (0.5cm);
\draw[line width=0.3mm] (0, 0) circle (0.5cm);
\draw[line width=0.3mm] (0.5*0.707, 0.5*0.707) -- (1, 1);
\draw[line width=0.3mm] (0.5*0.707, -0.5*0.707) -- (1, -1);
\end{tikzpicture}
\end{gathered}\hspace{0.1cm}\right) =
\begin{gathered}
\begin{tikzpicture}[scale=0.6]
\draw[line width=0.3mm] (-1, 1) -- (-0.5*0.707, 0.5*0.707);
\draw[line width=0.3mm] (-1, -1) -- (-0.5*0.707, -0.5*0.707);
\filldraw[pattern = north west lines] (0, 0) circle (0.5cm);
\draw[line width=0.3mm] (0, 0) circle (0.5cm);
\draw[line width=0.3mm] (0.5*0.866, 0.5*0.5) -- (2 - 0.5*0.866, 0.5*0.5);
\draw[line width=0.3mm] (0.5*0.866, -0.5*0.5) -- (2 - 0.5*0.866, -0.5*0.5);
\filldraw[pattern=north west lines] (2, 0) circle (0.5cm);
\draw[line width=0.3mm] (2, 0) circle (0.5cm);
\draw[line width=0.3mm] (2 + 0.5*0.707, 0.5*0.707) -- (3, 1);
\draw[line width=0.3mm] (2 + 0.5*0.707, -0.5*0.707) -- (3, -1);
\filldraw[orange!10, opacity=0.5] (1, 1.1) rectangle (3.2, -1.1);
\draw[line width=0.5mm, dashed, orange] (1, 1.1) -- (1, -1.1);
\end{tikzpicture}
\end{gathered} \: \: + \dots
\end{equation}
We can match powers of $\lambda$ in the unitarity equation, \autoref{eq:UnitarityEquation}, to find
\begin{equation}
\Im\left(\M_4^{\left(L\right)}\right) = \frac{16\pi^2}{2} \sum_{L^\prime = 0}^{L - 1}\int \M_4^{\left(L^\prime\right)} \left(\M^{\left(L - L^\prime - 1\right)}_4\right)^{\ast} \dd{\Phi}_2 + \dots \: ,
\end{equation}
where the ellipsis contains higher-particle cut contributions. With our simplified ansatz, \autoref{eq:3ch4PtAnsatzSimplified}, the leading logarithm contribution to the imaginary part is
\begin{equation}
\text{Im}\left(\M^{\left(L, L\right)}\right) = L \pi m_{L, L} \log^{L - 1}_s,
\end{equation}
where the superscript indicates $L + 1$ order in $\lambda$ and $L$ order in logarithm, i.e. the leading logarithmic contribution. This follows from expanding the logarithm as in \autoref{eq:ExpansionInLogs}. On the cut side of the calculation, we can similarly expand the logarithm using \autoref{eq:LeftMandelstam} and \autoref{eq:RightMandelstam}: 
\begin{subequations}
\begin{align}
\log^L_{-t_L} &= \left(\log_s + \log\left(\frac{1}{\sin^2\left(\theta_L/2\right)}\right)\right)^L,\\
&= \log^L_s + L\log\left(\frac{1}{\sin^2\left(\theta_L/2\right)}\right) \log^{L - 1}_s + \dots
\label{eq:LogExpansionIntegrand}
\end{align}
\end{subequations}
We have included the subleading logarithm for future convenience. Plugging the leading contribution into the 2-cut and simplifying, we obtain a recursion relation
\begin{equation}
\eval{\text{Im}\left(\M^{\left(L, L\right)}\right)}_{2-\text{cuts}} = \frac{16\pi^2}{2} \sum_{L^\prime = 0}^{L - 1} 9 m_{L^\prime, L^\prime} m_{L - L^\prime - 1, L - L^\prime - 1} \log^{L - 1}_s \Phi_2. 
\end{equation}
Using the 2-particle phase space result in \autoref{eq:2particlePhaseSpace}, we find that
\begin{equation}
m_{L, L} = \frac{9}{2L} \sum_{L^\prime = 0}^{L - 1} m_{L^\prime, L^\prime} m_{L - L^\prime - 1, L - L^\prime - 1}.
\end{equation}
Solving the recursion relation subject to the input data, $m_{0, 0} = 1/3$ (as fixed by the tree-level amplitude $-\lambda$), gives
\begin{empheq}[box=\sentencebox]{equation}
\label{eq:mLeadingLogUnitarity}
m_{L, L} = \frac{1}{3}\left(\frac{3}{2}\right)^L.
\end{empheq}
The first few terms are displayed in \autoref{fig:results}.

Having determined the coefficients of the leading 4-particle logarithms, we are now able to constrain the form of the leading non-trivial logarithms for the 2- and 6-particle amplitudes. The latter will be useful for computing the subleading coefficients of the 4-particle amplitude, while the former will be useful for matching to RG.

\subsection{4-Particle Subleading Log Constraints}
We now push our computation to subleading logarithmic order, where both 2-cuts and 4-cuts contribute:
\begin{equation}
2\: \Im\left(\hspace{0.1cm}
\begin{gathered}
\begin{tikzpicture}[scale=0.6]
\draw[line width=0.3mm] (-1, 1) -- (-0.5*0.707, 0.5*0.707);
\draw[line width=0.3mm] (-1, -1) -- (-0.5*0.707, -0.5*0.707);
\filldraw[pattern=north west lines] (0, 0) circle (0.5cm);
\draw[line width=0.3mm] (0, 0) circle (0.5cm);
\draw[line width=0.3mm] (0.5*0.707, 0.5*0.707) -- (1, 1);
\draw[line width=0.3mm] (0.5*0.707, -0.5*0.707) -- (1, -1);
\end{tikzpicture}
\end{gathered}\hspace{0.1cm}\right) =
\begin{gathered}
\begin{tikzpicture}[scale=0.6]
\draw[line width=0.3mm] (-1, 1) -- (-0.5*0.707, 0.5*0.707);
\draw[line width=0.3mm] (-1, -1) -- (-0.5*0.707, -0.5*0.707);
\filldraw[pattern = north west lines] (0, 0) circle (0.5cm);
\draw[line width=0.3mm] (0, 0) circle (0.5cm);
\draw[line width=0.3mm] (0.5*0.866, 0.5*0.5) -- (2 - 0.5*0.866, 0.5*0.5);
\draw[line width=0.3mm] (0.5*0.866, -0.5*0.5) -- (2 - 0.5*0.866, -0.5*0.5);
\filldraw[pattern=north west lines] (2, 0) circle (0.5cm);
\draw[line width=0.3mm] (2, 0) circle (0.5cm);
\draw[line width=0.3mm] (2 + 0.5*0.707, 0.5*0.707) -- (3, 1);
\draw[line width=0.3mm] (2 + 0.5*0.707, -0.5*0.707) -- (3, -1);
\filldraw[orange!10, opacity=0.5] (1, 1.1) rectangle (3.2, -1.1);
\draw[line width=0.5mm, dashed, orange] (1, 1.1) -- (1, -1.1);
\end{tikzpicture}
\end{gathered} \: \: + \: \:
\begin{gathered}
\begin{tikzpicture}[scale=0.6]
\draw[line width=0.3mm] (-1, 1) -- (-0.5*0.707, 0.5*0.707);
\draw[line width=0.3mm] (-1, -1) -- (-0.5*0.707, -0.5*0.707);
\draw[pattern=north west lines] (0, 0) circle (0.5cm);
\draw[line width=0.3mm] (0, 0) circle (0.5cm);
\draw[line width=0.3mm] (0.5*0.866, 0.5*0.5) -- (2 - 0.5*0.866, 0.5*0.5);
\draw[line width=0.3mm] (0.5*0.985, 0.5*0.174) -- (2 - 0.5*0.985, 0.5*0.174);
\draw[line width=0.3mm] (0.5*0.985, -0.5*0.174) -- (2 - 0.5*0.985, -0.5*0.174);
\draw[line width=0.3mm] (0.5*0.866, -0.5*0.5) -- (2 - 0.5*0.866, -0.5*0.5);
\draw[pattern = north west lines] (2, 0) circle (0.5cm);
\draw[line width=0.3mm] (2, 0) circle (0.5cm);
\draw[line width=0.3mm] (2 + 0.5*0.707, 0.5*0.707) -- (3, 1);
\draw[line width=0.3mm] (2 + 0.5*0.707, -0.5*0.707) -- (3, -1);
\filldraw[orange!10, opacity=0.5] (1, 1.1) rectangle (3.2, -1.1);
\draw[line width=0.5mm, dashed, orange] (1, 1.1) -- (1, -1.1);
\end{tikzpicture}
\end{gathered} \: \: \dots
\end{equation}
Matching powers of $\lambda$ in the unitarity equation, \autoref{eq:UnitarityEquation}, gives
\begin{equation}
\begin{aligned}
\Im\left(\M_4^{\left(L\right)}\right) &= \frac{16\pi^2}{2}\sum_{L^\prime = 0}^{L - 1} \int \M_4^{\left(L^\prime\right)} \left(\M_4^{\left(L - L^\prime - 1\right)}\right)^{\ast} \dd{\Phi}_2 \\
&\hspace{0.5cm} + \frac{\left(16\pi^2\right)^3}{2} \sum_{L^\prime = 0}^{L - 3} \int \M_6^{\left(L^\prime\right)} \left(\M_6^{\left(L - L^\prime - 1\right)}\right)^{\ast} \dd{\Phi}_4 + \dots
\end{aligned}
\end{equation}
In the 2-cut contribution, there are two possible terms at subleading order in logarithm. The first comes from keeping the subleading logarithm in \autoref{eq:LogExpansionIntegrand}, but multiplying the leading logarithms of both the amplitudes on the left and right of the cut. The second comes from multiplying the leading logarithm of the amplitude on the left of the cut and the subleading logarithm of the amplitude on the right of the cut and vice versa. Similar analysis for the 4-cut, however, would say that multiplying the leading logarithms of the two 6-particle amplitudes would give a sub-subleading contribution to the imaginary part. As is shown in \autoref{a:4ParticlePhaseSpace}, the four-particle cut integral will itself sometimes contribute an additional logarithmic power, called a \textit{log enhancement}, ensuring that this will still have some contribution to constraining subleading coefficients.

The left-hand side of the unitarity equation at sub-leading logarithmic order is
\begin{equation}
\Im\left(\M_4^{\left(L, L - 1\right)}\right) = \left(L - 1\right)\pi m_{L, L - 1} \log^{L - 2}_s. 
\end{equation}
This follows, as before, from expanding the logarithm as in \autoref{eq:ExpansionInLogs}. 

The two contributions to the imaginary part at subleading order from 2-cuts can be expressed as 
\begin{equation}
\begin{aligned}
\eval{\Im\left(\M_4^{\left(L, L - 1\right)}\right)}_{2-\text{cuts}} &= \frac{16\pi^2}{2} \sum_{L^\prime = 0}^{L - 1} 3m_{L^\prime, L^\prime} m_{L - L^\prime - 1, L - L^\prime - 1} \log^{L - 2}_s \\
&\hspace{1cm} \times 2L^\prime \int \left(\log\left(\frac{1}{\sin^2\left(\theta/2\right)}\right) + \log\left(\frac{1}{\cos^2\left(\theta/2\right)}\right) \right) \dd{\Phi}_2 \\
&\hspace{0.4cm} + \frac{16\pi^2}{2} \sum_{L^\prime = 0}^{L - 1} 18 m_{L^\prime, L^\prime - 1} m_{L - L^\prime - 1, L - L^\prime - 1} \log^{L - 2}_s \Phi_2. 
\end{aligned}
\end{equation}
In the first two lines, we simplified the integral by first noting that the sum is invariant under the variable change $L' \rightarrow L - L^\prime - 1$ to combine multiple terms. We additionally noted that since the angle $\theta$ is being integrated over, we are free to choose the azimuthal angle term-by-term to be $\theta_L$ or $\theta_R$, meaning each of the angular integrals contribute the same result. In the second line, there are two contributions which can be combined once again using the judicious interchange of $L^\prime$ with $L - L^\prime - 1$. Using \autoref{eq:logsinint} and \autoref{eq:logcosint}, we find
\begin{equation}
\begin{aligned}
\eval{\Im\left(\M_4^{\left(L, L - 1\right)}\right)}_{2-\text{cuts}} &= 6\pi \log^{L - 2}_s \sum_{L^\prime = 0}^{L - 1} L^\prime m_{L^\prime, L^\prime} m_{L - L^\prime - 1, L - L^\prime - 1} \\
&\hspace{0.4cm} + 9\pi \log^{L - 2}_s \sum_{L^\prime = 0}^{L - 1} m_{L^\prime, L^\prime - 1} m_{L - L^\prime - 1, L - L^\prime - 1}.
\end{aligned}
\end{equation}
Using the leading log result, \autoref{eq:mLeadingLogUnitarity}, this simplifies to
\begin{equation}
\label{eq:SL2cutContribution}
\eval{\Im\left(\M_4^{\left(L, L - 1\right)}\right)}_{2-\text{cuts}} = \left(\frac{3}{2}\right)^L 2\pi \left[\frac{L\left(L - 1\right)}{9} + \sum_{L^\prime = 1}^{L - 1} \left(\frac{2}{3}\right)^{L^\prime} m_{L^\prime, L^\prime - 1}\right] \log^{L - 2}_s.
\end{equation}
Note that in the last term, we start the sum from $L^\prime = 1$ since when $L^\prime = 0$, the sum involves $m_{0, -1} = 0$. 

Now we will turn our attention to the 4-cuts. At leading order, as shown in \autoref{eq:6ParticleFactorization}, the 6-particle amplitude factorizes into a product of 4-particle amplitudes, with explicit momentum dependence shown below:
\begin{equation}
\M_6 = \: \ \begin{gathered}
\begin{tikzpicture}[scale=0.6]
\draw[line width=0.3mm] (-1, 1) -- (-0.5*0.707, 0.5*0.707);
\draw[line width=0.3mm] (-1, -1) -- (-0.5*0.707, -0.5*0.707);
\filldraw[pattern=north west lines] (0, 0) circle (0.5cm);
\draw[line width=0.3mm] (0, 0) circle (0.5cm);
\draw[line width=0.3mm] (0.5*0.707, 0.5*0.707) -- (3.25, 1.25);
\draw[line width=0.3mm] (0.5, 0) -- (1.5, 0);
\filldraw[pattern=north west lines] (2, 0) circle (0.5cm);
\draw[line width=0.3mm] (2, 0) circle (0.5cm);
\draw[line width=0.3mm] (2 + 0.5*0.707, 0.5*0.707) -- (3.25, 0.75);
\draw[line width=0.3mm] (2 + 0.5, 0) -- (3.25, 0);
\draw[line width=0.3mm] (2 + 0.5*0.707, -0.5*0.707) -- (3.25, -0.75);
\node at (-1.4, -1) {$p_1$};
\node at (-1.4, 1) {$p_2$};
\node at (3.6, 1.4) {$\ell_1$};
\node at (3.6, 0.75) {$\ell_2$};
\node at (3.6, 0) {$\ell_3$};
\node at (3.6, -0.75) {$\ell_4$};
\end{tikzpicture}
\end{gathered} \: \: + \: \text{perms} + \ldots\, ,
\end{equation}
where there are nine additional permutations of external legs. In equations,
\begin{equation}
\M_6^{\left(L\right)} = \sum_{L^\prime = 0}^{L} \M_4^{\left(L - L^\prime\right)} \frac{-i}{\left(p_1 + p_2 - \ell_1\right)^2 - i\varepsilon} \left(\M_4^{\left(L^\prime\right)}\right)^{\ast} + \: \text{perms} + \ldots
\end{equation} 
We want the eventual contribution from this 6-particle amplitude to the subleading logarithm of the 4-particle amplitude. This will only depend on the leading logarithms of the constituent left and right 4-particle amplitudes: 
\begin{subequations}
\begin{align}
\M_4^{\left(L - L^\prime\right)} &= \frac{1}{3}\left(\frac{3}{2}\right)^{L - L^\prime} \left(\log^{L - L^\prime}_{\left(p_1 + p_2\right)^2} + \left(p_2 \leftrightarrow \ell_1\right) + \left(p_1 \leftrightarrow \ell_1\right) \right),\\
\M_4^{\left(L^\prime\right)} &= \frac{1}{3}\left(\frac{3}{2}\right)^{L^\prime} \left(\log^{L^\prime}_{\left(\ell_2 + \ell_3\right)^2} + \left(\ell_3\leftrightarrow \ell_4\right) + \left(\ell_2 \leftrightarrow \ell_4\right)\right).
\end{align}
\end{subequations}
Using the same expansion in \autoref{eq:LogExpansionIntegrand}, the left amplitude simplifies to 
\begin{equation}
\M_4^{\left(L - L^\prime\right)} = \left(\frac{3}{2}\right)^{L - L^\prime} \log^{L - L^\prime}_{s}.
\end{equation}
This argument does not apply to the right amplitude, so it will remain unchanged. Therefore, the general 6-particle amplitude we will consider is 
\begin{equation}
\label{eq:6-particle factorization}
\begin{aligned}
\M_6^{\left(L\right)} &= \! \sum_{L^\prime = 0}^{L} \! \frac{1}{3} \! \left(\frac{3}{2}\right)^L \! \log^{L - L^\prime}_s \! \frac{-i}{\left(p_1 \! + \! p_2 \! - \! \ell_1\right)^2 \! - \! i\varepsilon} \! \left(\log^{L^\prime}_{\left(\ell_2 + \ell_3\right)^2} + \left(\ell_3\! \leftrightarrow \! \ell_4\right) + \left(\ell_2 \! \leftrightarrow\! \ell_4\right)\right) + \text{perms}.
\end{aligned}
\end{equation} 

As mentioned before, when plugging this result into the 4-cut, we notice that the only way to obtain a logarithm of order $L - 2$ is if the phase space integral itself contributes an additional logarithm. Of the possible products of 6-particle amplitude, there is only one type of phase space integral that contributes an enhancement in the power of logarithm: it is when the propagators are the same and depend on only a single loop momentum (see \autoref{a:4ParticlePhaseSpace}). In particular, this means that we get four equivalent contributions (for each of the four permutations shown in \autoref{eq:6ParticleFactorization}), which combine into:
\begin{equation}
\eval{\Im\left(\M_4^{\left(L, L - 1\right)}\right)}_{4-\text{cuts}} = \frac{16}{81} \sum_{L' = 0}^{L - 3} \sum_{L_1 = 0}^{L^\prime} \sum_{L_2 = 0}^{L - L^\prime - 3} \left(\frac{3}{2}\right)^L \log^{L - L_1 - L_2 - 3}_s \left(I_1 + 2 I_2\right),
\end{equation}
where 
\begin{subequations}
\begin{align}
I_1 &= \left(16\pi^2\right)^3\int \log^{L_1 + L_2}_{\left(\ell_3 + \ell_4\right)^2} \dd{\Phi}_4,\\
I_2 &= \left(16\pi^2\right)^3\int \log^{L_1}_{\left(\ell_2 + \ell_3\right)^2} \log^{L_2}_{\left(\ell_3 + \ell_4\right)^2} \dd{\Phi}_4.
\end{align}
\end{subequations}
In the expressions above, it was helpful to note that in the phase space integral, one can relabel $\ell_1 \dots \ell_4$ to show that many of the integrals are the same. In fact, as shown in \autoref{eq:4ParticleLogEnhancement}, $I_1$ and $I_2$ also turn out to evaluate to the same result at leading logarithmic order,
\begin{equation}
I_1 = I_2 = -\frac{\pi}{24\left(L_1 + L_2 + 1\right)} \log^{L_1 + L_2 + 1}_s + \dots 
\end{equation}
As a result, 
\begin{equation}
\eval{\Im\left(\M_4^{\left(L, L - 1\right)}\right)}_{4-\text{cuts}} = -\frac{2\pi}{81} \left(\frac{3}{2}\right)^L \log^{L - 2}_s \sum_{L^\prime = 0}^{L - 3} \sum_{L_1 = 0}^{L^\prime} \sum_{L_2 = 0}^{L - L^\prime - 3}\frac{1}{L_1 + L_2 + 1}.
\end{equation}
Evaluating the triple sum, we find 
\begin{equation}
\label{eq:SL4cutContribution}
\eval{\Im\left(\M_4^{\left(L, L - 1\right)}\right)}_{4-\text{cuts}} = -\frac{\pi}{81} \left(\frac{3}{2}\right)^L \left(L - 1\right)\left(L - 2\right) \log^{L - 2}_s.
\end{equation}

\begin{figure}
\begin{tikzpicture}[remember picture]
\matrix (m) [
        matrix of math nodes, 
        nodes={text width=1cm, minimum height=1cm, anchor=center, align=center}, 
        row sep=-\pgflinewidth, 
        column sep=-\pgflinewidth,  
        label={[font=\bfseries]above: {\qquad\qquad Coefficients $m_{L,k}$}}
    ]
    {
        |[text width=1.3cm]| {\bf L=0} & \tfrac{1}{3} &  &  &  & \\
        |[text width=1.3cm]| {\bf L=1} & |[fill=orange!20]| \tfrac{1}{2}  &  m_{1,0} &  &  & \\
        |[text width=1.3cm]| {\bf L=2} & |[fill=orange!20]| \tfrac{3}{4} & |[fill=orange!20,text width=1.55cm]| { 1 {+} 3 m_{1,0}} &  m_{2,0} & \\
        |[text width=1.3cm]| {\bf L=3} & |[fill=orange!20]| \tfrac{9}{8} & |[fill=orange!20,text width=1.55cm]| \tfrac{89 + 162 m_{1,0}}{24}  & |[fill=gray!10]| m_{3,1} & m_{3,0} &  \\
        |[text width=1.3cm]|{\bf \vdots} & |[fill=orange!20]| \vdots & |[fill=orange!20,text width=1.55cm]| \vdots & |[fill=gray!10]| \vdots & |[fill=gray!10]| \vdots & \ddots  \\
        |[text width=1.3cm]| & \substack{{\bf k=L} \\ \autoref{eq:mLeadingLogUnitarity}}   &  \substack{{\bf k=L{-}1}\\ \autoref{eq:mSubLeadingUnitarity}} &  \dots &  &  \\
    };
    \draw [<-, orange!80!black, shorten <=-4pt, shorten >=-4pt] (m-1-2) -- (m-2-2);
    \draw [<-, orange!80!black, shorten <=-4pt, shorten >=-4pt] (m-2-2) -- (m-3-2);
    \draw [<-, orange!80!black, shorten <=-4pt, shorten >=-4pt] (m-3-2) -- (m-4-2);
    \draw [<-, orange!80!black, shorten <=-4pt, shorten >=-4pt] (m-4-2) -- (m-5-2);
    \draw [<-, orange!80!black, shorten <=-4pt, shorten >=-4pt] (m-2-3) -- (m-3-3);
    \draw [<-, orange!80!black, shorten <=-4pt, shorten >=-4pt] (m-3-3) -- (m-4-3);
    \draw [<-, orange!80!black, shorten <=-4pt, shorten >=-4pt] (m-4-3) -- (m-5-3);
    \draw [<-, orange!80!black, shorten <=-4pt, shorten >=-4pt] (m-3-3) -- (m-2-2);
    \draw [<-, gray, shorten <=-4pt, shorten >=-4pt] (m-3-4) -- (m-4-4);
    \draw [<-, gray, shorten <=-4pt, shorten >=-4pt] (m-4-4) -- (m-5-4);
    \draw [<-, gray, shorten <=-4pt, shorten >=-4pt] (m-4-4) -- (m-3-3);
    \draw [<-, gray, shorten <=-4pt, shorten >=-4pt] (m-4-5) -- (m-5-5);
    \draw [<-, gray, shorten <=-4pt, shorten >=-4pt] (m-4-4) -- (m-5-5);
\end{tikzpicture}
\begin{tikzpicture}[remember picture]
\matrix (n) [
        matrix of math nodes, 
        nodes={text width=1cm, minimum height=1cm, anchor=center, align=center}, 
        row sep=-\pgflinewidth, 
        column sep=-\pgflinewidth,  
        label={[font=\bfseries]above: {Coefficients $n_{L,k}$}}
    ]
    {
         0 &  &  &  & \\
        |[fill=orange!20]| 0  & 0 &  &  & \\
        |[fill=orange!20]| 0 & |[fill=orange!20,text width=1.55cm]| -\tfrac{1}{12} &  n_{2,0} & \\
        |[fill=orange!20]| 0 & |[fill=orange!20,text width=1.55cm]| -\tfrac{1}{8}  & |[fill=gray!10]| n_{3,1} & n_{3,0} &  \\
        |[fill=orange!20]| \vdots & |[fill=orange!20,text width=1.55cm]| \vdots & |[fill=gray!10]| \vdots & |[fill=gray!10]| \vdots & \ddots  \\
        \substack{{\bf k=L} \\ \autoref{eq:nLeadingUnitarity}}   &  \substack{{\bf k=L{-}1}\\ \autoref{eq:nSubleadingUnitarity}} &  \dots &  &  \\
    };
    \draw [<-, orange!80!black, shorten <=-4pt, shorten >=-4pt] (n-1-1) -- (n-2-1);
    \draw [<-, orange!80!black, shorten <=-4pt, shorten >=-4pt] (n-2-1) -- (n-3-1);
    \draw [<-, orange!80!black, shorten <=-4pt, shorten >=-4pt] (n-3-1) -- (n-4-1);
    \draw [<-, orange!80!black, shorten <=-4pt, shorten >=-4pt] (n-4-1) -- (n-5-1);
    \draw [<-, orange!80!black, shorten <=-4pt, shorten >=-4pt] (n-2-2) -- (n-3-2);
    \draw [<-, orange!80!black, shorten <=-4pt, shorten >=-4pt] (n-3-2) -- (n-4-2);
    \draw [<-, orange!80!black, shorten <=-4pt, shorten >=-4pt] (n-4-2) -- (n-5-2);
    \draw [<-, orange!80!black, shorten <=-4pt, shorten >=-4pt] (n-3-2) -- (n-2-1);
    \draw [<-, gray, shorten <=-4pt, shorten >=-4pt] (n-3-3) -- (n-4-3);
    \draw [<-, gray, shorten <=-4pt, shorten >=-4pt] (n-4-3) -- (n-5-3);
    \draw [<-, gray, shorten <=-4pt, shorten >=-4pt] (n-4-3) -- (n-3-2);
    \draw [<-, gray, shorten <=-4pt, shorten >=-4pt] (n-4-4) -- (n-5-4);
    \draw [<-, gray, shorten <=-4pt, shorten >=-4pt] (n-4-3) -- (n-5-4);
\end{tikzpicture}
\begin{tikzpicture}[remember picture, overlay]
\draw [->, orange!80!black, shorten <=-4pt, shorten >=-4pt] (m-1-2) -- (n-1-1);
\end{tikzpicture}
\caption{\label{fig:results}Summary of the unitarity constraints. The first values of $m_{L,k}$ and $n_{L,k}$ through subleading order are given in terms of $m_{1,0}$, which is the only initial condition (scheme dependence). The arrow pointing for the 4-particle coefficients to the 2-particle coefficients indicates that, at subleading order, the 2-particle coefficients are constrained in terms of 4-particle coefficients by the unitarity equation. The analogous recursions coming from renormalization are illustrated in \autoref{fig:resultsRG2} and \autoref{fig:resultsRG}.}
\end{figure}
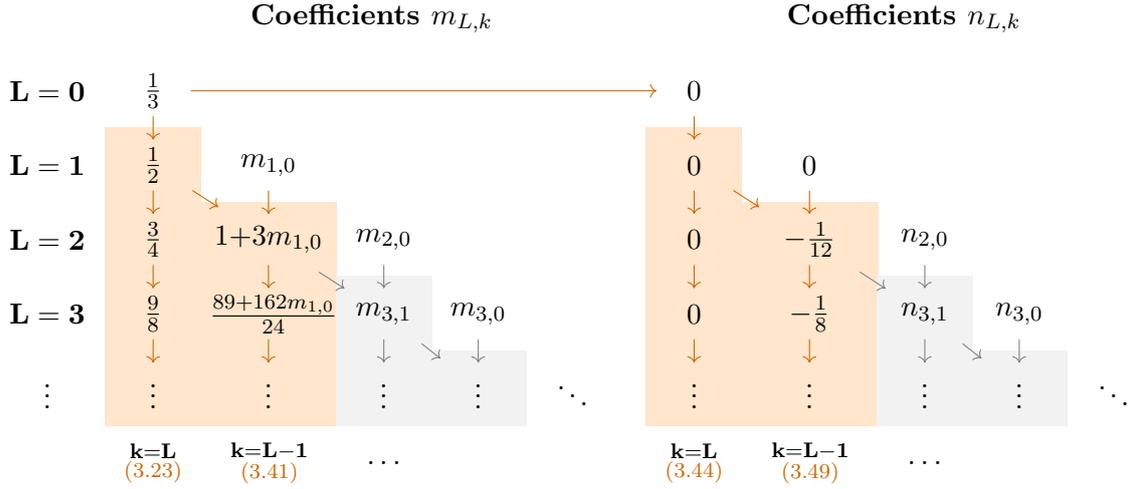

Combining both \autoref{eq:SL2cutContribution} and \autoref{eq:SL4cutContribution} and plugging them into the unitarity equation, we find a closed relationship for the subleading coefficients 
\begin{equation}
m_{L, L - 1} = \left(\frac{3}{2}\right)^L\left[-\frac{L - 2}{81} + \frac{2L}{9} + \frac{2}{L - 1}\sum_{L^\prime = 1}^{L - 1} \left(\frac{2}{3}\right)^{L^\prime} m_{L^\prime, L^\prime - 1}\right].
\end{equation}
The solution to this recursion relation for scheme-dependent input data $m_{1, 0}$ is 
\begin{empheq}[box=\sentencebox]{equation}
\label{eq:mSubLeadingUnitarity}
m_{L, L - 1} = \frac{2}{81}\left(\frac{3}{2}\right)^L\left[L\left(27 m_{1, 0} - 16\right) + 17 L H_L - 1\right],
\end{empheq}
where $H_L$ is the $L^{\text{th}}$ harmonic number. The first few expressions are displayed in \autoref{fig:results}.

\subsection{Constraints on the 2-Particle Amplitude}
With the leading coefficient for the 4-particle amplitude, we will be able to constrain the subleading coefficient for the 2-particle amplitude. To begin, recall that the leading contribution for the 2-particle amplitude, which by construction has no 1-particle cut, is given by a 3-cut 
\begin{equation}
2\: \Im\left(
\begin{gathered}
\begin{tikzpicture}[scale=0.6]
\draw[line width=0.3mm] (-1.5, 0) -- (-0.5, 0);
\draw[pattern= north west lines] (0, 0) circle (0.5cm);
\draw[line width=0.3mm] (0, 0) circle (0.5cm); 
\draw[line width=0.3mm] (0.5, 0) -- (1.5, 0);
\end{tikzpicture}
\end{gathered}
\right) = 
\begin{gathered}
\begin{tikzpicture}[scale=0.6]
\draw[line width=0.3mm] (-1.5, 0) -- (-0.5, 0);
\draw[pattern= north west lines] (0, 0) circle (0.5cm);
\draw[line width=0.3mm] (0, 0) circle (0.5cm);
\draw[line width=0.3mm] (0.5*0.866, 0.5*0.5) -- (2 - 0.5*0.866, 0.5*0.5);
\draw[line width=0.3mm] (0.5*0.866, -0.5*0.5) -- (2 - 0.5*0.866, -0.5*0.5);
\draw[pattern=north west lines] (2, 0) circle (0.5cm);
\draw[line width=0.3mm] (2, 0) circle (0.5cm);
\draw[line width=0.3mm] (0.5, 0) -- (1.5, 0);
\draw[line width=0.3mm] (2.5, 0) -- (3.5, 0);
\filldraw[orange!10, opacity=0.5] (1, 1) rectangle (3.7, -1);
\draw[line width=0.5mm, dashed, orange] (1, 1) -- (1, -1);
\end{tikzpicture}
\end{gathered} \: \: + \dots
\end{equation}
In an equation, 
\begin{equation}
\label{eq:2ParticleUnitarity}
\Im\left(\M^{\left(L\right)}_2\left(p^2\right)\right) = \frac{\left(16\pi^2\right)^2}{2}\sum_{L^\prime = 0}^{L - 2} \int \M_4^{\left(L^\prime\right)} \left(\M_4^{\left(L - L^\prime - 2\right)}\right)^{\ast} \dd{\Phi}_3 + \dots
\end{equation}
There are several aspects of this equation to note. First, the power in logarithm, as we will verify by explicit computation, will at most be subleading. This will constrain all the leading 2-particle coefficients, $n_{L, L}$, to be zero,
\begin{empheq}[box=\sentencebox]{equation}
\label{eq:nLeadingUnitarity}
n_{L, L} = 0.
\end{empheq}
Second, the subleading coefficients, $n_{L, L - 1}$, will be constrained in terms of the 4-particle leading coefficients. Third, the sum in \autoref{eq:2ParticleUnitarity} has no contribution from $L = 1$. This constrains the $L = 1$ subleading term, $n_{1, 0}$, to be zero, 
\begin{equation}
\label{eq:n10unitarity}
n_{1,0}=0.
\end{equation}

To use the unitarity equation, we first note that at subleading order in logarithm,
\begin{equation}
\Im\left(\M^{\left(L, L - 1\right)}_2\right) = \left(L - 1\right) \pi n_{L, L - 1} \log^{L - 2}_{p^2}.
\end{equation}
Plugging in the leading 4-particle amplitude and extracting the leading log contribution from each using similar manipulations to \autoref{eq:LogExpansionIntegrand}, we find that 
\begin{equation}
\eval{\text{Im}\left(\M^{\left(L, L - 1\right)}_2\left(p^2\right)\right)}_{3-\text{cuts}} = \frac{2\left(16\pi^2\right)^2}{9} \left(L - 1\right) \left(\frac{3}{2}\right)^{L} \log^{L - 2}_{p^2} \Phi_3. 
\end{equation}
Using the definition of $\Phi_3$ in \autoref{eq:3ParticlePhaseSpace}, we find 
\begin{equation}
\eval{\text{Im}\left(\M^{\left(L, L - 1\right)}_2\left(p^2\right)\right)}_{3-\text{cuts}} = -\frac{\pi p^2}{27} \left(L - 1\right) \left(\frac{3}{2}\right)^{L} \log^{L - 2}_{p^2}. 
\end{equation}
Matching both sides of the unitarity equation gives the recursion 
\begin{empheq}[box=\sentencebox]{equation}
\label{eq:nSubleadingUnitarity}
n_{L, L - 1} = -\frac{1}{27}\left(\frac{3}{2}\right)^L
\end{empheq}
for $L \geq 2$.
The results is constrained entirely by the leading coefficient of the 4-particle amplitude. The first few values are shown in \autoref{fig:results}.

\subsection{Resummation and Triviality}
With the help of unitarity, we were able to compute the leading and subleading amplitude coefficients in $\lambda \phi^4$ theory to all loop order. It is helpful to investigate what happens when the amplitude is then re-summed. We will commit to the Wilsonian picture with a cutoff $\Lambda$ to make the following discussion easier. Let's begin with the 4-particle amplitude. By summing over the leading coefficients and the subleading coefficients, we find
\begin{equation}
\begin{aligned}
\M_4 &= -\frac{\lambda}{3}\left(\frac{1}{1 + \frac{3\lambda}{32\pi^2}\log_{\abs{s}}} + \left(s\leftrightarrow t\right) + \left(s \leftrightarrow u\right)\right) \\
&\hspace{0.4cm} - \frac{2}{27} \frac{\lambda^2}{32\pi^2} \left(\frac{\frac{3\lambda}{32\pi^2} \log_{\abs{s}} + 17\log\left(1 + \frac{3\lambda}{32\pi^2}\log_{\abs{s}}\right) - 27 m_{1, 0}}{\left(1 + \frac{3\lambda}{32\pi^2}\log_{\abs{s}}\right)^2} + \left(s\leftrightarrow t\right) + \left(s \leftrightarrow u\right)\right). 
\end{aligned}
\end{equation}
Notice that when the cutoff $\Lambda$ is sufficiently small, i.e. $\Lambda^2 \ll \abs{s}$, a Landau pole develops at the location where the denominator goes to zero, 
\begin{equation}
1 + \frac{3\lambda}{32\pi^2} \log\left(\frac{\Lambda^2}{\abs{s}}\right) = 0.
\end{equation}
This would indicate a breakdown in perturbation theory since the value of the amplitude, and the renormalized coupling, would diverge. One avenue for resolving this potential inconsistency is to set $\lambda(\Lambda) = 0$ as $\Lambda \to 0$. In other words, the theory in the IR needs to be trivial, which is known as the statement of $\lambda \phi^4$ triviality \cite{Wolff:2014}.\footnote{We thank Nima Arkani-Hamed and Alberto Nicolis for discussions about this point.}

Similarly, the 2-particle amplitude can also be resummed at subleading logarithmic order 
\begin{equation}
\M_2 = p^2\left(1 + \frac{\lambda^2}{3\left(32\pi^2\right)^2} \frac{\log_{\abs{p^2}}}{1 + \frac{3\lambda}{32\pi^2}\log_{\abs{p^2}}}\right).
\end{equation}
This amplitude has the same structure in the denominator and hence the same Landau pole. 

\section{Renormalization Constraints on \texorpdfstring{$\lambda \phi^{4}$}{lambda phi 4}}
\label{s:RGconstraintsPhi4}

We now demonstrate that the recursion relations derived from the unitarity of the $S$-matrix in the previous section are equivalent to those generated by RG, again working within massless $\lambda\phi^{4}$ theory with the amplitude ans\"{a}tze defined in \autoref{eq:3ch4PtAnsatz} and \autoref{eq:3ch2PtAnsatz}. As in the O$\left(N\right)$ model, our starting point is the Callan--Symanzik equation \autoref{eq:CSonM}. The new feature of full $\lambda\phi^{4}$ theory that was not present for the O$\left(N\right)$ model analysis, however, is the appearance of the wavefunction anomalous dimension $\gamma$. The Taylor expansion of the $\beta$ function in the coupling $\lambda$ is given in \autoref{eq:betaO(N)model}, while the $\gamma$ function admits a similar expansion
\begin{equation}
\label{eq:gammaPhi4Theory}
\gamma\left(\lambda\right)=\sum_{L=1}^{\infty}\frac{\gamma_{L}\left(-\lambda\right)^{L}}{\left(16\pi^{2}\right)^{L}},
\end{equation}
where we have neglected a constant term in $\gamma$ since this is immediately set to zero by the CS equation.

Our method closely follows that of \autoref{s:ONModel}: upon inserting the amplitude ans\"{a}tze, \autoref{eq:3ch4PtAnsatz} and \autoref{eq:3ch2PtAnsatz}, along with the $\beta$ and $\gamma$ functions, \autoref{eq:betaO(N)model} and \autoref{eq:gammaPhi4Theory}, into the CS equation \autoref{eq:CSonM} and matching orders in $\lambda$, a whole tower of recursions will fall onto our lap. We do this for the 2-particle amplitude in \autoref{ss:2particleRG} and for the 4-particle amplitude in \autoref{ss:4particleRG}. We then compare the results of these computations to those derived in \autoref{s:UnitarityconstraintsPhi4} from unitarity and observe how $\beta$ and $\gamma$ act as the additional input data needed for the RG recursions, which start at one higher order in $\lambda$ than the unitarity recursions.

\subsection{2-Particle Amplitude}
\label{ss:2particleRG}

In order to derive the 2-particle recursions, we start by inputting our 2-particle amplitude ansatz \autoref{eq:3ch2PtAnsatz} into the CS
equation \autoref{eq:CSonM}, 
\begin{equation}
\label{eq:CS2particle}
\left[\mu\partial_{\mu}+\beta\left(\lambda\right)\partial_{\lambda}-2\gamma\left(\lambda\right)\right]\mathcal{M}_{2}\left(p^{2}\right)=0.
\end{equation}
For clarity, we proceed term-by-term. The first term in \autoref{eq:CS2particle} is given by
\begin{equation}
p^{-2}\mu\partial_{\mu}\mathcal{M}_{2}=-\sum_{L=1}^{\infty}\sum_{k=0}^{L-1}\frac{\left(-\lambda\right)^{L}}{\left(16\pi^{2}\right)^{L}}2\left(k+1\right)n_{L,k+1}\log^{k}_{p^2}.
\end{equation}
The second term in \autoref{eq:CS2particle} requires more work, but follows the same steps as outlined in \autoref{s:ONModel}. We first compute the derivative, 
\begin{equation}
p^{-2}\beta\left(\lambda\right)\partial_{\lambda}\mathcal{M}_{2} =\sum_{L=1}^{\infty}\sum_{L^{\prime}=1}^{\infty}\sum_{k=0}^{L}\frac{\left(-\lambda\right)^{L+L'}}{\left(16\pi^{2}\right)^{L+L^{\prime}}}L\beta_{L^{\prime}}n_{L,k}\log^{k}_{p^2},
\end{equation}
then shift the starting points of the sums so that we can apply the double sum identity given in \autoref{eq:DoubleSum} to give 
\begin{equation}
p^{-2}\beta\left(\lambda\right)\partial_{\lambda}\mathcal{M}_{2} = \sum_{L=0}^{\infty}\sum_{L^{\prime}=0}^{L}\sum_{k=0}^{L-L^{\prime}}\frac{\left(-\lambda\right)^{L+2}}{\left(16\pi^{2}\right)^{L+2}}\left(L-L^{\prime}+1\right)\beta_{L^{\prime}+1}n_{L-L^{\prime}+1,k}\log^{k}_{p^2},
\end{equation}
before finally swapping the order of the sums over $L^{\prime}$ and $k$:
\begin{equation}
\begin{aligned}
p^{-2}\beta\left(\lambda\right)\partial_{\lambda}\mathcal{M}_{2} & =\sum_{L=2}^{\infty}\sum_{L^{\prime}=0}^{L-2}\frac{\left(-\lambda\right)^{L}}{\left(16\pi^{2}\right)^{L}}\left(L-L^{\prime}-1\right)\beta_{L^{\prime}+1}n_{L-L^{\prime}-1,0}\\
 & \hphantom{=}+\sum_{L=2}^{\infty}\sum_{k=1}^{L-1}\sum_{L^{\prime}=0}^{L-k-1}\frac{\left(-\lambda\right)^{L}}{\left(16\pi^{2}\right)^{L}}\left(L-L^{\prime}-1\right)\beta_{L^{\prime}+1}n_{L-L^{\prime}-1,k}\log^{k}_{p^2}.
\end{aligned}
\end{equation}
Through equivalent manipulations, the third term in \autoref{eq:CS2particle} becomes 
\begin{equation}
\begin{aligned}-2\gamma\left(\lambda\right)p^{-2}\mathcal{M}_{2} & =-2\sum_{L=1}^{\infty}\frac{\gamma_{L}\left(-\lambda\right)^{L}}{\left(16\pi^{2}\right)^{L}}+2\sum_{L=2}^{\infty}\sum_{L^{\prime}=0}^{L-2}\frac{\left(-\lambda\right)^{L}}{\left(16\pi^{2}\right)^{L}}\gamma_{L^{\prime}+1}n_{L-L^{\prime}-1,0}\\
 & \hphantom{=}+2\sum_{L=2}^{\infty}\sum_{k=1}^{L-1}\sum_{L^{\prime}=0}^{L-k-1}\frac{\left(-\lambda\right)^{L}}{\left(16\pi^{2}\right)^{L}}\gamma_{L^{\prime}+1}n_{L-L^{\prime}-1,k}\log^{k}_{p^2}
\end{aligned}
\end{equation}

Putting this all together, the 2-particle Callan--Symanzik equation gives a new
constraint from each order in log, determined by the terms in the $k$ sum.
Setting $k=L-1$ isolates the leading log coefficients for $L\geq2$:
\begin{equation}
n_{L,L}=\frac{1}{2L}\left(\left(L-1\right)\beta_{1}+2\gamma_{1}\right)n_{L-1,L-1}.
\end{equation}
This recursion resums to give the 2-particle leading log coefficients:
\begin{equation}
\label{eq:nLeadingLogRG}
n_{L,L}=n_{1,1}\left(\frac{\beta_{1}}{2}\right)^{L-1}\frac{\Gamma\left(L+\frac{2\gamma_{1}}{\beta_{1}}\right)}{\Gamma\left(1+\frac{2\gamma_{1}}{\beta_{1}}\right)L!},
\end{equation}
where $\Gamma\left(\dots\right)$ denotes the gamma function and the equation holds for $L\geq2$.

The $k=L-2$ terms give the subleading log coefficients for $L\geq3$:
\begin{equation}
n_{L,L-1}=\frac{1}{2\left(L-1\right)}\left[\left(L-1\right)\beta_{1}+2\gamma_{1}\right]n_{L-1,L-2}+\frac{1}{2\left(L-1\right)}\left[\left(L-2\right)\beta_{2}+2\gamma_{2}\right]n_{L-2,L-2} .
\end{equation}
We can again perform resummation, which yields
\begin{equation}
\label{eq:nSubLeadingLogRG}
\begin{aligned}
&n_{L,L-1}=  \left(\frac{\beta_{1}}{2}\right)^{L-3}\frac{\Gamma\left(L+\frac{2\gamma_{1}}{\beta_{1}}\right)}{\Gamma\left(1+\frac{2\gamma_{1}}{\beta_{1}}\right)\left(L-1\right)!}\\
 & \qquad\times\left\{ \frac{\beta_{1}}{2\left(1+\frac{2\gamma_{1}}{\beta_{1}}\right)}n_{2,1}+n_{1,1}\left[\frac{1}{2}\beta_{2}\left[\psi\left(L-1+\frac{2\gamma_{1}}{\beta_{1}}\right)-\psi\left(1+\frac{2\gamma_{1}}{\beta_{1}}\right)\right]\right.\right.\\
 & \qquad\left.\left.\hphantom{\times\left\{ \right.}\;\;+\frac{1}{2}\left(\beta_{2}\left(1+\frac{2\gamma_{1}}{\beta_{1}}\right)-2\gamma_{2}\right)\left(\frac{1}{L-1+\frac{2\gamma_{1}}{\beta_{1}}}-\frac{\beta_{1}}{\beta_{1}+2\gamma_{1}}\right)\right]\right\} ,
\end{aligned}
\end{equation}
where $\psi\left(\dots\right)$ is the digamma function and this is valid for $L\geq3$. Note that we used the leading log 2-particle recursion \autoref{eq:nLeadingLogRG} in order to arrive at this subleading log result. For the purposes of this paper we truncate at subleading log order.

\begin{figure}
\centering
\begin{tikzpicture}
\matrix (m) [
        matrix of math nodes, 
        nodes={text width=1cm, minimum height=1cm, anchor=center, align=center}, 
        row sep=-\pgflinewidth, 
        column sep=-\pgflinewidth,  
        label={[font=\bfseries]above: {Coefficients $n_{L,k}$}}
    ]
    {
        |[text width=1.1cm]| {\bf L=0} & 0 & (\beta_1, \gamma_1) &  &  \\
        |[text width=1.1cm]| {\bf L=1} & |[fill=orange!20,text width=1.8cm]| -\gamma_1  & n_{1,0} & (\beta_2, \gamma_2) &  \\
        |[text width=1.1cm]| {\bf L=2} & |[fill=orange!20,text width=1.8cm]| -\tfrac{\gamma_1 (\beta_1 + 2 \gamma_1)}{4} & |[fill=orange!20,text width=2.5cm]| \tfrac{(\beta_1 + 2\gamma_1) n_{1,0} - 2\gamma_2}{2} &  n_{2,0} & (\beta_3, \gamma_3) \\
        |[text width=1.1cm]| \vdots &|[fill=orange!20,text width=1.8cm]| \vdots & |[fill=orange!20,text width=2.5cm]| \vdots & |[fill=orange!20]| \vdots & \ddots  \\
        & \substack{{\bf k=L} \\ \autoref{eq:nLeadingLogRGfinal}}   &  \substack{{\bf k=L{-}1}\\ \autoref{eq:nSubLeadingLogRGfinal}} &  \dots &  \\
    };
    \draw [<-, orange!80!black, shorten <=-4pt, shorten >=-4pt] (m-1-2) -- (m-2-2);
    \draw [<-, orange!80!black, shorten <=-4pt, shorten >=-4pt] (m-2-2) -- (m-3-2);
    \draw [<-, orange!80!black, shorten <=-4pt, shorten >=-4pt] (m-3-2) -- (m-4-2);
    \draw [<-, orange!80!black, shorten <=-4pt, shorten >=-4pt] (m-2-2) -- (m-1-3);
    \draw [<-, orange!80!black, shorten <=-4pt, shorten >=-4pt] (m-2-3) -- (m-3-3);
    \draw [<-, orange!80!black, shorten <=-4pt, shorten >=-4pt] (m-3-3) -- (m-4-3);
    \draw [<-, orange!80!black, shorten <=-4pt, shorten >=-4pt] (m-3-3) -- (m-2-4);
    \draw [<-, orange!80!black, shorten <=-4pt, shorten >=-4pt] (m-3-3) -- (m-2-2);
    \draw [<-, orange!80!black, shorten <=-4pt, shorten >=-4pt] (m-3-4) -- (m-4-4);
    \draw [<-, Maroon, shorten <=-4pt, shorten >=-4pt] (m-3-5) -- (m-4-4);
    \draw [<-, orange!80!black, shorten <=-4pt, shorten >=-4pt] (m-4-4) -- (m-3-3);
\end{tikzpicture}
\caption{\label{fig:resultsRG2}Summary of the renormalization constraints on $n_{L,k}$. The first few coefficients are displayed in terms of their dependence on the initial conditions. This table is to be compared with \autoref{fig:results} (right).}
\end{figure}
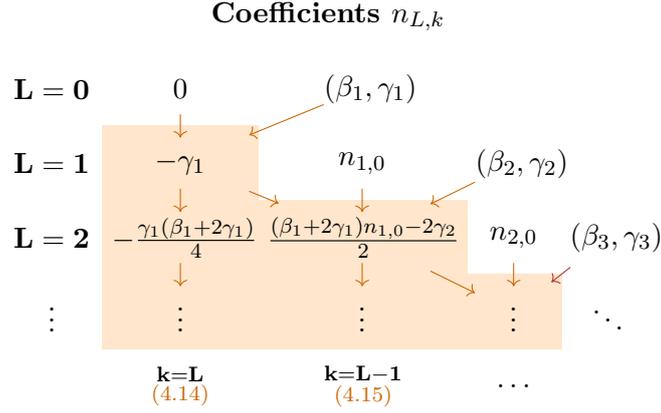

Finally, the $k=0$ constant terms give an expression for $\gamma_{L}$ for $L\geq1$:
\begin{equation}
\label{eq:gammaBC}
2\gamma_{L}=-2n_{L,1}+\sum_{L^{\prime}=0}^{L-2}\left[\left(L-L^{\prime}-1\right)\beta_{L^{\prime}+1}+2\gamma_{L^{\prime}+1}\right]n_{L-L^{\prime}-1,0} ,
\end{equation}
which serves as the boundary condition for the 2-particle amplitude
recursions. We demonstrate this for the first two terms, $L=1$ and $L=2$, which rearrange to
\begin{subequations}
\begin{equation}
\label{eq:gamma1BC}
n_{1,1}=-\gamma_{1},
\end{equation}
\begin{equation}
\label{eq:gamma2BC}
n_{2,1}=-\gamma_{2}+\frac{1}{2}\left(\beta_{1}+2\gamma_{1}\right)n_{1,0}.
\end{equation}
\end{subequations}
As we see, through subleading logarithmic order, the base of the RG recursion towers for the 2-particle amplitude coefficients depends on $\beta$ and $\gamma$ function coefficients. We can therefore substitute \autoref{eq:gamma1BC} into \autoref{eq:nLeadingLogRG} to arrive at 
\begin{empheq}[box=\sentencebox]{equation}
\label{eq:nLeadingLogRGfinal}
n_{L,L}=-\gamma_{1}\left(\frac{\beta_{1}}{2}\right)^{L-1}\frac{\Gamma\left(L+\frac{2\gamma_{1}}{\beta_{1}}\right)}{\Gamma\left(1+\frac{2\gamma_{1}}{\beta_{1}}\right)L!}
\end{empheq}
and substitute \autoref{eq:gamma1BC} and \autoref{eq:gamma2BC} into \autoref{eq:nSubLeadingLogRG} to find
\begin{empheq}[box=\sentencebox]{equation}
\label{eq:nSubLeadingLogRGfinal}
\begin{aligned}
&n_{L,L-1}=  \left(\frac{\beta_{1}}{2}\right)^{L-3}\frac{\Gamma\left(L+\frac{2\gamma_{1}}{\beta_{1}}\right)}{\Gamma\left(1+\frac{2\gamma_{1}}{\beta_{1}}\right)\left(L-1\right)!}\\
 & \!\!\!\!\times\!\left\{ \frac{\beta_{1}\left(-\gamma_{2}+\frac{1}{2}\left(\beta_{1}+2\gamma_{1}\right)n_{1,0}\right)}{2\left(1+\frac{2\gamma_{1}}{\beta_{1}}\right)}-\gamma_{1}\left[\frac{1}{2}\beta_{2}\left[\psi\left(L-1+\frac{2\gamma_{1}}{\beta_{1}}\right)-\psi\left(1+\frac{2\gamma_{1}}{\beta_{1}}\right)\right]\right.\right.\!\!\!\!\!\!\!\\
 & \left.\left.\hphantom{\times\left\{ \right.}\;\;+\frac{1}{2}\left(\beta_{2}\left(1+\frac{2\gamma_{1}}{\beta_{1}}\right)-2\gamma_{2}\right)\left(\frac{1}{L-1+\frac{2\gamma_{1}}{\beta_{1}}}-\frac{\beta_{1}}{\beta_{1}+2\gamma_{1}}\right)\right]\right\}.
\end{aligned}
\end{empheq}
The leading log recursion depends entirely on the input data $\gamma_{1}$ and $\beta_{1}$, while the subleading logarithm recursion depends on the input data $\beta_{1}$, $\gamma_{1}$, $\beta_{2}$, $\gamma_{2}$, and $n_{1,0}$. See \autoref{fig:resultsRG2} for examples and a summary table of the recursions.

\subsection{4-Particle Amplitude}
\label{ss:4particleRG}

We now repeat the analysis for the 4-particle amplitude.
Inserting our 4-particle amplitude ansatz \autoref{eq:3ch4PtAnsatz} into the CS
equation \autoref{eq:CSonM}, we aim to solve 
\begin{equation}
\label{eq:CS4particle}
\left[\mu\partial_{\mu}+\beta\left(\lambda\right)\partial_{\lambda}-4\gamma\left(\lambda\right)\right]\mathcal{M}_{4}\left(s,t,u\right)=0.
\end{equation}
The analysis proceeds in the same way as for the 2-particle amplitude.
The first term in \autoref{eq:CS4particle} is
\begin{equation}
\mu\partial_{\mu}\mathcal{M}_{4}=\sum_{L=1}^{\infty}\sum_{k=0}^{L-1}\frac{\left(-\lambda\right)^{L+1}}{\left(16\pi^{2}\right)^{L}}2\left(k+1\right)m_{L,k+1}\left(\log^{k}_{-s}+\log^{k}_{-t}+\log^{k}_{-u}\right).
\end{equation}
The second term in \autoref{eq:CS4particle}, upon similar steps as before that employ the double sum identity \autoref{eq:DoubleSum} and switch the orders of the $L^{\prime}$ and $k$ sums, becomes
\begin{equation}
\begin{aligned}
\beta\left(\lambda\right)\partial_{\lambda}\mathcal{M}_{4} & =-3\sum_{L=1}^{\infty}\sum_{L^{\prime}=0}^{L-1}\frac{\left(-\lambda\right)^{L+1}}{\left(16\pi^{2}\right)^{L}}\left(L-L^{\prime}\right)\beta_{L^{\prime}+1}m_{L-L^{\prime}-1,0}\\
 & \hphantom{=}-\sum_{L=1}^{\infty}\sum_{k=1}^{L-1}\sum_{L^{\prime}=0}^{L-k-1}\Bigg[\frac{\left(-\lambda\right)^{L+1}}{\left(16\pi^{2}\right)^{L}}\left(L-L^{\prime}\right)\beta_{L^{\prime}+1}m_{L-L^{\prime}-1,k}\left(\log^{k}_{-s}+\log^{k}_{-t}+\log^{k}_{-u}\right)\Bigg].
\end{aligned}
\end{equation}
Finally, the third term in \autoref{eq:CS4particle} is 
\begin{equation}
\begin{aligned}-4\gamma\left(\lambda\right)\mathcal{M}_{4} & =-12\sum_{L=1}^{\infty}\sum_{L^{\prime}=0}^{L-1}\frac{\left(-\lambda\right)^{L+1}}{\left(16\pi^{2}\right)^{L}}\gamma_{L^{\prime}+1}m_{L-L^{\prime}-1,0}\\
 & \hphantom{=}-4\sum_{L=1}^{\infty}\sum_{k=1}^{L-1}\sum_{L^{\prime}=0}^{L-k-1}\Bigg[\frac{\left(-\lambda\right)^{L+1}}{\left(16\pi^{2}\right)^{L}}\gamma_{L^{\prime}+1}m_{L-L^{\prime}-1,k}\left(\log^{k}_{-s}+\log^{k}_{-t}+\log^{k}_{-u}\right)\Bigg].
\end{aligned}
\end{equation}
We then combine these individual contributions to the CS equation and derive constraints.

We begin with the leading log terms, which correspond to $k=L-1$ for $L\geq2$:
\begin{equation}
m_{L,L}=\left(\frac{1}{2}\beta_{1}+\frac{2}{L}\gamma_{1}\right)m_{L-1,L-1}.
\end{equation}
The first couple of outputs are
\begin{subequations}
\begin{align}
m_{2,2}&=\left(\frac{\beta_{1}}{2}+\gamma_{1}\right)m_{1,1},\\
m_{3,3}&=\left(\frac{\beta_{1}}{2}+\frac{2\gamma_{1}}{3}\right)m_{2,2} = 
\left(\frac{\beta_{1}}{2}+\frac{2\gamma_{1}}{3}\right)\left(\frac{\beta_{1}}{2}+\gamma_{1}\right)m_{1,1}
,
\end{align}
\end{subequations}
and the whole tower resums, for $L\geq2$, to the closed form
\begin{equation}
\label{eq:mLeadingRG}
m_{L,L}=m_{1,1}\left(\frac{\beta_{1}}{2}\right)^{L-1}\frac{\Gamma\left(L+1+\frac{4\gamma_{1}}{\beta_{1}}\right)}{\Gamma\left(2+\frac{4\gamma_{1}}{\beta_{1}}\right)L!}.
\end{equation}

Next, $k=L-2$ isolates the subleading log terms for $L\geq3$: 
\begin{equation}
m_{L,L-1}=\frac{1}{2\left(L-1\right)}\left[L\beta_{1}+4\gamma_{1}\right]m_{L-1,L-2}+\frac{1}{2\left(L-1\right)}\left[\left(L-1\right)\beta_{2}+4\gamma_{2}\right]m_{L-2,L-2} .
\end{equation}
By plugging in the leading log 4-particle recursion \autoref{eq:mLeadingRG} we are again able to find a closed form solution to the recursion which holds for $L\geq3$:
\begin{equation}
\label{eq:mSubleadingRG}
\begin{aligned}
&m_{L,L-1} = \left(\frac{\beta_{1}}{2}\right)^{L-3}\frac{\Gamma\left(L+1+\frac{4\gamma_{1}}{\beta_{1}}\right)}{\Gamma\left(2+\frac{4\gamma_{1}}{\beta_{1}}\right)\left(L-1\right)!}\\
 &\!\!\!\! \times\left\{ \frac{\beta_{1}^{2}}{4\left(\beta_{1}+2\gamma_{1}\right)}m_{2,1}+\frac{1}{2}m_{1,1}\left[4\left(\gamma_{2}-\frac{\gamma_{1}\beta_{2}}{\beta_{1}}\right)\left[\psi\left(L+\frac{4\gamma_{1}}{\beta_{1}}\right)-\psi\left(2+\frac{4\gamma_{1}}{\beta_{1}}\right)\right]\right.\right.\!\!\!\!\!\! \\
 & \left.\left.\hphantom{\times\left\{ \right.}\;\;+\left(\beta_{2}\left(1+\frac{4\gamma_{1}}{\beta_{1}}\right)-4\gamma_{2}\right)\left[\psi\left(L+1+\frac{4\gamma_{1}}{\beta_{1}}\right)-\psi\left(3+\frac{4\gamma_{1}}{\beta_{1}}\right)\right]\right]\right\} .
\end{aligned}
\end{equation}

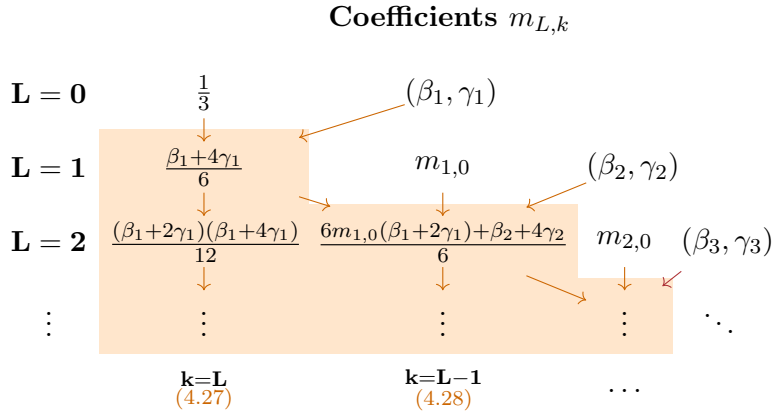
\begin{figure}
\centering
\begin{tikzpicture}
\matrix (m) [
        matrix of math nodes, 
        nodes={text width=1cm, minimum height=1cm, anchor=center, align=center}, 
        row sep=-\pgflinewidth, 
        column sep=-\pgflinewidth,  
        label={[font=\bfseries]above: {\qquad\qquad Coefficients $m_{L,k}$}}
    ]
    {
        |[text width=1.1cm]| {\bf L=0} & \tfrac{1}{3} & (\beta_1,\gamma_1) &  &  \\
        |[text width=1.1cm]| {\bf L=1} & |[fill=orange!20,text width=2.5cm]| \tfrac{\beta_1 + 4 \gamma_1}{6}  &  m_{1,0} & (\beta_2,\gamma_2) &  \\
        |[text width=1.1cm]| {\bf L=2} & |[fill=orange!20,text width=2.5cm]| \tfrac{(\beta_1 + 2\gamma_1)(\beta_1 + 4\gamma_1)}{12} & |[fill=orange!20,text width=3.3cm]|  \tfrac{6m_{1,0}(\beta_1 + 2 \gamma_1) + \beta_2 + 4\gamma_2}{6} &  m_{2,0} & (\beta_3, \gamma_3) \\
        |[text width=1.1cm]|{\bf \vdots} & |[fill=orange!20,text width=2.5cm]| \vdots & |[fill=orange!20,text width=3.3cm]| \vdots & |[fill=orange!20]| \vdots & \ddots  \\
        |[text width=1.1cm]| & \substack{{\bf k=L} \\ \autoref{eq:mLeadingLogRGfinal}}   &  \substack{{\bf k=L{-}1}\\ \autoref{eq:mSubLeadingLogRGfinal}} &  \dots &   \\
    };
    \draw [<-, orange!80!black, shorten <=-4pt, shorten >=-4pt] (m-1-2) -- (m-2-2);
    \draw [<-, orange!80!black, shorten <=-4pt, shorten >=-4pt] (m-2-2) -- (m-3-2);
    \draw [<-, orange!80!black, shorten <=-4pt, shorten >=-4pt] (m-3-2) -- (m-4-2);
    \draw [<-, orange!80!black, shorten <=-4pt, shorten >=-4pt] (m-2-2) -- (m-1-3);
    \draw [<-, orange!80!black, shorten <=-4pt, shorten >=-4pt] (m-2-3) -- (m-3-3);
    \draw [<-, orange!80!black, shorten <=-4pt, shorten >=-4pt] (m-3-3) -- (m-4-3);
    \draw [<-, orange!80!black, shorten <=-4pt, shorten >=-4pt] (m-3-3) -- (m-2-4);
    \draw [<-, orange!80!black, shorten <=-4pt, shorten >=-4pt] (m-3-3) -- (m-2-2);
    \draw [<-, orange!80!black, shorten <=-4pt, shorten >=-4pt] (m-3-4) -- (m-4-4);
    \draw [<-, Maroon, shorten <=-4pt, shorten >=-4pt] (m-3-5) -- (m-4-4);
    \draw [<-, orange!80!black, shorten <=-4pt, shorten >=-4pt] (m-4-4) -- (m-3-3);
\end{tikzpicture}
\caption{\label{fig:resultsRG}Summary of the renormalization constraints on $m_{L,k}$. The first few coefficients are displayed in terms of their dependence on the initial conditions. This table is to be compared with \autoref{fig:results} (left).}
\end{figure}

Finally, we isolate the $k=0$ constant terms that give the boundary condition for the 4-particle amplitude recursions for $L\geq1$: 
\begin{equation}
\label{eq:BetaBC}
\beta_{L}=\frac{1}{m_{0,0}}\Bigg[2m_{L,1}-4m_{0,0}\gamma_{L}-\sum_{L^{\prime}=0}^{L-2}\left(\left(L-L^{\prime}\right)\beta_{L^{\prime}+1}+4\gamma_{L^{\prime}+1}\right)m_{L-L^{\prime}-1,0}\Bigg].
\end{equation}
The first two terms for $L=1$ and $L=2$ rearrange to 
\begin{subequations}
\begin{equation}
\label{eq:beta1BC}
m_{1,1}=\left(\frac{1}{2}\beta_{1}+2\gamma_{1}\right)m_{0,0},
\end{equation}
\begin{equation}
\label{eq:beta2BC}
m_{2,1}=\frac{1}{2}m_{0,0}\beta_{2}+\left(\beta_{1}+2\gamma_{1}\right)m_{1,0}+2m_{0,0}\gamma_{2}.
\end{equation}
\end{subequations}
We find, through subleading order, that the link between the base of the RG recursion towers for the 4-particle amplitude coefficients, $m_{1,1}$ and $m_{2,1}$, and the lowest possible level, $m_{0,0}$ and $m_{1,0}$, is provided by the $\beta$ and $\gamma$ function coefficients. We substitute \autoref{eq:beta1BC} into \autoref{eq:mLeadingRG}, which gives
\begin{empheq}[box=\sentencebox]{equation}
\label{eq:mLeadingLogRGfinal}
m_{L,L}=\left(\frac{1}{2}\beta_{1}+2\gamma_{1}\right)m_{0,0}\left(\frac{\beta_{1}}{2}\right)^{L-1}\frac{\Gamma\left(L+1+\frac{4\gamma_{1}}{\beta_{1}}\right)}{\Gamma\left(2+\frac{4\gamma_{1}}{\beta_{1}}\right)L!},
\end{empheq}
and \autoref{eq:beta1BC} and \autoref{eq:beta2BC} into \autoref{eq:mSubleadingRG}, yielding
\begin{empheq}[box=\sentencebox]{equation}
\label{eq:mSubLeadingLogRGfinal}
\begin{aligned}
& m_{L,L-1}= \left(\frac{\beta_{1}}{2}\right)^{L-3}\frac{\Gamma\left(L+1+\frac{4\gamma_{1}}{\beta_{1}}\right)}{\Gamma\left(2+\frac{4\gamma_{1}}{\beta_{1}}\right)\left(L-1\right)!}\\
 & \times\left\{ \frac{\beta_{1}^{2}}{4\left(\beta_{1}+2\gamma_{1}\right)}\left(\frac{1}{2}m_{0,0}\beta_{2}+\left(\beta_{1}+2\gamma_{1}\right)m_{1,0}+2m_{0,0}\gamma_{2}\right)\right.\\
 & \begin{aligned} & \hspace{1.05cm} +m_{0,0}\left(\frac{\beta_{1}}{4}+\gamma_{1}\right)\left[4\left(\gamma_{2}-\frac{\gamma_{1}\beta_{2}}{\beta_{1}}\right)\left[\psi\left(L+\frac{4\gamma_{1}}{\beta_{1}}\right)-\psi\left(2+\frac{4\gamma_{1}}{\beta_{1}}\right)\right]\right.\\
 & \left.\left.\hspace{1cm} +\left(\beta_{2}\left(1+\frac{4\gamma_{1}}{\beta_{1}}\right)-4\gamma_{2}\right)\left[\psi\left(L+1+\frac{4\gamma_{1}}{\beta_{1}}\right)-\psi\left(3+\frac{4\gamma_{1}}{\beta_{1}}\right)\right]\right]\right\} ,
\end{aligned}
\end{aligned}
\end{empheq}
The leading log recursion is completely specified by the input data $\gamma_{1}$, $\beta_{1}$, and $m_{0,0}$, while the subleading log recursion depends entirely on the input data $\gamma_{1}$, $\beta_{1}$, $m_{0,0}$, $\beta_{2}$, $\gamma_{2}$, and $m_{1,0}$. The structure of the recursion and a few examples are given in \autoref{fig:resultsRG}.

\subsection{Comparing Renormalization to Unitarity}

We now compare the results of \autoref{s:UnitarityconstraintsPhi4} to those of \autoref{s:RGconstraintsPhi4}. We will provide two different, equivalent perspectives. The first perspective is that the recursion relations resulting from unitarity are equivalent to those from the Callan--Symanzik equation. As summarized in \autoref{f:RecursionTowers}, the unitarity recursions begin at the lowest order in the coupling and thus require one input, namely the lowest order amplitude coefficient. The CS recursions begin at one higher order in the coupling and therefore require additional input data, which is provided by $\beta_{L}$ and $\gamma_{L}$. As a second perspective, we will then reverse the logic and show that the unitarity recursions can be inserted into the Callan--Symanzik equation to solve for the $\beta$ and $\gamma$ functions. 

We start with the leading log terms. From the explicit forms of  \autoref{eq:nLeadingLogRGfinal} and \autoref{eq:mLeadingLogRGfinal}, it is clear that the required input data is $\gamma_{1}$ and $\beta_{1}$ in the case of the 2-particle amplitude, and $\gamma_{1}$, $\beta_{1}$, and $m_{0,0}$ in the case of the 4-particle amplitude. In \autoref{a:ExactResults} we directly compute this input data to be $\gamma_{1}=0$ and $\beta_{1}=3$. Additionally, as in the unitarity case, we set the input datum $m_{0,0}=1/3$ to ensure that the tree-level term is $-\lambda$. With these inputs, we can now compare the leading log recursions from RG and unitarity. Since every level of the 2-particle leading recursion in \autoref{eq:nLeadingLogRGfinal} is proportional to $\gamma_{1}$, specifying the input data $\gamma_{1}=0$ immediately collapses the entire tower to 
\begin{equation}
n_{L,L}=0.
\end{equation}
This is precisely the condition we found from the lack of 1-particle unitarity cuts of the 2-particle amplitude in \autoref{eq:nLeadingUnitarity}.
Next, we turn to the leading log 4-particle amplitude coefficients given in \autoref{eq:mLeadingLogRGfinal}.
Inserting  the input data $\gamma_{1}=0$,
$\beta_{1}=3$, and $m_{0,0}=1/3$ returns
\begin{equation}
\label{eq:LeadingLogsRG}
m_{L,L}=\frac{1}{3}\left(\frac{3}{2}\right)^{L}.
\end{equation}
This agrees with the unitarity computation in  \autoref{eq:mLeadingLogUnitarity}. 

We can find similar agreement for the subleading log constraints. As can be seen from \autoref{eq:nSubLeadingLogRGfinal} and \autoref{eq:mSubLeadingLogRGfinal}, at the subleading log order there is a need for additional input data in the form of $\gamma_{2}$, $\beta_{2}$ and $n_{1,0}$ in the 2-particle case and $\gamma_{2}$, $\beta_{2}$ and $m_{1,0}$ in the 4-particle case. In \autoref{a:ExactResults}, the (scheme-independent) input data is computed to be $\gamma_{2}=1/12$, $\beta_{2}=17/3$, and $n_{1,0}=0$, while we leave $m_{1,0}$ as an unspecified scheme-dependent input. We begin again by comparing the the 2-particle unitarity and RG results. We previously found that the 2-particle subleading log recursion resums to \autoref{eq:nSubLeadingLogRGfinal}. Substituting the leading log result $n_{1,1}=0$, the leading log inputs $\gamma_{1}=0$ and $\beta_{1}=3$, and the subleading log inputs, $\gamma_{2}=1/12$, $\beta_{2}=17/3$, and $n_{1,0}=0$, into \autoref{eq:nSubLeadingLogRGfinal}, we find 
\begin{equation}
n_{L,L-1}=-\frac{1}{27}\left(\frac{3}{2}\right)^{L},
\end{equation}
in agreement with
unitarity in \autoref{eq:nSubleadingUnitarity}. Now for the subleading 4-particle amplitude
coefficients \autoref{eq:mSubLeadingLogRGfinal}.
Substituting the leading log result \autoref{eq:LeadingLogsRG}, the leading log inputs $\gamma_{1}=0$, $\beta_{1}=3$, and $m_{0,0}=1/3$, and the scheme-independent subleading log inputs, $\gamma_{2}=1/12$, $\beta_{2}=17/3$, and $m_{1,0}=0$, into \autoref{eq:mSubLeadingLogRGfinal}, the recursion drastically simplifies to 
\begin{equation}
m_{L,L-1}=\left(\frac{3}{2}\right)^{L}\frac{L\left(54m_{1,0}-32\right)+34LH_{L}-2}{81},
\end{equation}
precisely the unitarity result in \autoref{eq:mSubLeadingUnitarity}.

The difference in complexity between the RG and unitarity approaches is stark. While the phase space integrals on the unitarity side are labor-intensive, the fact that this gives access to the tree-level amplitude coefficient through a 1-loop cut means that the corresponding recursion tower starts at one lower order in the coupling than the RG side, leading to significantly simpler recursion relations. We have therefore shown, up to subleading order in logs, that the recursion relations resulting from unitarity are the same as those from RG, where the $\beta$ and $\gamma$ functions act as additional input data for the RG recursions. Furthermore, note that the equivalence between the unitarity and Callan--Symanzik is independent of scheme up to subleading order. As part of the Unitarity Flow Conjecture \cite{Chavda:2025}, we conjecture that these facts remain true for all orders.

We close this subsection by remarking that the logic above can be reversed: we can input the recursion relations obtained from unitarity into the Callan--Symanzik equation to find the values of $\gamma_{L}$ and $\beta_{L}$. To subleading order, this means determining the values of $\gamma_{1}$, $\gamma_{2}$, $\beta_{1}$, and $\beta_{2}$. The equations that determine the values of these parameters are \autoref{eq:gamma1BC}, \autoref{eq:gamma2BC}, \autoref{eq:beta1BC}, and \autoref{eq:beta2BC}, which followed from the CS equation and which we reproduce here in a form more convenient for extracting the $\beta$ and $\gamma$ function coefficients. In contrast to the previous argument, the values of $\gamma_{1}$, $\gamma_{2}$, $\beta_{1}$, and $\beta_{2}$ are outputs; the inputs are the recursion relations determined from unitarity (we do not assume any knowledge of the RG recursions now and exclusively use the recursions following from unitarity in \autoref{s:UnitarityconstraintsPhi4}).

We begin by determining the 1-loop $\beta$ and $\gamma$ function coefficients. The CS constraint on $\gamma_{1}$ is
\begin{equation}
\label{eq:gamma1Output}
\gamma_{1}=-n_{1,1}.
\end{equation}
In \autoref{eq:nLeadingUnitarity} we found that, since there are no 1-cuts of the 1PI diagrams, unitarity enforces $n_{1,1}=0$. Plugging this into \autoref{eq:gamma1Output} determines the first gamma function coefficient as $\gamma_{1}=0$. Next, the $\beta_{1}$ constraint from the CS equation is 
\begin{equation}
\label{eq:beta1Output}
3m_{0,0}\beta_{1}=6m_{1,1}-12m_{0,0}\gamma_{1}.
\end{equation}
With the lowest order input to the 4-particle amplitude, $m_{0,0}=1/3$, which follows from the demand that the tree-level amplitude be $-\lambda$, the leading log unitarity recursion in \autoref{eq:mLeadingLogUnitarity} gives $m_{1,1}=1/2$. Inputting this into \autoref{eq:beta1Output} correctly sets $\beta_{1}=3$.

Now we turn to the 2-loop $\beta$ and $\gamma$ function coefficients. The 2-loop $\gamma$ function obeys the RG constraint 
\begin{equation}
\label{eq:gamma2Output}
\gamma_{2}=-n_{2,1}+\frac{1}{2}\left(\beta_{1}+2\gamma_{1}\right)n_{1,0}.
\end{equation}
In \autoref{eq:n10unitarity}, unitarity sets $n_{1,0}=0$, while the subleading log recursion \autoref{eq:nSubleadingUnitarity} returns $n_{2,1}=-1/12$ for $L=2$. Together these constraints give $\gamma_{2}=1/12$. 
Finally, we consider the 2-loop $\beta$ function, which RG constrains to be 
\begin{equation}
\label{eq:beta2Output}
\beta_{2}=2\frac{m_{2,1}}{m_{0,0}}-6\beta_{1}m_{1,0}-4\gamma_{2}-12\gamma_{1}\frac{m_{1,0}}{m_{0,0}}.
\end{equation}
The 4-particle subleading log unitarity recursion in \autoref{eq:mSubLeadingUnitarity} sets $m_{2,1}=3m_{1,0}+1$. Inserting this, along with the now known values of $m_{0,0}=1/3$, $\gamma_{1}=0$, $\beta_{1}=3$ from the leading analysis and $\gamma_{2}=1/12$ as computed before, into \autoref{eq:beta2Output} we find that any appearance of the scheme-dependent term $m_{1,0}$ exactly cancels and we are left with the scheme-independent result $\beta_{2}=17/3$. Therefore, through subleading log order, although the individual amplitude coefficients depend on the choice of scheme, the recursion relations do not and we reproduce the known universality of the $\beta$ and $\gamma$ functions through 2-loops \cite{Arnone:2003pa}.

\section{Connection to Renormalized Perturbation Theory}
\label{s:ConenctionToRPT}
In the previous sections we worked with the finite, renormalized 2-particle and 4-particle amplitude ans\"{a}tze and obtained recursion relations among leading and subleading coefficients from unitarity and RG. We showed that these each give equivalent recursions, up to the selection of initial data. As we shall now demonstrate, using a renormalized amplitude is not a requirement. We can instead work with the unrenormalized, dimensionally regularized amplitude in $d = 4 - 2\epsilon$ dimensions, where we instead organize our perturbative expansion in loop and pole order in $\epsilon$.

The most general unrenormalized 4-particle amplitude ansatz consistent with the assumptions in \autoref{Introduction} (massless, marginal, IR finite, hard scattering) is  
\begin{equation}
\label{eq:UnrenormalizedAnsatz4pt}
\widetilde{\mathcal{M}}_4 = \tilde{\mu}^{2\epsilon} \left[-\lambda_0 + \sum_{L = 1}^{\infty} \frac{\left(-\lambda_0\right)^{L + 1}}{\left(16\pi^2\right)^L} \widetilde{\M}^{\left(L\right)}_4\right],
\end{equation}
where $\mu^2 = 4\pi e^{-\gamma_E} \tilde{\mu}^2$ and
\begin{equation}
\label{eq:4ptAnsatzUnexpanded}
\widetilde{\M}^{\left(L\right)}_4 = \sum_{a + b + c = L} \tilde{m}_{L, \left\{a, b, c\right\}}\left(\epsilon\right) \left(\frac{\mu^2}{-s}\right)^{a\epsilon} \left(\frac{\mu^2}{-t}\right)^{b\epsilon} \left(\frac{\mu^2}{-u}\right)^{c\epsilon} 
\end{equation}
That the powers are fixed to be $a + b + c = L$ follows from dimensional analysis. We can further assume permutation symmetry, implying that $\tilde{m}_{L, abc}$ is totally symmetric in the indices $a$, $b$, and $c$. These coefficients admit their own expansion in $\epsilon$:
\begin{equation}
\tilde{m}_{L, \left\{a, b, c\right\}} = \frac{\tilde{m}_{L, \left\{a, b, c\right\}, L}}{\epsilon^{L}} + \frac{\tilde{m}_{L, \left\{a, b, c\right\}, L - 1}}{\epsilon^{L - 1}} + \dots = \sum_{L^\prime = -\infty}^{L} \frac{\tilde{m}_{L, \left\{a, b, c\right\}, L^\prime}}{\epsilon^{L'}}. 
\end{equation}
Plugging this into \autoref{eq:UnrenormalizedAnsatz4pt} and expanding to subleading order in $\epsilon$, we find 
\begin{equation}
\label{eq:4ptAnsatz}
\widetilde{\M}_4^{\left(L\right)} = \frac{M_{L, L}}{\epsilon^L} + \frac{M_{L, L - 1}}{\epsilon^{L - 1}} + \frac{L M_{L, L}}{3\epsilon^{L - 1}}\left(\log_{-s} + \log_{-t} + \log_{-u}\right) + \dots ,
\end{equation}
where 
\begin{equation}
M_{L, L^\prime} \equiv \sum_{a + b + c = L} \tilde{m}_{L, \left\{a, b, c\right\}, L^\prime}.
\end{equation}
To derive the coefficients of the logarithms, we used the symmetry of $\tilde{m}_{L, \left\{a, b, c\right\}, L^\prime}$ to note that
\begin{equation}
\sum_{a + b + c = L} a \tilde{m}_{L, \left\{a, b, c\right\}, L^\prime} = \frac{1}{3} \sum_{a + b + c = L} \left(a + b + c\right) \tilde{m}_{L, \left\{a, b, c\right\}, L^\prime} = \frac{L}{3} M_{L, L^\prime}. 
\end{equation}
For the unrenormalized amplitude, we will refer to the $\epsilon^{-L}$ term as the \textit{leading divergence}, the $\epsilon^{-L + 1}$ term as the \textit{subleading divergence}, etc. Then $M_{L,L}$ is the \textit{leading $\epsilon$ coefficient} and $M_{L,L-1}$ is the \textit{subleading $\epsilon$ coefficient} for the 4-particle amplitude. 
Similarly, the 2-particle amplitude is 
\begin{equation}
\widetilde{\M}_2 = p^2 - \sum_{L = 1}^{\infty} \frac{\left(-\lambda_0\right)^L}{\left(16\pi^2\right)^L} \widetilde{\M}_2^{\left(L\right)}, 
\end{equation}
where 
\begin{equation}
\label{eq:2ptAnsatz}
p^{-2} \widetilde{\mathcal{M}}_{2}^{(L)} = \frac{N_{L, L}}{\epsilon^L} + \frac{N_{L, L - 1}}{\epsilon^{L - 1}} + \frac{L N_{L, L}}{\epsilon^{L - 1}}\log_{p^2} + \dots
\end{equation}
We will similarly refer to $N_{L,L}$ as the \textit{leading $\epsilon$ coefficient} and $N_{L,L-1}$ as the \textit{subleading $\epsilon$ coefficient} for the 2-particle amplitude. In the subsequent subsections we will demonstrate that unitarity can be applied to the unrenormalized amplitude ans\"atze to derive recursion relations among amplitude coefficients that are the same as those implied by finiteness constraints on the renormalized amplitude.

\subsection{Constraints from Unitarity}
We will approach the constraints on the leading and subleading $\epsilon$ coefficients in the same way we approached the leading and subleading logarithm coefficients, meaning it will be organized by the number of exchanged particles on the cut. As shown in \autoref{eq:4ParticlePicture} and \autoref{eq:2ParticlePicture}, the leading $\epsilon$ coefficients of the 4-particle amplitude can be computed just from the 2-cuts. Then, both the 2-cuts and 4-cuts will be sufficient to constrain the subleading $\epsilon$ coefficients. As a byproduct of having constrained the leading 4-particle divergences, we will also be able to constrain the subleading $\epsilon$ coefficients of the 2-particle amplitude.

\subsubsection{Leading 4-particle Constraints}
We will begin with the leading $\epsilon$ constraints on the 4-particle amplitude. The unitarity equation truncated at 2-cuts implies 
\begin{equation}
\Im\left(\widetilde{\M}^{\left(L\right)}_4\right) = \frac{16\pi^2 \tilde{\mu}^{2\epsilon}}{2} \sum_{L^\prime = 0}^{L - 1} \int \widetilde{\M}_4^{\left(L^\prime\right)} \left(\widetilde{\M}_4^{\left(L - L^\prime - 1\right)}\right)^\ast \dd{\Phi}_2 + \dots,
\end{equation}
where the ellipsis contains higher-particle cuts. Noting that in physical kinematics, where $s > 0$ and $-s < t, u < 0$, 
\begin{equation}
\text{Im}\left(\left(\frac{\mu^2}{-s}\right)^{a\epsilon \:} \right) = \left(\frac{\mu^2}{s}\right)^{a\epsilon} \sin\left(a\pi\epsilon\right). 
\end{equation}
Multiplying this by $\tilde{m}_{L,\left\{a, b, c\right\}}$ and expanding, we find that the imaginary part of \autoref{eq:4ptAnsatzUnexpanded} is 
\begin{equation}
\text{Im}\left(\mathcal{M}_4^{\left(L\right)}\right) = \frac{L \pi M_{L, L}}{3 \epsilon^{L - 1}} + \frac{L \pi M_{L, L - 1}}{3\epsilon^{L - 2}} + \frac{M^{a^2}_{L, L}}{\epsilon^{L - 2}} \log_s + \frac{M^{ab}_{L, L}}{\epsilon^{L - 2}} \left(\log_{-t} + \log_{-u}\right),
\end{equation}
where we introduce the weighted sum notation
\begin{equation}
M^A_{L, L^\prime} = \sum_{a + b + c = L} A\left(a, b, c\right) m_{L, \left\{a, b, c\right\}, L^\prime}. 
\end{equation}

The leading contribution to the 2-cuts comes from the product of the two leading coefficients:
\begin{equation}
\eval{\Im\left(\widetilde{\M}^{\left(L\right)}_4\right)}_{2-\text{cuts}} = \frac{16\pi^2 \tilde{\mu}^2}{2 \epsilon^{L - 1}} \sum_{L^\prime = 0}^{L - 1} M_{L^\prime, L^\prime} M_{L - L^\prime - 1, L - L^\prime - 1} \Phi_2.
\end{equation}
Utilizing \autoref{eq:2particlePhaseSpace} and expanding to next-to-leading order in $\epsilon$, we find 
\begin{equation}
\label{eq:1stComponent}
\eval{\Im\left(\widetilde{\M}^{\left(L\right)}_4\right)}_{2-\text{cuts}} = \frac{\pi}{2\epsilon^{L - 1}} \sum_{L^\prime = 0}^{L - 1} M_{L^\prime, L^\prime} M_{L - L^\prime - 1, L - L^\prime - 1}\left(1 + \left(2 + \log_s\right)\epsilon + \mathcal{O}\left(\epsilon^2\right)\right).
\end{equation}
The term linear in $\epsilon$ will be useful for the computation of the subleading $\epsilon$ coefficients. Matching at order $1/\epsilon^{L - 1}$, we find
\begin{equation}
\frac{L}{3} M_{L, L} = \frac{1}{2} \sum_{L^\prime = 0}^{L - 1} M_{L^\prime, L^\prime} M_{L - L^\prime - 1, L - L^\prime - 1},
\end{equation}
with the initial data fixed by the tree-level result, $M_{0, 0} = 1$. The solution to this recursion relation is then 
\begin{equation}
\label{eq:LeadingDivUnitarity}
M_{L, L} = \left(\frac{3}{2}\right)^L.
\end{equation}
Note that the only integral we needed to compute is the same as in the case of leading logarithms for the renormalized amplitude, the full 2-particle phase space integral. 

\subsubsection{Subleading 4-particle Constraints}
The subleading $\epsilon$ constraints will now be fixed by both 2- and 4-cuts, 
\begin{equation}
\begin{aligned}
\Im\left(\widetilde{\M}^{\left(L\right)}_4\right) &= \frac{16\pi^2 \tilde{\mu}^{2\epsilon}}{2} \sum_{L^\prime = 0}^{L - 1} \int \widetilde{\M}_4^{\left(L^\prime\right)} \left(\widetilde{\M}_4^{\left(L - L^\prime - 1\right)}\right)^{\ast} \dd{\Phi}_2 \\
&\hspace{0.4cm} + \frac{\left(16\pi^2\right)^3 \tilde{\mu}^{2\epsilon}}{2} \sum_{L^\prime = 0}^{L - 3} \int \widetilde{\M}_6^{\left(L^\prime\right)} \left(\widetilde{\M}_6^{\left(L - L^\prime - 3\right)} \right)^{\ast} \dd{\Phi}_4 + \dots
\end{aligned}
\end{equation}
The three contributions to the subleading divergence come from the subleading piece of the 2-particle phase space integral shown in \autoref{eq:1stComponent}, the product of leading and subleading $\epsilon$ coefficients of the amplitudes on the left and right of the cut, and the 4-cut.

Focusing first on the 2-cuts, as we just stated, one contribution to the 2-cut is from \autoref{eq:1stComponent}, 
\begin{equation}
\eval{\Im\left(\widetilde{\M}^{\left(L, L - 1\right)}_4\right)}_{2-\text{cuts}, 1} = \frac{\pi \left(L - 1\right)}{3\epsilon^{L - 1}} \left(\frac{3}{2}\right)^L \left(2 + \log_s\right). 
\end{equation}
Here, we used the result from the leading $\epsilon$ coefficients, \autoref{eq:LeadingDivUnitarity}. The subscript $1$ indicates this is the first contribution from the 2-cut and the superscript $\left(L, L - 1\right)$ indicates the order in coupling and that we are studying the subleading divergence. The second contribution from the 2-cut after simplifying and using the phase space integrals \autoref{eq:2particlePhaseSpace} and \autoref{eq:logsinint} 
\begin{equation}
\label{eq:2ndComponent}
\eval{\Im\left(\widetilde{\M}^{\left(L, L - 1\right)}_4\right)}_{2-\text{cuts}, 2} = \frac{2\pi}{3\epsilon^{L - 2}} \left(\frac{3}{2}\right)^{L} \left[\frac{L\left(L - 1\right)}{3}\left(\! 1 \! + \! \frac{3}{2}\log_s \!\right) + \sum_{L^\prime = 1}^{L - 1}\left(\frac{2}{3}\right)^{L^\prime} \! M_{L^\prime, L^\prime - 1}\right].
\end{equation}

The last contribution comes from the 4-cut. The procedure is very similar to the computation of the 4-cut for the renormalized amplitude. We first note that the 6-particle amplitude factorizes into products of 4-particle amplitudes. The analog of \autoref{eq:6-particle factorization} is
\begin{equation}
\widetilde{\M}_6^{\left(L\right)} \! = \! \frac{\tilde{\mu}^{2\epsilon}}{3\epsilon^L} \left(\frac{3}{2}\right)^L \! \frac{-i}{\left(p_1 \! + \! p_2 \! - \! \ell_1\right)^2 \! - \! i\varepsilon} \! \sum_{L^\prime = 0}^{L} \left[ \! \left(\frac{\mu^2}{\left(\ell_2 + \ell_3\right)^2}\right)^{L^\prime} \! + \! \left(\ell_3 \leftrightarrow \ell_4\right) \! + \! \left(\ell_2 \leftrightarrow \ell_4\right) \! \right] + \text{perms}.
\end{equation}
By explicit computation, discussed in \autoref{a:4ParticlePhaseSpace}, the argument shown in \autoref{eq:PropCorrectionCuts} still holds. Only products of the above permutation of external legs in the 6-particle amplitude with itself will yield an enhancement. In this instance, the enhancement is not in logarithmic order but rather in divergence. Only some 4-particle phase space integrals have an additional $\epsilon$-divergence. The leading divergence of the 4-particle phase integrals computed using the above amplitude is shown in \autoref{eq:EhnancedLeadingDiv}. Therefore, 
\begin{equation}
\eval{\Im\left(\widetilde{\M}^{\left(L, L - 1\right)}_4\right)}_{4-\text{cuts}} = -\frac{4\pi}{81\epsilon^{L - 2}} \left(\frac{3}{2}\right)^{L} \sum_{L^\prime = 0}^{L - 3} \sum_{L_1 = 0}^{L^\prime} \sum_{L_2 = 0}^{L - L^\prime - 3} \frac{1}{L_1 + L_2 + 2}.
\end{equation}
Evaluating the sum then yields
\begin{equation}
\label{eq:3rdComponent}
\eval{\Im\left(\widetilde{\M}^{\left(L, L - 1\right)}_4\right)}_{4-\text{cuts}} = -\frac{2\pi}{81\epsilon^{L - 2}} \left(\frac{3}{2}\right)^{L} \left(L - 2\right)\left(LH_{L - 1} + 2\right).
\end{equation}
Combining the results in \autoref{eq:1stComponent}, \autoref{eq:2ndComponent}, and \autoref{eq:3rdComponent}, we find that the recursion relation among the subleading $M$ coefficients is
\begin{equation}
\label{eq:UnitaritySubMRecursion}
M_{L, L - 1} = \left(\frac{3}{2}\right)^{L - 1}\! \left[L + 2 - \frac{L^2 + L - 2 - 2 LH_{L - 1}}{18L} + \! \frac{3}{L}\sum_{L^\prime = 0}^{L - 1} \! \left(\frac{2}{3}\right)^{L^\prime}\! M_{L^\prime, L^\prime - 1}\right].
\end{equation}
The recursion relation is valid for all $L \geq 2$, with $M_{1, 0}$ as input data. With that said $M_{1,0}$ can be determined from the unitarity cut and computation of the 2-particle phase space integral to be $M_{1, 0} = 3$. The solution to the recursion is then just
\begin{equation}
\label{eq:SubLeadingDivUnitarityMs}
M_{L, L - 1} = \frac{2}{27}\left(\frac{3}{2}\right)^L \left[\left(17 L + 16\right) H_{L} - 6 L\right]. 
\end{equation}

We additionally note that at subleading order, we can also match powers of log. Since there are no $\log_t$ or $\log_u$ terms that arose from computing the cuts, we find that
\begin{equation}
M^{ab}_{L, L} = \sum_{a + b + c = L} ab \: \tilde{m}_{L, \left\{a, b, c\right\}, L} = 0.
\end{equation}
This means that, at this order, there are no terms that involve mixing channels in logarithm, reminiscent of the Steinmann relations as discussed in \autoref{ss:Steinmann}. The coefficient of $\log_s$, on the other hand, is non-trivial but yields the same constraint as the leading divergence. 

\subsubsection{Subleading 2-particle Constraints}
As in the renormalized case, from the coefficients of the leading 4-particle divergences we are able to constrain the form of the leading non-trivial divergences for the 2-particle amplitude. Since the 2-particle amplitude by definition is 1PI, there are no 1-cuts, implying once again that
\begin{equation}
\label{eq:LeadingDivUnitarityNs}
N_{L, L} = 0.
\end{equation}
The leading divergence comes from 3-cuts, meaning the coefficients of leading non-trivial divergences in the 2-particle amplitude are related to coefficients in the 4-particle amplitude,
\begin{equation}
\text{Im}\left(\widetilde{\M}_2^{\left(L\right)}\right) = \frac{\left(16\pi^2\right)^2 \left(\tilde{\mu}^2\right)^2}{2} \sum_{L' = 0}^{L - 2} \int \widetilde{\M}_{4}^{\left(L'\right)} \left(\widetilde{\M}_{4}^{\left(L - L' - 2\right)}\right)^{\ast} \dd{\Phi}_3.
\end{equation}
At $L$-loop order, the left-hand side of the unitarity equation using \autoref{eq:2ptAnsatz} is: 
\begin{equation}
\text{Im}\left(\widetilde{\M}_2^{(L)}\right) = \frac{p^2 \pi L N_{L, L - 1}}{\epsilon^{L - 2}}. 
\end{equation}
On the right-hand side of the unitarity equation, using the leading $\epsilon$ coefficient of the 4-particle amplitude, \autoref{eq:LeadingDivUnitarity}, and summing over $L^\prime$, we find 
\begin{equation}
\eval{\text{Im}\left(\widetilde{\M}_2^{(L)}\right)}_{\text{3-cuts}} = \frac{\left(16\pi^2\right)^2}{\epsilon^{L - 2}} \left(\frac{3}{2}\right)^{L - 2} \left(L - 1\right) \tilde{\mu}^{4\epsilon}\Phi_3.
\end{equation}
Using \autoref{eq:3ParticlePhaseSpace}, we find 
\begin{equation}
\eval{\text{Im}\left(\widetilde{\M}_2^{(L)}\right)}_{\text{3-cuts}} = -\frac{p^2 \pi}{\epsilon^{L - 2}} \left(\frac{3}{2}\right)^{L - 2} \frac{\left(L - 1\right)}{12}.
\end{equation}
Hence, the coefficient of the leading non-trivial divergence for the 2-particle amplitude is
\begin{equation}
\label{eq:SubLeadingDivUnitarityNs}
N_{L, L - 1} = -\left(\frac{3}{2}\right)^{L - 2} \frac{L - 1}{12L}.
\end{equation}
Hence, analogously to the renormalized amplitude, knowledge of the leading coefficients of the 4-particle amplitude are sufficient to reconstruct the $L$-order subleading divergences of the 2-particle amplitude. 

\subsection{Constraints from Finiteness}

Because we work with unrenormalized amplitude ans\"{a}tze, \autoref{eq:2ptAnsatz} and \autoref{eq:4ptAnsatz}, the Callan--Symanzik equation is no longer applicable. Instead, we will be able to derive constraints through imposing finiteness on the unrenormalized amplitude, upon substitution of the counterterms. This method most closely matched how renormalized perturbation theory is done in practice. We define the counterterms through the relationship between the bare and renormalized parameters. Starting with the action
\begin{equation}
S = \int -\left(\frac{1}{2}\partial_\mu \phi_0 \partial^\mu \phi_0 + \frac{\tilde{\mu}^{2\epsilon} \lambda_0}{4!} \phi_0^4\right) \dd[d]{x},
\end{equation}
the bare (which carry a subscript $0$) and renormalized parameters are related by
\begin{equation}
\label{eq:RenormalizationDefinitions}
\phi_0 = Z_\phi^{1/2} \phi, \qquad \lambda_0 = Z_\lambda \lambda, \qquad \mathcal{M}_n = Z_\phi^{n/2} \widetilde{\M}_n,
\end{equation}
where $\M_n$ is a finite renormalized amplitude and $\widetilde{\M}_n$ is an unrenormalized amplitude. The counterterms admit a power series expansion in the coupling and a Laurent expansion in $\epsilon$ poles:
\begin{subequations}
\begin{align}
\label{eq:ZlambdaDefinition}
Z_\lambda &= 1 + \sum_{L^\prime = 0}^{\infty} \sum_{k = 0}^{L} \frac{z_{L^\prime, k}}{\epsilon^k}\frac{\lambda^{L^\prime}}{\left(16\pi^2\right)^{L^\prime}},\\
Z_\phi &= 1 + \sum_{L^\prime = 0}^{\infty} \sum_{k = 0}^{L} \frac{y_{L^\prime, k}}{\epsilon^k}\frac{\lambda^{L^\prime}}{\left(16\pi^2\right)^{L^\prime}},
\label{eq:ZphiDefinition}
\end{align}
\end{subequations}
where we remain agnostic about scheme dependence by including $k=0$ counterterms (these are zero for the $\overline{\mathrm{MS}}$ scheme, for example).

Our goal will be to constrain the leading and subleading $\epsilon$ coefficients. Upon substitution of \autoref{eq:RenormalizationDefinitions}, \autoref{eq:ZlambdaDefinition} and \autoref{eq:ZphiDefinition} into \autoref{eq:2ptAnsatz} and \autoref{eq:4ptAnsatz}, the amplitudes admit a power series in $\lambda$ and a Laurent series in $\epsilon$. Imposing that the counterterms must be such that all $\epsilon$ poles vanish in the renormalized amplitude places non-trivial constraints among the counterterms and amplitude coefficients. These constraints can be solved recursively order-by-order in the loop expansion, with each order giving simultaneous equations between the 2-particle amplitude and 4-particle amplitude whose solutions are then plugged in to the next order constraints. With some work, a pattern emerges. The $z_{L,k}$ counterterms form closed recursion relations among themselves both at leading and subleading $\epsilon$ order, as do the $y_{L,k}$ counterterms, the $n_{L,k}$ 2-particle amplitude coefficients and the $M_{L,k}$ 4-particle amplitude coefficients.

Given the intricate nature of nested divergences that arise as we go to higher loop order, reflected in the many appearances of the coupling $\lambda$ inside sums and powers in \autoref{eq:ZlambdaDefinition} and \autoref{eq:ZphiDefinition}, this derivation of recursion relations is highly non-trivial. Our approach reflects this difficulty: we analyzed low loop explicit outputs and deduced a pattern of recursion relations which we then conjectured to all orders. We then verified our recursion relations are correct up to 7-loop order. Indeed, this is one of the key benefits of the unitarity approach: we are able to obtain analytic all-orders results directly by computing phase space integrals rather than noticing a pattern as we will do here. We state these final formulae for the recursions below and then compare them to the predictions from unitarity.

The leading coupling constant counterterms $z_{L,L}$ obey the recursion
\begin{equation}
z_{L, L} = z_{1, 1}^{L}
\end{equation}
for $L\geq1$ subject to the boundary condition that ensures finiteness: 
\begin{equation}
z_{1, 1} = M_{1, 1} + 2N_{1, 1}.
\end{equation}
The leading wavefunction counterterms $y_{L,L}$ obey the constraint, for $L\geq1$,
\begin{equation}
y_{L, L} = -N_{L, L},
\end{equation}
where the leading logarithmic 2-particle amplitude coefficients satisfy the recursion 
\begin{equation}
\label{eq:LeadingDivRGNs}
N_{L,L}=\frac{N_{1,1}}{L!}\left(M_{1,1}+2N_{1,1}\right)^{L-1}\frac{\Gamma\left(L-\frac{N_{1,1}}{M_{1,1}+2N_{1,1}}\right)}{\Gamma\left(1-\frac{N_{1,1}}{M_{1,1}+2N_{1,1}}\right)},
\end{equation}
valid for $L\geq1$. Finally, the leading logarithm 4-particle amplitude coefficients follow the recursion
\begin{equation}
\label{eq:LeadingDivRGMs}
M_{L,L}=\frac{\left(M_{1,1}+2N_{1,1}\right)^{L}}{L!}\frac{\Gamma\left(L+1-\frac{2N_{1,1}}{M_{1,1}+2N_{1,1}}\right)}{\Gamma\left(1-\frac{2N_{1,1}}{M_{1,1}+2N_{1,1}}\right)},
\end{equation}
for $L\geq1$.

In a similar way to what was found in the Callan--Symanzik analysis of the renormalized amplitude in \autoref{s:RGconstraintsPhi4}, the finiteness recursions start at one higher order in the coupling than those coming from unitarity, requiring additional inputs to completely specify the whole tower. As we saw in the previous subsection, the only input needed for the leading divergence unitarity recursions \autoref{eq:LeadingDivUnitarity} and \autoref{eq:LeadingDivUnitarityNs} was the tree-level data $M_{0,0}=1$. For \autoref{eq:LeadingDivRGNs} and \autoref{eq:LeadingDivRGMs}, on the other hand, we also need to know $N_{1,1}$ and $M_{1,1}$. This extra initial data is computed in \autoref{a:ExactResults}. We can read off $M_{1,1}=3/2$ by expanding \autoref{eq:alpha1expansion} in small $\epsilon$ and multiplying by 3 for the three channels, while by inspection we see that \autoref{eq:UnrenormalizedM2exact} sets $N_{1,1} = 0$. Plugging these inputs into \autoref{eq:LeadingDivRGNs} and \autoref{eq:LeadingDivRGMs}, immediately returns
\begin{equation}
M_{L,L}=\left(\frac{3}{2}\right)^{L},
\end{equation}
in agreement with \autoref{eq:LeadingDivUnitarity}, and 
\begin{equation}
N_{L,L}=0,
\end{equation}
as was found in \autoref{eq:LeadingDivUnitarityNs} from the lack of 1-particle unitarity cuts.

We now push the finiteness analysis further to the subleading divergence order, for which the analysis becomes increasingly involved. In order to make the following computations doable, we used the leading recursions. We additionally inserted the initial data that $n_{1, 1} = 0$, meaning $n_{L, L} = 0$ for all $L$ as well. With this, we find the following recursion for the subleading coupling constant counterterms $z_{L,L-1}$:
\begin{equation}
\label{eq:zSubLeading}
z_{L, L - 1} = M_{1,1}^{L - 2} \left(2\left(H_L - L\right)\left(M_{2, 1}  + 2N_{2, 1} - 2M_{1, 1} \left(M_{1, 0} + N_{1, 0}\right)\right) + L M_{1, 1} z_{1, 0}\right),
\end{equation}
while for $y_{L,L-1}$, the subleading wavefunction counterterms, we find the recursion:
\begin{equation}
y_{L, L - 1} = -\frac{2}{L} \left(M_{1, 1}^{L - 1} N_{1, 0} - M_{1, 1}^{L - 2} N_{2, 1}\right).
\end{equation}
The subleading 2-particle amplitude coefficients in \autoref{eq:2ptAnsatz} obey
\begin{equation}
\label{eq:SubLeadingDivRGNs}
N_{L, L - 1} = \frac{1}{L}\left(-\left(L - 2\right) M_{1, 1}^{L - 1} N_{1, 0} + 2\left(L - 1\right) M_{1, 1}^{L - 2} N_{2, 1}\right).
\end{equation}
Finally, we are able to find a closed recursion among the subleading 4-particle amplitude coefficients defined in \autoref{eq:4ptAnsatz}: 
\begin{equation}
\begin{aligned}
\label{eq:SubLeadingDivRGMs}
M_{L, L - 1} = -\left(-M_{1, 1}\right)^L&\bigg[\frac{4\left(M_{2, 1} - 2 M_{1, 0} M_{1, 1} - \left(L - 4\right)\left(N_{2, 1} - M_{1, 1} N_{1, 0}\right)\right)}{M_{1, 1}^2 L\left(L - 1\right)\left(L - 2\right)} \\
&\hspace{0.5cm} + \sum_{L^\prime = 1}^{L - 1} \left(-M_{1, 1}\right)^{-L^\prime} \binom{L - 1}{L^\prime - 1} M_{L^\prime, L^\prime - 1}
\bigg].
\end{aligned}
\end{equation}
The solution to this recursion relation is
\begin{equation}
\label{eq:SubLeadingDivMFinal}
\begin{aligned}
M_{L, L - 1} = M_{1, 1}^{L - 2}\bigg[&2\left(\left(L + 1\right) H_L - 2L \right) M_{2, 1} - \left(4\left(L + 1\right) H_L - 9 L\right) M_{1, 1} M_{1, 0} \\
&+ 4\left(\left(L + 2\right) H_L - 3L\right)\left(N_{2, 1} - M_{1, 1} N_{1, 0}\right)\bigg].
\end{aligned}
\end{equation}
Note that the recursions for $N_{L,L-1}$ and $M_{L,L-1}$ are scheme-independent, with no dependence on the finite $z_{k,0}$ nor $y_{k,0}$. This followed from a highly non-trivial cancellation of scheme-dependent $z_{1,0}$ terms that appear in the counterterm recursion \autoref{eq:zSubLeading} but drop out of the decoupled amplitude coefficient recursions.

We again see a dependence on new input data in \autoref{eq:SubLeadingDivRGNs} and \autoref{eq:SubLeadingDivMFinal} that was not needed in the unitarity computation of \autoref{eq:SubLeadingDivUnitarityNs} and \autoref{eq:SubLeadingDivUnitarityMs}, namely a dependence on $M_{1,0}$ (which could be computed directly in unitarity), $N_{1,0}$, $M_{2,1}$, and $N_{2,1}$. We again compute this input data in \autoref{a:ExactResults}. Taking the $\epsilon\rightarrow0$ limit of \autoref{eq:alpha1expansion} and \autoref{eq:alpha2expansion} and multiplying by 3 for three channels gives us that $M_{1,0}=3$ and $M_{2,1}=21/2$. Taking the small-$\epsilon$ expansion of \autoref{eq:UnrenormalizedM2exact} gives $N_{1,0}=0$ and $N_{2,1}=-1/24$. With this input data, along with the leading divergence inputs $M_{1,1}=3/2$ and $N_{1,1}=0$, the subleading 2-particle amplitude coefficient recursion \autoref{eq:SubLeadingDivRGNs} simplifies to 
\begin{equation}
N_{L, L - 1} = -\left(\frac{3}{2}\right)^{L - 2} \frac{L - 1}{12L},
\end{equation}
in agreement with the unitarity recursion \autoref{eq:SubLeadingDivUnitarityNs}. Similarly, with the given input data, the expression for $M_{L, L - 1}$, \autoref{eq:SubLeadingDivMFinal}, becomes 
\begin{equation}
M_{L, L - 1} = \frac{2}{27} \left(\frac{3}{2}\right)^L\left(\left(17 L + 16\right) H_L - 6 L\right),
\end{equation}
in agreement with the unitarity prediction, \autoref{eq:SubLeadingDivUnitarityMs}.

Although upon insertion of the initial data the final answers agree, it is not obvious that this had to be the case. This is perhaps most striking when taking one step back and looking at the unsolved recursion relation, for example \autoref{eq:SubLeadingDivRGMs} from RG versus \autoref{eq:UnitaritySubMRecursion} from unitarity. In the former case, the recursion involves binomial coefficients and different powers of $L$ in the denominator, which is not present in the unitarity case, and the two recursions begin at different starting points. 

Finally, we state that in this subsection we have used the lowest loop values of the leading and subleading amplitude coefficients as input data for the recursion relations. An alternative point of view that we adopted in the previous section was to instead use the $\beta$ and $\gamma$ functions as input data. For example, $\beta_2$ was an input as opposed to $m_{2, 1}$ in \autoref{eq:mSubLeadingLogRGfinal}. A similar approach can be taken here, but one would need to compute the $\beta$ and $\gamma$ functions in terms of the counterterms using the $\mu$ independence of the bare coupling, $\tilde{\mu}^{2\epsilon} \lambda_0$, as done in \cite[Ch.~10]{Kleinert:2001}. 

\section{Discussion}
\label{s:Discussion}

In this paper, we bootstrapped the structure of renormalization at leading and subleading orders, based on the assumptions of symmetry, analyticity, and unitarity. Working in the massless quartic scalar theory, we found that imposing unitarity of the $S$-matrix leads to a web of identities that match those that would have normally been obtained as a consequence of the renormalization group equations. The only difference is in the initial conditions, which on the unitarity side are sourced by the constant coefficients of the amplitude (scheme dependence), and on the renormalization side are sourced additionally by the beta function and the anomalous dimension. The unitarity recursions are summarized in \autoref{fig:results} and the renormalization ones in \autoref{fig:resultsRG2} and \autoref{fig:resultsRG}, as well as the equations quoted there. Matching these initial conditions therefore gives a purely on-shell derivation of the beta function and the anomalous dimension. 

There are a number of natural directions for future research. First, the Unitarity Flow Conjecture states that our findings extend to all orders in the logarithmic expansion and all orders in perturbation theory \cite{Chavda:2025}. For example, our expectation is that the subsubleading order will require computing unitarity cuts over 6 intermediate states and up to $8$-particle amplitudes. It would be interesting to do these computations by recognizing which parts of the phase space integration lead to the most important contributions, perhaps employing a version of the method of regions \cite{Beneke:1997zp,Jantzen:2012mw}.

Along similar lines, it is worth understanding how to obtain the differential equation governing renormalization, the Callan--Symanzik equation, directly from unitarity. In this work, we showed that both of them structurally lead to the same recursion relations, except for the initial conditions. It would be interesting to close the circle and instead derive the CS equation from scratch, which could serve as a starting point for reformulating RG flows based on unitarity.

Beyond the massless $\lambda \phi^4$ theory, the obvious direction is to include the mass and study to what extent the mass anomalous dimension can be constrained using unitarity. Naively, this is a more difficult feat because mass would appear through logarithms of the type $\log(\frac{\mu^2}{m^2})$ which cannot be cut. Here ideas of generalized unitarity might prove useful \cite{Bern:1994cg}. Another natural extension is to include effective field theory corrections, whereby we extend the amplitude ansatz to include polynomial terms at all multiplicities. At least at tree-level and one-loop, it is known that unitarity constraints can lead to certain bounds on the Wilson coefficients \cite{deRham:2022hpx}. It would be interesting to understand the interplay between such bounds and our work.

Towards the direction of applying our techniques to more realistic theories such as QCD, it will be necessary to understand how infrared divergences modify our constraints. Two natural theories to consider are $\mathcal{N}=4$ super Yang--Mills and $g\phi^3$ theory in six dimensions, which can serve as toy model theories with non-trivial infrared dynamics. Here, one may employ Regge-physics techniques along the lines of \cite{Rothstein:2023,Rothstein:2024fpx}, where one split logarithms further into ``rapidity logarithms''. 

Lastly, multi-particle unitarity seems to give a new handle on studying the analyticity structure of the S-matrix. For example, we have shown that at the leading and subleading order, overlapping logarithms of the type $\log_{-s}^a \log_{-t}^b$ with $a+b=L$ and $L-1$ cannot occur. This result has conceptually different origins from the proof of Steinmann relations which relies on causality as the underlying principle \cite{Steinmann1960a,Steinmann1960b} (though it is known that the proof does not apply to massless theories and 4-particle amplitudes \cite{Britto:2024}). Therefore, it would be interesting to extend our proof to higher orders. Along the same vein, the structure of more complicated iterated discontinuities, and hence the monodromy group of the $S$-matrix across Riemann sheets, should be constrained by the same principles. We leave the exploration of these fascinating topics until future work.

\acknowledgments

We thank Nima Arkani-Hamed, Miguel Correia, Frederik Denef, Mathieu Giroux, and Giulio Salvatori for useful comments and discussions. A.C. is supported by the National Science Foundation Graduate Research Fellowship under Grant No. DGE-2036197. The work of D.M. and J.S. is supported by the US Department of Energy grant DE-SC011941.


\appendix

\section{On-Shell Large \texorpdfstring{$N$}{N} Structure}
\label{a:onShellLargeON}
In the main text, we used the O$\left(N\right)$ model as a helpful toy model since its flavor-ordered amplitude had no 4-cuts, no wavefunction renormalization, and only depended on one Mandelstam variable as opposed to three. In this appendix, we provide an on-shell argument for why the amplitude must have these properties in the large $N$ limit. 

The most general amplitude involving four scalars with flavor, labeled by Latin letters $i, j, k, l$, can be decomposed into three unique tensor structures: 
\begin{equation}
\M_{ij; kl}\left(s, t, u\right) = \frac{1}{N}\left[\M_s\left(s, t, u\right) \delta_{ij} \delta_{kl} + \M_t\left(s, t, u\right) \delta_{jk} \delta_{il} + \M_u\left(s, t, u\right) \delta_{ik} \delta_{jl}\right], 
\end{equation}
where the factor of $N$ ensures that 4-particle amplitudes at higher order do not diverge in the large $N$ limit and that at lowest order in $\lambda$ we have $\M_s = -\lambda$. The first term is called the $s$-flavor ordered amplitude, the second is the $t$-flavor-ordered amplitude, and the third is the $u$-flavor-ordered amplitude. This can be expressed diagrammatically, with equivalent colors indicating a Kronecker $\delta$:
\begin{equation}
\M_{ij; kl} = \: \: 
\frac{1}{N} \: \begin{gathered}
\begin{tikzpicture}[scale=0.6]
\draw[line width=0.3mm, color=RoyalBlue] (-1, 1) -- (-0.5*0.707, 0.5*0.707);
\node[RoyalBlue] at (-1.3, -1) {$i$};
\node[RoyalBlue] at (-1.3, 1) {$j$};
\draw[line width=0.3mm, color=RoyalBlue] (-1, -1) -- (-0.5*0.707, -0.5*0.707);
\filldraw[pattern=north west lines] (0, 0) circle (0.5cm);
\draw[line width=0.3mm] (0, 0) circle (0.5cm);
\draw[line width=0.3mm, color=Maroon] (0.5*0.707, 0.5*0.707) -- (1, 1);
\draw[line width=0.3mm, color=Maroon] (0.5*0.707, -0.5*0.707) -- (1, -1);
\node[Maroon] at (1.3, 1) {$k$};
\node[Maroon] at (1.3, -1) {$l$};
\end{tikzpicture}
\end{gathered} \: \: + \: \: 
\frac{1}{N} \: \begin{gathered}
\begin{tikzpicture}[scale=0.6]
\draw[line width=0.3mm, color=RoyalBlue] (-1, 1) -- (-0.5*0.707, 0.5*0.707);
\node[Maroon] at (-1.3, -1) {$i$};
\node[RoyalBlue] at (-1.3, 1) {$j$};
\draw[line width=0.3mm, color=Maroon] (-1, -1) -- (-0.5*0.707, -0.5*0.707);
\filldraw[pattern=north west lines] (0, 0) circle (0.5cm);
\draw[line width=0.3mm] (0, 0) circle (0.5cm);
\draw[line width=0.3mm, color=RoyalBlue] (0.5*0.707, 0.5*0.707) -- (1, 1);
\draw[line width=0.3mm, color=Maroon] (0.5*0.707, -0.5*0.707) -- (1, -1);
\node[RoyalBlue] at (1.3, 1) {$k$};
\node[Maroon] at (1.3, -1) {$l$};
\end{tikzpicture}
\end{gathered} \: \: + \: \: 
\frac{1}{N} \: \begin{gathered}
\begin{tikzpicture}[scale=0.6]
\draw[line width=0.3mm, color=Maroon] (-1, 1) -- (-0.5*0.707, 0.5*0.707);
\node[RoyalBlue] at (-1.3, -1) {$i$};
\node[Maroon] at (-1.3, 1) {$j$};
\draw[line width=0.3mm, color=RoyalBlue] (-1, -1) -- (-0.5*0.707, -0.5*0.707);
\filldraw[pattern=north west lines] (0, 0) circle (0.5cm);
\draw[line width=0.3mm] (0, 0) circle (0.5cm);
\draw[line width=0.3mm, color=RoyalBlue] (0.5*0.707, 0.5*0.707) -- (1, 1);
\draw[line width=0.3mm, color=Maroon] (0.5*0.707, -0.5*0.707) -- (1, -1);
\node[RoyalBlue] at (1.3, 1) {$k$};
\node[Maroon] at (1.3, -1) {$l$};
\end{tikzpicture}
\end{gathered}\:\:.
\end{equation}
To clarify, while external lines with the same color leg are the same flavor, particles of different colors are not necessarily different flavors either. The amplitude will be constrained by the unitarity equation \autoref{eq:UnitarityEquation}, which is organized by the number of internal lines that are cut. 

We first argue that the unitarity equation for the full amplitude implies unitarity for each of the flavor-ordered amplitudes individually. To see this, we can compare the contributions on either side of a 2-cut. In particular,
\begin{equation}
\label{eq:DiagramFlavor2Cuts}
\eval{2 \: \Im \left(\M_{ij; kl}\right)}_{2-\text{cuts}} \supset \: \: 
\frac{\textcolor{green!60!black}{N}}{N^2} \: \begin{gathered}
\begin{tikzpicture}[scale=0.65]
\draw[line width=0.3mm, RoyalBlue] (-1, 1) -- (-0.5*0.707, 0.5*0.707);
\node[RoyalBlue] at (-1.3, -1) {$i$};
\node [RoyalBlue] at (-1.3, 1) {$j$};
\draw[line width=0.3mm, RoyalBlue] (-1, -1) -- (-0.5*0.707, -0.5*0.707);
\filldraw[pattern = north west lines] (0, 0) circle (0.5cm);
\draw[line width=0.3mm] (0, 0) circle (0.5cm);
\draw[line width=0.3mm, green!60!black] (0.5*0.866, 0.5*0.5) -- (2 - 0.5*0.866, 0.5*0.5);
\draw[line width=0.3mm, green!60!black] (0.5*0.866, -0.5*0.5) -- (2 - 0.5*0.866, -0.5*0.5);
\node[green!60!black] at (0.65, 0.5) {$a$};
\node[green!60!black] at (0.65, -0.6) {$b$};
\node[green!60!black] at (1.35, 0.5) {$c$};
\node[green!60!black] at (1.35, -0.6) {$d$};
\filldraw[pattern=north west lines] (2, 0) circle (0.5cm);
\draw[line width=0.3mm] (2, 0) circle (0.5cm);
\draw[line width=0.3mm, Maroon] (2 + 0.5*0.707, 0.5*0.707) -- (3, 1);
\draw[line width=0.3mm, Maroon] (2 + 0.5*0.707, -0.5*0.707) -- (3, -1);
\node[Maroon] at (3.3, 1) {$k$};
\node[Maroon] at (3.3, -1) {$l$};
\draw[line width=0.5mm, dashed, orange] (1, 1.1) -- (1, -1.1);
\end{tikzpicture}
\end{gathered} \: \: + \: \: 
\frac{1}{N^2} \: \begin{gathered}
\begin{tikzpicture}[scale=0.65]
\draw[line width=0.3mm, RoyalBlue] (-1, 1) -- (-0.5*0.707, 0.5*0.707);
\draw[line width=0.3mm, RoyalBlue] (-1, -1) -- (-0.5*0.707, -0.5*0.707);
\node[RoyalBlue] at (-1.3, -1) {$i$};
\node[RoyalBlue] at (-1.3, 1) {$j$};
\filldraw[pattern = north west lines] (0, 0) circle (0.5cm);
\draw[line width=0.3mm] (0, 0) circle (0.5cm);
\draw[line width=0.3mm, Maroon] (0.5*0.866, 0.5*0.5) -- (2 - 0.5*0.866, 0.5*0.5);
\draw[line width=0.3mm, Maroon] (0.5*0.866, -0.5*0.5) -- (2 - 0.5*0.866, -0.5*0.5);
\node[Maroon] at (3.3, 1) {$k$};
\node[Maroon] at (3.3, -1) {$l$};
\node[Maroon] at (0.65, 0.5) {$a$};
\node[Maroon] at (0.65, -0.6) {$b$};
\node[Maroon] at (1.35, 0.5) {$c$};
\node[Maroon] at (1.35, -0.6) {$d$};
\filldraw[pattern=north west lines] (2, 0) circle (0.5cm);
\draw[line width=0.3mm] (2, 0) circle (0.5cm);
\draw[line width=0.3mm, Maroon] (2 + 0.5*0.707, 0.5*0.707) -- (3, 1);
\draw[line width=0.3mm, Maroon] (2 + 0.5*0.707, -0.5*0.707) -- (3, -1);
\draw[line width=0.5mm, dashed, orange] (1, 1.1) -- (1, -1.1);
\end{tikzpicture}
\end{gathered} \: \: + \dots 
\end{equation}
Note that, while we are dropping the orange shading that we use to denote complex conjugation in order to make the different flavors more visible, the amplitude on the right of the cut is still complex conjugated. The first diagram on the right-hand side arises from the product of two $s$-flavor-ordered amplitudes. The loop in the middle, colored green, could generically be \textit{any} color and indeed is summed over. Hence, this gives an additional factor of $N$, shown in green. The second diagram, on the other hand, arises from an $s$-flavor-ordered amplitude multiplying a $t$-flavor-ordered amplitude. The flavor of the cut lines are then restricted based on the external lines and no sum occurs, indicated by the fact that the internal lines are of the same color as the external $k$ and $l$ legs. This can equivalently be seen from the tensor structure. The first diagram will have the following multiplication along the cut: 
\begin{equation}
\delta^{ij} \delta^{ab} \delta_{ac} \delta_{bd} \delta^{cd} \delta^{kl} = N \delta^{ij} \delta^{kl}.
\end{equation}
The second 2-cut, on the other hand, involves
\begin{equation}
\delta^{ij} \delta^{ab} \delta_{ac} \delta_{bd} \delta^{ck} \delta^{dl} = \delta^{ij} \delta^{kl}. 
\end{equation}
While both will contribute to the $s$-flavor-ordered amplitude in general, only the first diagram will contribute in the large $N$ limit. This implies, at the level of 2-cuts, that leading contributions arise from products of one flavor-ordered amplitude with another flavor-ordered amplitude of the same flavor. It additionally implies that amplitudes in the large $N$ limit must only have cuts through flavor loops. 

The argument above relied on 2-cuts, which raises the question as to whether 4-cuts affect this argument. To see this, we note that, just as the 4-particle amplitude $\M_4$ scales as $\lambda/N$, the 6-particle amplitude $\M_6$ scales as $\left(\lambda/N\right)^2$. One way to see this is from unitarity. As argued in the main text, the 6-particle amplitude at leading order factorizes into products of 4-particle amplitudes, and so the 6-particle amplitude scales with twice the power in $N$ that the 4-particle amplitude does. With that said, the largest 4-cut contribution arises from a cut through 2 flavor loops: 
\begin{equation}
\eval{2 \: \Im \left(\M_{ij; kl}\right)}_{4-\text{cuts}} \supset \: \:
\frac{\textcolor{green!60!black}{N} \textcolor{yellow!75!black}{N}}{N^4} \: \begin{gathered}
\begin{tikzpicture}[scale=0.65, remember picture]
\draw[line width=0.3mm, RoyalBlue] (-1, 1) -- (-0.5*0.707, 0.5*0.707);
\draw[line width=0.3mm, RoyalBlue] (-1, -1) -- (-0.5*0.707, -0.5*0.707);
\node[RoyalBlue] at (-1.3, -1) {$i$};
\node[RoyalBlue] at (-1.3, 1) {$j$};
\node[Maroon] at (3.3, 1) {$k$};
\node[Maroon] at (3.3, -1) {$l$};
\draw[pattern=north west lines] (0, 0) circle (0.5cm);
\draw[line width=0.3mm] (0, 0) circle (0.5cm);
\draw[line width=0.3mm, green!60!black] (0.5*0.866, 0.5*0.5) -- (2 - 0.5*0.866, 0.5*0.5);
\draw[line width=0.3mm, green!60!black] (0.5*0.985, 0.5*0.174) -- (2 - 0.5*0.985, 0.5*0.174);
\draw[line width=0.3mm, yellow!75!black] (0.5*0.985, -0.5*0.174) -- (2 - 0.5*0.985, -0.5*0.174);
\draw[line width=0.3mm, yellow!75!black] (0.5*0.866, -0.5*0.5) -- (2 - 0.5*0.866, -0.5*0.5);
\draw[pattern = north west lines] (2, 0) circle (0.5cm);
\draw[line width=0.3mm] (2, 0) circle (0.5cm);
\draw[line width=0.3mm, Maroon] (2 + 0.5*0.707, 0.5*0.707) -- (3, 1);
\draw[line width=0.3mm, Maroon] (2 + 0.5*0.707, -0.5*0.707) -- (3, -1);
\draw[line width=0.5mm, dashed, orange] (1, 1.1) -- (1, -1.1);
\end{tikzpicture}
\end{gathered} \: \:. 
\end{equation}
Each flavor loop contributes a factor of $N$. However, each 6-particle amplitude contributes an additional factor of $1/N^2$ compared to the 4-particle amplitudes in the 2-cut. Hence, this 4-cut is subleading compared to the first 2-cut on the right-hand side of \autoref{eq:DiagramFlavor2Cuts}. The same pattern will persist for cuts through a larger number of particles. Therefore, in the large $N$ limit, we need only worry about 2-cuts and, as a result, we can apply the unitarity equation onto the $s$-flavor-ordered amplitude by itself. 

The same argument will also exclude corrections to the 2-particle amplitude in the large $N$ limit. The leading non-trivial contribution to the 2-particle amplitude arises from 3-cuts. The highest possible contribution arises from one flavor loop: 
\begin{equation}
\eval{2 \: \Im \left(\M_{i;j}\right)}_{3-\text{cuts}} \supset \frac{1}{N^2} \: \begin{gathered}
\begin{tikzpicture}[scale=0.6]
\draw[line width=0.3mm, RoyalBlue] (-1.5, 0) -- (-0.5, 0);
\draw[pattern= north west lines] (0, 0) circle (0.5cm);
\draw[line width=0.3mm] (0, 0) circle (0.5cm);
\draw[line width=0.3mm, green!60!black] (0.5*0.866, 0.5*0.5) -- (2 - 0.5*0.866, 0.5*0.5);
\draw[line width=0.3mm, RoyalBlue] (0.5*0.866, -0.5*0.5) -- (2 - 0.5*0.866, -0.5*0.5);
\draw[pattern=north west lines] (2, 0) circle (0.5cm);
\draw[line width=0.3mm] (2, 0) circle (0.5cm);
\draw[line width=0.3mm, green!60!black] (0.5, 0) -- (1.5, 0);
\draw[line width=0.3mm, RoyalBlue] (2.5, 0) -- (3.5, 0);
\draw[line width=0.5mm, dashed, orange] (1, 1) -- (1, -1);
\end{tikzpicture}
\end{gathered} \: .
\end{equation}
The flavor loop contributes one additional factor of $N$, giving an overall scaling of, at best, $1/N$. This is therefore suppressed compared to the bare propagator's contribution of $p^2$. As a result, in the large $N$ limit, there are no corrections to the propagator and, as a result, the CS equation on the 2-particle amplitude becomes
\begin{equation}
\gamma = 0.
\end{equation}
In other words, there is no wavefunction renormalization. 

With this, we have reduced the problem of studying the amplitude $\M_{ij;kl}$ to just one flavor-ordered amplitude, $\M_s$. We further argued that we need only consider 2-cuts through flavor loops and that there is no wavefunction renormalization. However, the ansatz we used for the $s$-flavor-ordered amplitude had a further simplification: it only depended on one Mandelstam variable, $s$: 
\begin{equation}
\M_s = \sum_{L = 0}^{\infty} \frac{\left(-\lambda\right)^{L + 1}}{\left(16\pi^2\right)^L} \sum_{k = 0}^{L} m_{L, k} \log^k_{-s}.
\end{equation}
One might wonder why we did not have a symmetric combination of logs in the other Mandelstam variables, much like what was done in full $\lambda \phi^4$. This, however, assumes that each flavor-ordered amplitude is permutation symmetric. Cuts are related to discontinuities of amplitudes in particular channels, see \cite{Britto:2024} for a review. In turn, these discontinuities would compute the coefficients of $\log_{-s}$ and $\log_{-t}$ in the amplitude. In the case at hand: 
\begin{equation}
\text{Disc}_s \M_s \propto \: \: \begin{gathered}
\begin{tikzpicture}[scale=0.6]
\draw[line width=0.3mm, color=RoyalBlue] (-1, 1) -- (-0.5*0.707, 0.5*0.707);
\node[RoyalBlue] at (-1.3, -1) {$i$};
\node[RoyalBlue] at (-1.3, 1) {$j$};
\draw[line width=0.3mm, color=RoyalBlue] (-1, -1) -- (-0.5*0.707, -0.5*0.707);
\filldraw[pattern=north west lines] (0, 0) circle (0.5cm);
\draw[line width=0.3mm] (0, 0) circle (0.5cm);
\draw[line width=0.3mm, color=Maroon] (0.5*0.707, 0.5*0.707) -- (1, 1);
\draw[line width=0.3mm, color=Maroon] (0.5*0.707, -0.5*0.707) -- (1, -1);
\node[Maroon] at (1.3, 1) {$k$};
\node[Maroon] at (1.3, -1) {$l$};
\draw[line width=0.5mm, dashed, orange] (0, 1.25) -- (0, -1.25);
\end{tikzpicture}
\end{gathered} \qquad \text{and} \qquad \text{Disc}_t \M_s \propto \: \: 
\begin{gathered}
\begin{tikzpicture}[scale=0.6]
\draw[line width=0.3mm, color=RoyalBlue] (-1, 1) -- (-0.5*0.707, 0.5*0.707);
\node[RoyalBlue] at (-1.3, -1) {$i$};
\node[RoyalBlue] at (-1.3, 1) {$j$};
\draw[line width=0.3mm, color=RoyalBlue] (-1, -1) -- (-0.5*0.707, -0.5*0.707);
\filldraw[pattern=north west lines] (0, 0) circle (0.5cm);
\draw[line width=0.3mm] (0, 0) circle (0.5cm);
\draw[line width=0.3mm, color=Maroon] (0.5*0.707, 0.5*0.707) -- (1, 1);
\draw[line width=0.3mm, color=Maroon] (0.5*0.707, -0.5*0.707) -- (1, -1);
\node[Maroon] at (1.3, 1) {$k$};
\node[Maroon] at (1.3, -1) {$l$};
\draw[line width=0.5mm, dashed, orange] (-1.25, 0) -- (1.25, 0);
\end{tikzpicture}
\end{gathered} \: .
\end{equation}
Due to the tensor structure of each constituent amplitude, the discontinuities are not symmetric under the interchange of $s$ and $t$. Therefore, at leading order in $N$, we cannot include additional symmetric combinations of logarithms. The leading $N$ contribution instead arises from logarithms of just $s$, with permutation symmetry reserved for the full amplitude $\M_{ij;kl}$.

\section{Consistency with the Loop Expansion}
\label{a:ExactResults}
To validate the results derived from unitarity, we will compute the renormalized 2- and 4-particle amplitudes for $\lambda\phi^{4}$ theory using renormalized perturbation theory up to 3-loop order. This calculation enables a direct comparison of both the divergence coefficients in the unrenormalized amplitude and the logarithmic coefficients in the renormalized amplitude. For the 4-particle amplitude, we extend this computation to 4-loop order and up to subleading order in divergences. 

The bare action for massless $\lambda \phi^4$ theory is 
\begin{equation}
S = - \int \left(\frac{1}{2} \partial_\mu \phi_0 \partial^\mu \phi_0 + \frac{\tilde{\mu}^{2\epsilon}\lambda_0}{4!} \phi_0^4\right) \dd[d]{x}, 
\end{equation}
where $d = 4 - 2\epsilon$. The bare and renormalized quantities are related as
\begin{equation}
\label{eq:BareToRen}
\phi_0 = Z_\phi^{1/2} \phi, \qquad \lambda_0 = Z_\lambda \lambda, \qquad \mathcal{M}_n = Z_\phi^{n/2} \widetilde{\mathcal{M}}_n, 
\end{equation}
where $\M_n$ is the renormalized amplitude and $\widetilde{\M}_n$ is the unrenormalized amplitude. In order to obtain explicit results, we will use the $\overline{\text{MS}}$ renormalization scheme. Note that $\tilde{\mu}$ defined in the Lagrangian above is related to the $\mu$ that has and will appear in the amplitudes by 
\begin{equation}
\mu^2 = 4\pi e^{-\gamma_E} \tilde{\mu}^2. 
\end{equation}

All integrals were computed using the parametric representation (for a review of different representations of Feynman integrals, see \cite[Ch.~2]{Weinzierl:2022}) or the sector decomposition program \verb!pySecDec! \cite{Borowka:2017, Borowka:2018, Heinrich:2021, Heinrich:2023}. The input information for the Feynman integrals will be a graph $G$ that is described by its edges $E$ and loops $L$ in $d$ dimensions. Each edge will moreover contain a weight given by $\nu_e$. We place all the kinematic dependence in terms of Mandelstam variables $s_{ij}$. In other words, 
\begin{equation}
G \equiv \left(\nu_1, \dots, \nu_E; L, d; s_{ij}\right).
\end{equation}
In the \textit{momentum} representation, the loop integral of a graph $G$ is given by
\begin{equation}
\label{eq:GenericGraphI}
I\left[G\right] = \mu^{2L\epsilon} e^{L\gamma_E\epsilon} \int \prod_{e \in E} \frac{1}{\left(k_e^2 + m_e^2 - i\varepsilon\right)^{\nu_e}} \prod_{i \in L} \frac{\dd[d]{\ell_i}}{i\pi^{d/2}}, 
\end{equation}
where $\gamma_E$ is the Euler--Mascheroni constant. The equivalent parametric representation is
\begin{equation}
\label{eq:GraphIntegral}
I\left[G\right] = \frac{\mu^{2L\epsilon} e^{L\gamma_E\epsilon} \Gamma\left(d_G\right)}{\prod_{e \in E}\Gamma\left(\nu_e\right)} \int \frac{\prod_{e\in E} \alpha_e^{\nu_e - 1}}{\mathcal{U}^{d/2 - d_G} \mathcal{F}^{d_G}} \frac{\dd[E]{\alpha_E}}{\text{GL}\left(1\right)},
\end{equation}
where $\mathcal{U}$ and $\mathcal{F}$ are the Symanzik polynomials
\begin{equation}
\label{eq:SymanzikPolynomials}
\mathcal{U} = \sum_{\substack{\text{spanning}\\ \text{trees }T}} \prod_{e\notin T} \alpha_e, \qquad \mathcal{F} = \sum_{\substack{\text{spanning}\\ \text{two-trees} \\T_L\sqcup T_R}} p_L^2 \prod_{e\notin T_L \sqcup T_R} \alpha_e + \mathcal{U}\sum_{e = 1}^{E} m_e^2 \alpha_e, 
\end{equation}
and $d_G$ is the superficial degree of divergence,
\begin{equation}
\label{eq:DegOfDiv}
d_G = \sum_{e = 1}^{E} \nu_e - \frac{Ld}{2}.
\end{equation}
The GL(1) in the parametric representation is a freedom in choice of constraint that allows us to set any linear combination of the Schwinger parameters, $\alpha_i$, to $1$. This freedom is also called the \textit{Cheng--Wu theorem}. We refer to \cite[App.~A]{Hannesdottir:2022a} for details.

In what follows, we will list the results $I\left[G\right]$ computed in this way in tables organized by the number of external particles and number of loops. The general amplitude is written as a sum over all Feynman diagrams $G_i$:
\begin{equation}
\widetilde{\M}_n = \sum_{L = 0}^{\infty} \frac{\left(-\lambda_0\right)^{L + 1}}{\left(16\pi^2\right)^L} \sum_i \frac{M_i}{S_i}I\left[G_i\right],
\end{equation}
where $M_i$ is the multiplicity (the number of topologically distinct diagrams giving the same answer) and $S_i$ is the symmetry factor (the number of ways of swapping internal edges without changing the vertices that leaves the diagram unchanged). Note that many of the diagrams for which we were able to find analytic solutions are reducible to a 1-loop integral by replacing bubbles with propagators given by the 1-loop result. The price of reducing the $L$-loop integral to a 1-loop integral is a minimal one: changing the power of the denominator from $1$ to an $\epsilon$-dependent power $\nu_e$. All diagrams in $\lambda \phi^4$ theory were determined from the \verb!Phi4tools! Mathematica package \cite{Sberveglieri:2023}.

\subsection{Loop Integral Results}
We will start by computing the 2-particle amplitude. To do this, we must first compute the fully dressed propagator, $i\tilde{G}\left(p\right)$. This can be done by repeatedly inserting $\Sigma$, the sum of all $1PI$ diagrams:
\begin{equation}
i\tilde{G}\left(p\right) = \: 
\begin{gathered}
\vspace{0.15cm}
\begin{tikzpicture}[scale=0.9]
\draw[line width=0.3mm] (-1.5, 0) -- (1, 0);
\end{tikzpicture}
\end{gathered} \: + \:
\begin{gathered}
\vspace{-0.15cm}
\begin{tikzpicture}[scale=0.9]
\draw[line width=0.3mm] (-1.5, 0) -- (-0.5, 0);
\filldraw[orange!15, opacity=0.5] circle (0.5cm);
\draw[line width=0.3mm] circle (0.5cm);
\draw[line width=0.3mm] (0.5, 0) -- (1.5, 0);
\node at (0, 0) {$i\Sigma$};
\end{tikzpicture}
\end{gathered} \: + \: 
\begin{gathered}
\vspace{-0.15cm}
\begin{tikzpicture}[scale=0.9]
\draw[line width=0.3mm] (-1.5, 0) -- (-0.5, 0);
\filldraw[orange!15, opacity=0.5] circle (0.5cm);
\draw[line width=0.3mm] circle (0.5cm);
\node at (0, 0) {$i\Sigma$};
\draw[line width=0.3mm] (0.5, 0) -- (1.5, 0);
\filldraw[orange!15, opacity=0.5] (2, 0) circle (0.5cm);
\draw[line width=0.3mm] (2, 0) circle (0.5cm);
\node at (2, 0) {$i\Sigma$};
\draw[line width=0.3mm] (2.5, 0) -- (3.5, 0);
\end{tikzpicture}
\end{gathered} \: + \: \dots.
\end{equation}
Performing the geometric series, we find
\begin{equation}
i\tilde{G}\left(p\right) = \frac{-i}{p^2 - \Sigma\left(p\right) - i\varepsilon}.
\end{equation}
The denominator is, by definition, the 2-particle amplitude
\begin{equation}
\M_2\left(p^2\right) = p^2 - \Sigma\left(p^2\right).
\end{equation}

\begin{table}
\begin{center}
\begin{tabular}{c | c | c} 
 $G$ & $S$ & $I\left[G\right]$ \\ 
 \hline
 \adjustbox{valign=m}{
\begin{tikzpicture}[scale=0.4]
\draw[line width=0.3mm] (-2, 0) -- (0, 0);
\draw[line width=0.3mm] (0, 0) -- (2, 0);
\draw[line width=0.3mm] (2, 0) -- (4, 0);
\draw[line width=0.3mm] (0, 0) arc(180:0:1cm and 0.8cm);
\draw[line width=0.3mm] (2, 0) arc(0:-180:1cm and 0.8cm);
\end{tikzpicture}}  & 6 & \adjustbox{valign=m, margin=0pt 5pt}{$\displaystyle{\frac{e^{2\gamma_E \epsilon} \Gamma\left(2\epsilon - 1\right)\Gamma^3\left(1 - \epsilon\right)}{\Gamma\left(3 - 3\epsilon\right)} \left(\frac{\mu^2}{p^2}\right)^{2\epsilon}}$} \\ 
 \hline
\adjustbox{valign=m}{
\begin{tikzpicture}[scale=0.4]
\draw[line width=0.3mm] (-2, 0) -- (0, 0);
\draw[line width=0.3mm] (0, 0) arc(180:360:0.75cm);
\draw[line width=0.3mm] (1.5, 0) arc(0:180:0.75cm);
\draw[line width=0.3mm] (1.5, 0) arc(180:360:0.75cm);
\draw[line width=0.3mm] (3, 0) arc(0:180:0.75cm);
\draw[line width=0.3mm] (0, 0) arc(180:0:1.5cm and 1.5cm);
\draw[line width=0.3mm] (3, 0) -- (5, 0);
\end{tikzpicture}} & 4 & \adjustbox{valign=m, margin=0pt 5pt}{$\displaystyle{\frac{e^{3\gamma_E \epsilon} \Gamma^2\left(\epsilon\right)\Gamma\left(3\epsilon - 1\right)\Gamma\left(2 - 3\epsilon\right)\Gamma^5\left(1 - \epsilon\right)}{\Gamma^2\left(2 - 2\epsilon\right)\Gamma\left(2\epsilon\right)\Gamma\left(3 - 4\epsilon\right)} \left(\frac{\mu^2}{p^2}\right)^{3\epsilon}}$} 
\end{tabular}
\end{center}
\caption{2-particle diagrams. The first column, $G$, gives the diagram. The first is sometimes referred to as the \textit{sunrise} and the second as the \textit{goggles}. The second column, $S$, gives the symmetry factor. The last column gives the value of $I\left[G\right]$ after integration.}
\label{tab:2ParticleDiagrams}
\end{table}

For the 2-particle amplitude, the only non-zero loop diagrams are the \textit{sunrise} and \textit{goggles} diagrams shown in \autoref{tab:2ParticleDiagrams}. Any diagrams not included up to 3-loop order are scaleless in dimensional regularization, such as the 1-loop tadpole correction to the propagator. The resulting expression for the 2-particle amplitude is then:
\begin{equation}
\label{eq:UnrenormalizedM2exact}
\begin{aligned}
p^{-2}\widetilde{\mathcal{M}}_2\left(p^2\right) &= 1 - \frac{\left(-\lambda_0\right)^2 e^{2\gamma_E \epsilon}}{6\left(16\pi^2\right)^2} \left(\frac{\Gamma\left(2\epsilon - 1\right)\Gamma^3\left(1 - \epsilon\right)}{\Gamma\left(3 - 3\epsilon\right)}\right)\left(\frac{\mu^2}{p^2}\right)^{2\epsilon} \\
&\hspace{0.5cm}- \frac{\left(-\lambda_0\right)^3 e^{3\gamma_E \epsilon}}{4\left(16\pi^2\right)^3} \left(\frac{\Gamma^2\left(\epsilon\right)\Gamma\left(3\epsilon - 1\right)\Gamma\left(2 - 3\epsilon\right)\Gamma^5\left(1 - \epsilon\right)}{\Gamma^2\left(2 - 2\epsilon\right)\Gamma\left(2\epsilon\right)\Gamma\left(3 - 4\epsilon\right)}\right)\left(\frac{\mu^2}{p^2}\right)^{3\epsilon}.
\end{aligned}
\end{equation}

\begin{table}
\begin{center}
\begin{tabular}{c | c | c | c} 
 $G$ & $S$ & $M$ & $I\left[G\right]$ \\
 \hline
 \adjustbox{valign=m}{
\begin{tikzpicture}[scale=0.5]
\draw[line width=0.3mm] (-1, 0) -- (-2, 1);
\draw[line width=0.3mm] (-2, -1) -- (-1, 0);
\draw[line width=0.3mm] (2, 1) -- (1, 0);
\draw[line width=0.3mm] (1, 0) -- (2, -1);
\draw[line width=0.3mm] (1, 0) arc(0:180:1 cm and 0.5cm);
\draw[line width=0.3mm] (-1, 0) arc(180:360:1 cm and 0.5cm);
\end{tikzpicture}} & 2 & 1 & \adjustbox{valign=m, margin=0pt 5pt}{$\displaystyle{\frac{e^{\gamma_E \epsilon}\Gamma\left(\epsilon\right)\Gamma^2\left(1 - \epsilon\right)}{\Gamma\left(2 - 2\epsilon\right)} \left(\frac{\mu^2}{-s}\right)^{\epsilon} }$}
\end{tabular}
\end{center}
\caption{4-particle 1-loop diagram. The only distinct 1-loop diagram is the bubble, which has a symmetry factor of $2$, multiplicity $1$, and integral result shown in the final column.}
\label{tab:4Particle1LoopDiagrams}
\end{table}

\begin{table}
\begin{center}
\begin{tabular}{c | c | c | c} 
 $G$ & $S$ & $M$ & $I\left[G\right]$ \\
 \hline
 
\adjustbox{valign=m}{
\begin{tikzpicture}[scale=0.4]
\draw[line width=0.3mm] (-1, 0) -- (-2, 1);
\draw[line width=0.3mm] (-2, -1) -- (-1, 0);
\draw[line width=0.3mm] (4, 1) -- (3, 0);
\draw[line width=0.3mm] (3, 0) -- (4, -1);
\draw[line width=0.3mm] (1, 0) arc(0:180:1 cm and 0.5cm);
\draw[line width=0.3mm] (-1, 0) arc(180:360:1 cm and 0.5cm);
\draw[line width=0.3mm] (3, 0) arc(0:180:1 cm and 0.5cm);
\draw[line width=0.3mm] (1, 0) arc(180:360:1 cm and 0.5cm);
\end{tikzpicture}} & 4 & 1 & \adjustbox{valign=m, margin=0pt 5pt}{$\displaystyle{\left(\frac{e^{\gamma_E \epsilon}\Gamma\left(\epsilon\right)\Gamma^2\left(1 - \epsilon\right)}{\Gamma\left(2 - 2\epsilon\right)}\right)^2 \left(\frac{\mu^2}{-s}\right)^{2 \epsilon} }$} \\
\hline 
\adjustbox{valign=m}{
\begin{tikzpicture}[scale=0.4]
\draw[line width=0.3mm] (-1, 0) -- (-2, 1);
\draw[line width=0.3mm] (1, 1) -- (-1, 0);
\draw[line width=0.3mm] (2, 1.5) -- (1, 1);
\draw[line width=0.3mm] (-2, -1) -- (-1, 0);
\draw[line width=0.3mm] (-1, 0) -- (1, -1);
\draw[line width=0.3mm] (1, -1) -- (2, -1.5);
\draw[line width=0.3mm] (1, 1) arc(90:270:0.35cm and 1cm);
\draw[line width=0.3mm] (1, -1) arc(270:450:0.35cm and 1cm);
\end{tikzpicture}} & 2 & 2 & \adjustbox{valign=m, margin=0pt 5pt}{$\displaystyle{\frac{e^{2\gamma_E \epsilon} \Gamma\left(\epsilon\right)\Gamma\left(2\epsilon\right)\Gamma^2\left(1 - \epsilon\right)\Gamma^2\left(1 - 2\epsilon\right)}{\Gamma\left(2 - 2\epsilon\right)\Gamma\left(2 - 3\epsilon\right)} \left(\frac{\mu^2}{-s}\right)^{2\epsilon}}$} 
\end{tabular}
\end{center}
\caption{4-particle 2-loop diagrams. The first diagram, the double bubble, is the square of the 1-loop bubble. The second diagram is sometimes referred to as a \textit{parachute}.}
\label{tab:4Particle2LoopDiagrams}
\end{table}

\begin{table}
\begin{center}
\begin{tabular}{c | c | c | c} 
 $G$ & $S$ & $M$ & $I\left[G\right]$ \\  
 \hline
 
\adjustbox{valign=m}{
\begin{tikzpicture}[scale=0.35]
\draw[line width=0.3mm] (-1, 0) -- (-2, 1);
\draw[line width=0.3mm] (-2, -1) -- (-1, 0);
\draw[line width=0.3mm] (6, 1) -- (5, 0);
\draw[line width=0.3mm] (5, 0) -- (6, -1);
\draw[line width=0.3mm] (1, 0) arc(0:180:1 cm and 0.5cm);
\draw[line width=0.3mm] (-1, 0) arc(180:360:1 cm and 0.5cm);
\draw[line width=0.3mm] (3, 0) arc(0:180:1 cm and 0.5cm);
\draw[line width=0.3mm] (1, 0) arc(180:360:1 cm and 0.5cm);
\draw[line width=0.3mm] (5, 0) arc(0:180:1 cm and 0.5cm);
\draw[line width=0.3mm] (3, 0) arc(180:360:1cm and 0.5cm);
\end{tikzpicture}} & 8 & 1 & \adjustbox{valign=m, margin=0pt 5pt}{$\displaystyle{\left(\frac{e^{\gamma_E \epsilon}\Gamma\left(\epsilon\right)\Gamma^2\left(1 - \epsilon\right)}{\Gamma\left(2 - 2\epsilon\right)}\right)^3 \left(\frac{\mu^2}{-s}\right)^{3 \epsilon} }$} \\
\hline 
\adjustbox{valign=m, margin=0pt 5pt}{
\begin{tikzpicture}[scale=0.35]
\draw[line width=0.3mm] (-1.5, 0) -- (-2.5, 1);
\draw[line width=0.3mm] (1, 1.5) -- (-1.5, 0);
\draw[line width=0.3mm] (2, 2) -- (1, 1.5);
\draw[line width=0.3mm] (-2.5, -1) -- (-1.5, 0);
\draw[line width=0.3mm] (-1.5, 0) -- (1, -1.5);
\draw[line width=0.3mm] (1, -1.5) -- (2, -2);
\draw[line width=0.3mm] (1, 1.5) arc(90:270:0.3cm and 0.75cm);
\draw[line width=0.3mm] (1, 0) arc(270:450:0.3cm and 0.75cm);
\draw[line width=0.3mm] (1, 0) arc(90:270:0.3cm and 0.75cm);
\draw[line width=0.3mm] (1, -1.5) arc(270:450:0.3cm and 0.75cm);
\end{tikzpicture}} & 4 & 2 & \adjustbox{valign=m, margin=0pt 5pt}{$\displaystyle{\frac{e^{3\gamma_E \epsilon} \Gamma\left(3\epsilon\right)\Gamma^2\left(1 - 3\epsilon\right)}{\Gamma\left(2 - 4\epsilon\right)} \left(\frac{\Gamma\left(\epsilon\right)\Gamma^2\left(1 - \epsilon\right)}{\Gamma\left(2 - 2\epsilon\right)}\right)^2 \left(\frac{\mu^2}{-s}\right)^{3\epsilon}}$} \\
\hline 
\adjustbox{valign=m}{
\begin{tikzpicture}[scale=0.35]
\draw[line width=0.3mm] (-3, 0) -- (-4, 1);
\draw[line width=0.3mm] (1, 1) -- (-1, 0);
\draw[line width=0.3mm] (2, 1.5) -- (1, 1);
\draw[line width=0.3mm] (-4, -1) -- (-3, 0);
\draw[line width=0.3mm] (-1, 0) -- (1, -1);
\draw[line width=0.3mm] (1, -1) -- (2, -1.5);
\draw[line width=0.3mm] (1, 1) arc(90:270:0.35cm and 1cm);
\draw[line width=0.3mm] (1, -1) arc(270:450:0.35cm and 1cm);
\draw[line width=0.3mm] (-1, 0) arc(0:180:1 cm and 0.5cm);
\draw[line width=0.3mm] (-3, 0) arc(180:360:1cm and 0.5cm);
\end{tikzpicture}} & 4 & 2 & \adjustbox{valign=m, margin=0pt 5pt}{$\displaystyle{\frac{e^{3\gamma_E \epsilon} \Gamma\left(2\epsilon\right)\Gamma^2\left(1 - 2\epsilon\right)}{\Gamma\left(2 - 3\epsilon\right)} \left(\frac{\Gamma\left(\epsilon\right)\Gamma^2\left(1 - \epsilon\right)}{\Gamma\left(2 - 2\epsilon\right)}\right)^2 \left(\frac{\mu^2}{-s}\right)^{3\epsilon}}$} \\
\hline 
\adjustbox{valign=m}{
\begin{tikzpicture}[scale=0.4]
\draw[line width=0.3mm] (0, 0) -- (-1, 1);
\draw[line width=0.3mm] (-1, -1) -- (0, 0);
\draw[line width=0.3mm] (1, 1) -- (0, 0);
\draw[line width=0.3mm] (0, 0) -- (1, -1);
\draw[line width=0.3mm] (1, -1) -- (1, 1);
\draw[line width=0.3mm] (3, 0) -- (2, 0);
\draw[line width=0.3mm] (2, 0) -- (1, -1);
\draw[line width=0.3mm] (1, -1) -- (3, -1);
\begin{scope}[shift={(1.5,0.5)}, rotate=-45]
    \draw[line width=0.3mm]
      (-0.707cm,0) arc[
        x radius=0.707cm,
        y radius=0.2cm,
        start angle=180,
        end angle=0
      ];
\end{scope}
\begin{scope}[shift={(1.5,0.5)}, rotate=-45]
    \draw[line width=0.3mm]
      (0.707cm,0) arc[
        x radius=0.707cm,
        y radius=0.2cm,
        start angle=0,
        end angle=-180
      ];
\end{scope}
\end{tikzpicture}} & 2 & 4 & \adjustbox{valign=m, margin=0pt 5pt}{$\displaystyle{
\left(\frac{1}{6\epsilon^3} + \frac{3}{2\epsilon^2} + \frac{1}{\epsilon}\left(\frac{55}{6} + \frac{3}{4}\zeta_2\right) + \mathcal{O}\left(\epsilon^0\right) \right)\left(\frac{\mu^2}{-s}\right)^{3\epsilon}}$} \\
\hline 
\adjustbox{valign=m}{
\begin{tikzpicture}[scale=0.4]
\draw[line width=0.3mm] (-1.25, 0) -- (-2.25, 1);
\draw[line width=0.3mm] (-2.25, -1) -- (-1.25, 0);
\draw[line width=0.3mm] (1.25, 0) arc(0:90:1.25cm and 1cm);
\draw[line width=0.3mm] (0, 1) arc(90:180:1.25cm and 1cm);
\draw[line width=0.3mm] (-1.25, 0) arc(180:270:1.25cm and 1cm);
\draw[line width=0.3mm] (0, -1) arc(270:360:1.25cm and 1cm);
\draw[line width=0.3mm] (2.25, 1) -- (1.25, 0);
\draw[line width=0.3mm] (1.25, 0) -- (2.25, -1);
\draw[line width=0.3mm] (0, 1) arc(90:270:0.4cm and 1cm);
\draw[line width=0.3mm] (0, -1) arc(270:450:0.4cm and 1cm);
\end{tikzpicture}} & 4 & 1 & \adjustbox{valign=m, margin=0pt 5pt}{$\displaystyle{
\left(\frac{1}{3\epsilon^3} + \frac{7}{3\epsilon^2} + \frac{1}{\epsilon}\left(\frac{31}{3} - \frac{\zeta_2}{2}\right) + \mathcal{O}\left(\epsilon^0\right)\right)\left(\frac{\mu^2}{-s}\right)^{3\epsilon} }$} \\
\hline 
\adjustbox{valign=m, margin=0pt 5pt}{
\begin{tikzpicture}[scale=0.6]
\draw[line width=0.3mm] (-0.5, 0.5) -- (-1, 1);
\draw[line width=0.3mm] (0.5, 0.5) -- (-0.5, 0.5);
\draw[line width=0.3mm] (1, 1) -- (0.5, 0.5);
\draw[line width=0.3mm] (-1, -1) -- (-0.5, -0.5);
\draw[line width=0.3mm] (-0.5, -0.5) -- (0.5, -0.5);
\draw[line width=0.3mm] (0.5, -0.5) -- (1, -1);
\draw[line width=0.3mm] (-0.5, 0.5) arc(90:270:0.175cm and 0.5cm);
\draw[line width=0.3mm] (-0.5, -0.5) arc(270:450:0.175cm and 0.5cm);
\draw[line width=0.3mm] (0.5, 0.5) arc(90:270:0.175cm and 0.5cm);
\draw[line width=0.3mm] (0.5, -0.5) arc(270:450:0.175cm and 0.5cm);
\end{tikzpicture}} & 4 & 2 & \adjustbox{valign=m, margin=0pt 5pt}{$\displaystyle{\left(\frac{1}{3\epsilon^3} + \frac{8}{3\epsilon^2} + \frac{1}{\epsilon}\left(\frac{44}{3} + \frac{3}{2}\zeta_2\right) + \mathcal{O}\left(\epsilon^0\right)\right)\left(\frac{\mu^2}{-s}\right)^{3\epsilon}}$} \\
\hline 
\adjustbox{valign=m, margin=0pt 5pt}{
\begin{tikzpicture}[scale=0.6]
\draw[line width=0.3mm] (-1, 1) -- (-0.5, 0.5) -- (0.5, 0.5) -- (1, 1);
\draw[line width=0.3mm] (-1, -1) -- (-0.5, -0.5) -- (0.5, -0.5) -- (1, -1);
\draw[line width=0.3mm] (-0.5, 0.5) -- (-0.5, -0.5);
\draw[line width=0.3mm] (0.5, 0.5) -- (0.5, -0.5);
\draw[line width=0.3mm] (-0.5, -0.5) -- (0.5, 0.5);
\draw[line width=1mm, white] (-0.4, 0.4) -- (0.4, -0.4);
\draw[line width=0.3mm] (-0.5, 0.5) -- (0.5, -0.5);
\end{tikzpicture}} & 1 & 1 & \adjustbox{valign=m, margin=0pt 5pt}{$\displaystyle{\frac{2\zeta_3}{\epsilon} + \mathcal{O}\left(\epsilon^0\right)}$} \\
\hline 
\adjustbox{valign=m}{
\begin{tikzpicture}[scale=0.4]
\draw[line width=0.3mm] (0, 0) -- (-1, 1);
\draw[line width=0.3mm] (-1, -1) -- (0, 0);
\draw[line width=0.3mm] (0, 0) arc(180:0:1cm and 0.8cm);
\draw[line width=0.3mm] (2, 0) arc(0:-180:1cm and 0.8cm);
\draw[line width=0.3mm] (3, 1) -- (2, 0);
\draw[line width=0.3mm] (2, 0) -- (3, -1);
\draw[line width=0.3mm] (0.5, 0.75) arc(180:0:0.5cm and 0.4cm);
\draw[line width=0.3mm] (1.5, 0.75) arc(0:-180:0.5cm and 0.4cm);
\end{tikzpicture}} & 6 & 1 & \adjustbox{valign=m, margin=0pt 5pt}{$\displaystyle{\frac{e^{3\gamma_E \epsilon} \Gamma\left(3\epsilon\right)\Gamma\left(1 - 3\epsilon\right)}{\Gamma\left(2\epsilon + 1\right)\Gamma\left(2 - 4\epsilon\right)}\left(\frac{\Gamma\left(2\epsilon - 1\right)\Gamma^4\left(1 - \epsilon\right)}{\Gamma\left(3 - 3\epsilon\right)}\right) \left(\frac{\mu^2}{-s}\right)^{3\epsilon}}$}
\end{tabular}
\end{center}
\caption{4-particle 3-loop diagrams. Any result not stated to all orders in $\epsilon$ was reconstructed using \texttt{pySecDec} and PSLQ. Note that for the envelope, there are no additional channels and so there will be a difference by a factor of three for its contribution to the sub-subleading divergence compared to all the other diagrams. Here, $\zeta_k = \zeta(k)$ is the Riemann zeta function evaluated at integers, $k$.}
\label{tab:4Particle3LoopDiagrams}
\end{table}

We now quote the results of the explicit computation of the 4-particle amplitude up to three-loop order and up to finite order in $\epsilon$. The results of each diagram are contained in \autoref{tab:4Particle1LoopDiagrams}, \autoref{tab:4Particle2LoopDiagrams}, and \autoref{tab:4Particle3LoopDiagrams}. There, the multiplicity corresponds to the number of $s$-channel diagrams that are equivalent under interchange of external legs. We then include $t$ and $u$ by permutation symmetry when appropriate. Writing the 4-particle amplitude as
\begin{equation}
\tilde{\mu}^{-2\epsilon}\widetilde{\M}_4 = \sum_{L = 0}^{\infty} \frac{\left(-\lambda_0\right)^{L + 1}}{\left(16\pi^2\right)^L} \alpha_L\left(\epsilon\right)\left[\left(\frac{\mu^2}{-s}\right)^{L\epsilon} + \left(\frac{\mu^2}{-t}\right)^{L\epsilon} + \left(\frac{\mu^2}{-u}\right)^{L\epsilon\:}\right], 
\end{equation}
we find by summing over all the diagrams that
\begin{subequations}
\begin{align}
\alpha_0\left(\epsilon\right) &= \frac{1}{3},\\
\label{eq:alpha1expansion}
\alpha_1\left(\epsilon\right) &= \frac{e^{\gamma_E \epsilon} \Gamma\left(\epsilon\right)\Gamma^2\left(1 - \epsilon\right)}{2\Gamma\left(2 - 2\epsilon\right)},\\
\label{eq:alpha2expansion}
\alpha_2\left(\epsilon\right) &= \frac{e^{2\gamma_E \epsilon} \Gamma\left(\epsilon\right)\Gamma^2\left(1 - \epsilon\right)}{\Gamma\left(2 - 2\epsilon\right)} \left(\frac{\Gamma\left(\epsilon\right)\Gamma^2\left(1 - \epsilon\right)}{4\Gamma\left(2 - 2\epsilon\right)} + \frac{\Gamma\left(2\epsilon\right)\Gamma^2\left(1 - 2\epsilon\right)}{\Gamma\left(2 - 3\epsilon\right)}\right),\\
\alpha_3\left(\epsilon\right) &= \frac{9}{8\epsilon^3} + \frac{629}{72\epsilon^2} + \frac{747 + 45\zeta_2 + 64 \zeta_3}{16\epsilon} + \mathcal{O}\left(\epsilon^0\right), 
\end{align}
\end{subequations}
where $\zeta_k = \zeta(k)$ is the Riemann zeta function evaluated at integers. It is straightforward to read off, from an expansion in $\epsilon$, the leading and subleading divergences of the 4-particle amplitude. For example, at 3-loop order, we can see that $M_{3, 3} = 27/8$ and $M_{3, 2} = 629/24$, where the additional factor of 3 comes from the 3 channels. This matches the results from unitarity, \autoref{eq:LeadingDivUnitarity} and \autoref{eq:SubLeadingDivUnitarityMs}. We mention here that, by including additional four-particle cuts, we were able to reconstruct the sub-subleading divergences up to three-loop order. We additionally mention that by extracting the leading and subleading divergences of all 4-loop divergences, we found that
\begin{equation}
\alpha_4\left(\epsilon\right) = \frac{27}{16\epsilon^4} + \frac{151}{8\epsilon^3} + \mathcal{O}\left(\epsilon^{-2}\right).
\end{equation}
From here, we can read off that the leading divergence is $M_{4, 4} = 81/16$ and the subleading divergence is $M_{4, 3} = 453/8$. This once again agrees with the all $L$-order result found from unitarity in \autoref{eq:LeadingDivUnitarity} and \autoref{eq:SubLeadingDivUnitarityMs}.

\subsection{Perturbative Renormalization}
In order to compute the renormalized amplitude and its coefficients, we relate the bare and renormalized quantities as in \autoref{eq:BareToRen}. Working in the $\overline{\text{MS}}$ scheme, we find that the counterterms up to order $\lambda^3$ are
\begin{equation}
Z_\phi = 1 - \frac{1}{24\epsilon} \frac{\lambda^2}{\left(16\pi^2\right)^2} - \left(\frac{1}{24\epsilon^2} - \frac{1}{48\epsilon}\right) \frac{\lambda^3}{\left(16\pi^2\right)^3} + \mathcal{O}\left(\lambda^4\right)
\end{equation}
and 
\begin{equation}
\begin{aligned}
Z_\lambda &= 1 + \frac{3}{2\epsilon} \frac{\lambda}{16\pi^2} + \left(\frac{9}{4\epsilon^2} - \frac{17}{12\epsilon}\right) \frac{\lambda^2}{\left(16\pi^2\right)^2} \\
&\hspace{0.4cm} + \left(\frac{27}{8\epsilon^3} - \frac{119}{24\epsilon^2} + \frac{145 + 96\zeta_3}{48\epsilon}\right) \frac{\lambda^3}{\left(16\pi^2\right)^3} + \mathcal{O}\left(\lambda^4\right).
\end{aligned}
\end{equation}
With these counterterms, we can find the coefficients of logarithms in the renormalized amplitude. Recall that the coefficients of the renormalized 2- and 4-particle amplitudes are denoted by $n_{L, k}$ and $m_{L, k}$ respectively with $L$ describing loop order and $k$ describing logarithmic order, as shown in \autoref{eq:3ch2PtAnsatz} and \autoref{eq:3ch4PtAnsatz}. For the renormalized 2-particle amplitude, 
\begin{equation}
n_{2, 1} = -\frac{1}{12}, \quad n_{2, 0} = -\frac{13}{48}, \quad n_{3, 2} = -\frac{1}{8}, \quad n_{3, 1} = -\frac{7}{8}, \quad n_{3, 0} = -\frac{167}{96}
\end{equation}
and $n_{1, 1} = n_{1, 0} = n_{2, 2} = n_{3, 3} = 0$. For the renormalized 4-particle amplitude, we find that in addition to $m_{0, 0} = 1/3$, 
\begin{equation}
\begin{aligned}
&m_{1, 1} = \frac{1}{2}, \quad m_{1, 0} = 1, \quad m_{2, 2} = \frac{3}{4}, \quad m_{2, 1} = 4, \quad m_{2, 0} = \frac{13}{2} + \zeta_2, \\
&m_{3, 3} = \frac{9}{8}, \quad m_{3, 2} = \frac{251}{24}, \quad m_{3, 1} = \frac{1831}{48} + \frac{9}{2}\zeta_2 + 6 \zeta_3.
\end{aligned}
\end{equation}

Either using the renormalized coefficients (as in \autoref{eq:BetaBC} and \autoref{eq:gammaBC}) or the counterterms (as in \cite[Ch.~10]{Kleinert:2001}), we find that the $\beta$-function is
\begin{equation}
\beta\left(\lambda\right) = 3\frac{\left(-\lambda\right)^2}{16\pi^2} + \frac{17}{3} \frac{\left(-\lambda\right)^3}{\left(16\pi^2\right)^2} + \left(\frac{145}{8} + 12\zeta_3\right) \frac{\left(-\lambda\right)^4}{\left(16\pi^2\right)^3} + \mathcal{O}\left(\lambda^5\right)
\end{equation}
and that the wavefunction anomalous dimension is
\begin{equation}
\gamma\left(\lambda\right) = \frac{1}{12}\frac{\left(-\lambda\right)^2}{\left(16\pi^2\right)^2} + \frac{1}{16} \frac{\left(-\lambda\right)^3}{\left(16\pi^2\right)^3} + \mathcal{O}\left(\lambda^4\right).
\end{equation}
These expressions match with \cite[Eq.~(10.59--10.60)]{Kleinert:2001}.

\section{Phase Space Integration}
\label{a:PhaseSpaceIntegrals}
The principal technical computation in the main text is the integration over the 2-, 3-, and 4-particle Lorentz invariant phase spaces. In this appendix, we compute all the phase space integrals used elsewhere in the paper. The higher particle phase space integrals can be efficiently computed by shrewd insertions of 1 in the form of delta function integrals, which decompose an $n$-particle phase space into sequential $1\rightarrow2$ decays of massive fictitious particles integrated over intermediate 2-particle phase spaces \cite{Byckling:1971vca}.

The massless $n$-particle phase space measure is defined as
\begin{equation}
\dd{\Phi}_n \equiv \frac{1}{n!} \left(2\pi\right)^d \delta^{(d)}\left(\ell_1 + \dots + \ell_n - p\right) \frac{1}{2\omega_1} \frac{\dd[d - 1]{\ell_1}}{\left(2\pi\right)^{d - 1}} \dots \frac{1}{2\omega_n} \frac{\dd[d - 1]{\ell_n}}{\left(2\pi\right)^{d - 1}},
\end{equation}
where we take $\omega_i \equiv \ell_i^{\:0} > 0$ by the positive energy condition across a cut and $p$ is the total external timelike momentum crossing the cut with $p^2 < 0$. The additional $n!$ is a symmetry factor associated to the indistinguishability of particles along the cut. The integral over the $n$ particle phase space, weighted by a function $f$ is defined as
\begin{equation}
\Phi_n\left(f\right) \equiv \int f\left(\ell_1, \dots, \ell_n; p\right) \dd{\Phi}_n.
\end{equation}

\subsection{2-Particle Phase Space}
The full two-particle phase space is defined as
\begin{equation}
\label{eq:2ParticlePhaseSpaceStartingPoint}
\Phi_2 = \frac{1}{2}\int \left(2\pi\right)^d\delta^{\left(d\right)}\left(\ell_1 + \ell_2 - p\right) \frac{1}{2\omega_1} \frac{\dd[d - 1]{\ell_1}}{\left(2\pi\right)^{d - 1}} \frac{1}{2\omega_2} \frac{\dd[d - 1]{\ell_2}}{\left(2\pi\right)^{d - 1}}.
\end{equation}
If we select the rest frame $p^\mu = \left(E, \vb{0}\right)$ with $E > 0$ and integrate over $\ell_2$, then we find that $\vb*{{\ell}}_2 = -\vb*{{\ell}}_1$ and, as a result, $\omega_1 = \omega_2$. To simplify notation, let $\ell_1 \equiv \ell$. This yields
\begin{subequations}
\begin{align}
\Phi_2 &= \frac{1}{2} \int 2\pi \delta\left(2\omega - E\right) \frac{1}{\left(2\omega\right)^2}\frac{\dd[d - 1]{\ell}}{\left(2\pi\right)^{d - 1}},\\
&= \frac{\pi^{\epsilon}\Gamma\left(1 - \epsilon\right)}{16\pi\Gamma\left(2 - 2\epsilon\right)} \int \omega^{-2\epsilon} \delta\left(\omega - E/2\right) \dd{\omega}.
\end{align}
\end{subequations}
In going to the last line we noted that $\abs{\vb*{{\ell}}\:} = \omega$, allowing us to work in spherical coordinates for which the $d - 1$ solid angle is
\begin{equation}
\frac{\Omega_{d - 1}}{\left(2\pi\right)^{d - 1}} = \frac{\pi^{\epsilon}\:\Gamma\left(1 - \epsilon\right)}{2\pi^2 \Gamma\left(2 - 2\epsilon\right)},
\end{equation}
where we recall that $d = 4 - 2\epsilon$. Integrating over $\omega$, the final result is
\begin{equation}
\label{eq:2particlePhaseSpace}
\Phi_2 = \frac{\Gamma\left(1 - \epsilon\right)}{16\pi\Gamma\left(2 - 2\epsilon\right)} \left(\frac{4\pi}{-p^2}\right)^{\epsilon},
\end{equation}
where we have covariantized from the rest frame $E^2 \mapsto -p^2$.

A result that will be important for the subsequent evaluations of higher particle phase space integrals is the 2-particle phase space in which one of the cut momenta is massive. We will denote it by $q$ with $q^2 = -m^2$:
\begin{subequations}
\begin{align}
\Phi_2^{\left(m\right)} &= \frac{1}{2}\int \left(2\pi\right)^d \delta^{\left(d\right)}\left(\ell + q - p\right) \frac{1}{2\omega_{\ell}} \frac{\dd[d - 1]{\ell}}{\left(2\pi\right)^{d - 1}} \frac{1}{2\omega_q} \frac{\dd[d - 1]{q}}{\left(2\pi\right)^{d - 1}},\\
&= \int \frac{2\pi\delta\left(\omega_{\ell} + \sqrt{\omega_{\ell}^2 + m^2} - E\right)}{8\omega_{\ell} \sqrt{\omega_{\ell}^2 + m^2}} \frac{\dd[d - 1]{\ell}}{\left(2\pi\right)^{d - 1}},\\
&= \frac{\pi^{\epsilon} \Gamma\left(1 - \epsilon\right)}{8\pi \Gamma\left(2 - 2\epsilon\right)} \int \frac{\delta\left(\omega + \sqrt{\omega^2 + m^2} - E\right)}{\omega^{2\epsilon - 1} \sqrt{\omega^2 + m^2}} \dd{\omega}.
\end{align}
\end{subequations}
The manipulations up to the last line are the same as before, except we keep track of the mass dependence. The final answer, after performing the $\delta$-integration and putting it in Lorentz-invariant form, is
\begin{equation}
\label{eq:2ParticleMassivePhaseSpace}
\Phi_2^{\left(m\right)} = \frac{\Gamma\left(1 - \epsilon\right)}{16\pi\Gamma\left(2 - 2\epsilon\right)}\left(1 + \frac{m^2}{p^2}\right)^{1 - 2\epsilon} \left(\frac{4\pi}{-p^2}\right)^{\epsilon}\Theta\left(-p^2 - m^2\right). 
\end{equation}

It will also be important in the computation of subleading logarithms and divergences to consider 
\begin{equation}
\Phi_2\left(\log\left(\frac{1}{\sin^2\theta/2}\right)\right) = \int \log\left(\frac{1}{\sin^2\theta/2}\right) \dd{\Phi}_2.
\end{equation}
The steps are the same as before, except we select the azimuthal angle in the final $\ell$ integral to align with the angle in the integrand. Using the solid angle measure, 
\begin{equation}
\frac{\dd{\Omega}_{d - 1}}{\left(2\pi\right)^{d - 1}} = \frac{\left(4\pi\right)^{\epsilon}}{4\pi^2\Gamma\left(1 - \epsilon\right)} \sin^{1-2\epsilon}\theta \dd{\theta},
\end{equation}
and the angular integral, 
\begin{equation}
\int_{0}^{\pi} \log\left(\frac{1}{\sin^2\theta/2}\right) \sin^{1 - 2\epsilon} \theta \dd{\theta} = \frac{2\Gamma^2\left(1 - \epsilon\right)}{4^{\epsilon}\Gamma\left(2 - 2\epsilon\right)}\left(\psi\left(2 - 2\epsilon\right) - \psi\left(1 - \epsilon\right)\right),
\end{equation}
where $\psi$ is the digamma function, we find
\begin{equation}
\label{eq:logsinint}
\Phi_2\left(\log\left(\frac{1}{\sin^2\theta/2}\right)\right) = \frac{\Gamma\left(1 - \epsilon\right)}{16\pi\Gamma\left(2 - 2\epsilon\right)}\left(\psi\left(2 - 2\epsilon\right) - \psi\left(1 - \epsilon\right)\right) \left(\frac{4\pi}{-p^2}\right)^{\epsilon}.
\end{equation}
Carrying through the same computation, we also find that
\begin{equation}
\label{eq:logcosint}
\Phi_2\left(\log\left(\frac{1}{\cos^2\theta/2}\right)\right) = \Phi_2\left(\log\left(\frac{1}{\sin^2\theta/2}\right)\right).
\end{equation}
It is interesting to note that both of these results reduce to just $\Phi_2$ (i.e. with no integrand) when $\epsilon \rightarrow 0$. 

The computations of the remaining phase space integrals will involve some knowledge of more general 2-particle phase space integrals. For example, following the same steps outlined for previous integrals, including working in the rest frame of the external momentum $p$ and covariantizing at the end, we find that 
\begin{equation}
\label{eq:NecessaryFor4cut}
\Phi_2\left(\left(\frac{4\pi}{\left(\ell_2 + \ell_3\right)^2}\right)^{L \epsilon}\right) = \frac{\Gamma\left(1 - \left(L + 1\right)\epsilon\right)}{16\pi\Gamma\left(2 - \left(L + 2\right)\epsilon\right)} \left(\frac{4\pi}{-p^2}\right)^{\left(L + 1\right)\epsilon}.
\end{equation}
where $L \geq 0$ is an integer and $\ell_3$ is a separate massless momentum not being integrated over. In other words, the measure is still given as in \autoref{eq:2ParticlePhaseSpaceStartingPoint}. 

\subsection{\texorpdfstring{$n$}{n}-Particle Phase Space Reduction}

Written more compactly, the massless $n$-particle phase space weighted by an arbitrary function, $f$, is 
\begin{equation}
\Phi_n\left(f\right) = \frac{1}{n!}\int f\left(\ell_i, p\right) \left(2\pi\right)^d \delta^{\left(d\right)}\left(p - \sum_{i = 1}^{n} \ell_i\right) \prod_{i = 1}^{n} \frac{1}{2\omega_i} \frac{\dd[d - 1]{\ell_i}}{\left(2\pi\right)^{d - 1}}. 
\end{equation}
We notice that the following one-particle phase space integrated over the mass of the particle is 1:
\begin{equation}
\int_{0}^{\infty} \int \left(2\pi\right)^d \delta^{\left(d\right)}\left(q - \tilde{\ell}\right) \frac{1}{2\omega_q} \frac{\dd[d - 1]{q}}{\left(2\pi\right)^{d - 1}} \frac{\dd{m^2}}{2\pi} = 1. 
\end{equation}
where $q^2 = -m^2$ and $\omega_q = \sqrt{\abs{\vb{q}^2} + m^2}$. Letting $q = q_{n - 1}$ and $\tilde{\ell}$ be $\ell_1 + \dots + \ell_{n - 1}$, we can insert this into $\Phi_n\left(f\right)$ and find
\begin{equation}
\Phi_n\left(f\right) = \frac{2}{n!}\! \int \! f \left(2\pi\right)^d \delta^{\left(d\right)}\left(q_{n - 1}\! - \! \sum_{i = 1}^{n - 1} \ell_i\right) \dd{\Phi}_2\left(p; \ell_n \! + \! q_{n - 1}\right) \left(\prod_{i = 1}^{n - 1} \frac{1}{2\omega_i} \frac{\dd[d - 1]{\ell_i}}{\left(2\pi\right)^{d - 1}}\right) \frac{\dd{m}_{n - 1}^2}{2\pi}, 
\end{equation}
where the intermediate 2-particle phase space measure is 
\begin{equation}
\dd{\Phi}_2\left(p; \ell_n \! + \! q_{n - 1}\right) = \frac{1}{2}\left(2\pi\right)^d \delta^{\left(d\right)}\left(p - \ell_n - q_{n - 1}\right) \frac{1}{2\omega_{\ell_n}} \frac{\dd[d - 1]{\ell_n}}{\left(2\pi\right)^{d - 1}} \frac{1}{2\omega_q} \frac{\dd[d - 1]{q_{n - 1}}}{\left(2\pi\right)^{d - 1}}.
\end{equation}
This can be thought of as breaking up a decay $p \rightarrow \ell_1 + \dots + \ell_n$ into a process involving an intermediate step involving $q_{n - 1}$: 
\begin{equation}
p \rightarrow q_{n - 1} + \ell_n \rightarrow \ell_1 + \dots + \ell_n. 
\end{equation}
This procedure can be carried through to continue breaking up the now $\left(n - 1\right)$-particle phase space measure into 2-particle phase space measures. This is equivalent to writing the decay process 
\begin{equation}
p \rightarrow \ell_n + q_{n - 1} \rightarrow \ell_n + \ell_{n - 1} + q_{n - 2} \rightarrow \dots \rightarrow \ell_1 + \dots + \ell_n. 
\end{equation}
In exchange for reducing the complexity of the integral, we must integrate over the masses of the fictitious particles of momentum $q_i$. Therefore, 
\begin{equation}
\Phi_n\left(f\right) = \frac{2^{n - 1}}{n!} \! \int \! f \dd{\Phi}_2\left(p; \ell_n \! + \! q_{n - 1}\right) \dd{\Phi}_2\left(q_{n - 1}; \ell_{n - 1} \! + \! q_{n - 2}\right) \dots \dd{\Phi}_2\left(q_2; \ell_2 \! + \! \ell_1\right) \prod_{i = 1}^{n - 2} \frac{\dd{m}_i^2}{2\pi}.
\end{equation}
Note that the final integral over the masses will in general have restrictions on the range of allowed values. Using the physical multi-step decay process, we see that $q_{n - 1}$ cannot have $m_{n - 1}^2 > \abs{p}^2$, the energy that we eject into the system. Similarly, when $q_{n - 1}$ decays into $q_{n - 2} + \ell_{n - 1}$, the mass of $q_{n - 2}$ cannot exceed the mass of $q_{n - 1}$. This will be implemented automatically through $\Theta$ distributions as in \autoref{eq:2ParticleMassivePhaseSpace}, but for notational simplicity, we now use 
\begin{equation}
\int \prod_{i = 1}^{n - 2} \frac{\dd{m}_i^2}{2\pi} \equiv \int_{0}^{s} \frac{\dd{m_{n - 2}^2}}{2\pi} \int_{0}^{m_{n -2}^2} \frac{\dd{m_{n - 1}^2}}{2\pi} \dots \int_{0}^{m_2^2} \frac{\dd{m_1^2}}{2\pi}. 
\end{equation}
In other words, the region is restricted to $m_1^2 \leq m_2^2 \leq \dots \leq s$. 

\subsection{3-Particle Phase Space}
We now apply the $n$-particle phase space reduction to the 3-particle phase space integral of massless particles: 
\begin{equation}
\Phi_3 = \frac{2}{3}\int \dd{\Phi}_2\left(p; \ell_3 + q\right) \dd{\Phi}_2\left(q; \ell_2 + \ell_1\right) \frac{\dd{m^2}}{2\pi}.
\end{equation}
The 2-particle phase spaces completely decouple. Using \autoref{eq:2particlePhaseSpace} for the massless 2-particle phase space and \autoref{eq:2ParticleMassivePhaseSpace} for the massive 2-particle phase space, we find that the integral reduces to one over the mass 
\begin{equation}
\Phi_3 = \frac{\pi \Gamma^2\left(1 - \epsilon\right)}{3\left(16\pi^2\right)^2\Gamma^2\left(2 - 2\epsilon\right)} \left(\frac{4\pi}{-p^2}\right)^{\epsilon} \! \int_{0}^{-p^2} \! \left(1 + \frac{m^2}{p^2}\right)^{1 - 2\epsilon} \left(\frac{4\pi}{m^2}\right)^{\epsilon} \Theta\left(-p^2 - m^2\right) \dd{m^2}. 
\end{equation}
Evaluating the mass integral, we find the result 
\begin{equation}
\label{eq:3ParticlePhaseSpace}
\Phi_3\left(p\right) = -\frac{p^2 \pi \Gamma^3\left(1 - \epsilon\right)}{3\left(16\pi^2\right)^2 \Gamma\left(2 - 2\epsilon\right)\Gamma\left(3 - 3\epsilon\right)} \left(\frac{4\pi}{-p^2}\right)^{2\epsilon}.
\end{equation}

\subsection{4-Particle Phase Space}
\label{a:4ParticlePhaseSpace}
The same reduction can be used to evaluate 4-particle phase space integrals. In this section, we will start by considering two integrands:
\begin{subequations}
\begin{align}
\label{eq:enhancedDiv1}
\Sigma_1 &= \left(\frac{4\pi}{\left(\ell_1 + \ell_2\right)^2}\right)^{L\epsilon} \frac{1}{\left(\ell_4 - p\right)^4},\\
\label{eq:enhancedDiv2}
\Sigma_2 &= \left(\frac{4\pi}{\left(\ell_1 + \ell_2\right)^2}\right)^{L_1 \epsilon} \left(\frac{4\pi}{\left(\ell_2 + \ell_3\right)^2}\right)^{L_2 \epsilon} \frac{1}{\left(\ell_4 - p\right)^4}. 
\end{align}
\end{subequations}
Reduced into nested 2-particle phase space integrals, 
\begin{equation}
\Phi_4\left(\Sigma\right) = \frac{1}{3} \int \Sigma \dd{\Phi}_2\left(p; \ell_4 + q_2\right)\dd{\Phi}_2\left(q_2; \ell_3 + q_1\right)\dd{\Phi}_2\left(q_1; \ell_2 + \ell_1\right) \frac{\dd{m}_1^2}{2\pi} \frac{\dd{m}_2^2}{2\pi}.
\end{equation}
On the support of the $\delta$-distributions in the 2-particle phase spaces, the integrands simplify: 
\begin{equation}
\Sigma_1 \rightarrow \left(\frac{4\pi}{m_1^2}\right)^{L\epsilon} \frac{1}{m_2^4} \quad \text{and} \quad \Sigma_2 \rightarrow \left(\frac{4\pi}{m_1^2}\right)^{L_1\epsilon} \left(\frac{4\pi}{\left(\ell_2 + \ell_3\right)^2}\right)^{L_2 \epsilon} \frac{1}{m_2^4}. 
\end{equation}

For $\Sigma_1$, the remaining 2-particle phase spaces decouple, so using \autoref{eq:2particlePhaseSpace} and \autoref{eq:2ParticleMassivePhaseSpace}, we find 
\begin{equation}
\label{eq:Sigma1integral}
\begin{aligned}
\Phi_4\left(\Sigma_1\right) = \frac{\pi \Gamma^3\left(1 - \epsilon\right)}{12\left(16\pi^2\right)^3 \Gamma^3\left(2 - 2\epsilon\right)} \left(\frac{4\pi}{s}\right)^{\epsilon} \! \int \! & \left(1 - \frac{m_2^2}{s}\right)^{1 - 2\epsilon} \! \left(1 - \frac{m_1^2}{m_2^2}\right)^{1 - 2\epsilon} \\
&\hspace{-0.4cm} \times \left(\frac{4\pi}{m_1^2}\right)^{\left(L + 1\right)\epsilon} \left(\frac{4\pi}{m_2^2}\right)^{\epsilon} \frac{1}{m_2^4} \dd{m_1^2} \dd{m_2^2}, 
\end{aligned}
\end{equation}
where we let $s = -p^2$ for simplicity. Recall that the mass integrals are restricted so that $0 \leq m_1^2 \leq m_2^2 \leq s$. Evaluating these integrals, we find 
\begin{equation}
\Phi_4\left(\Sigma_1\right) = \frac{\pi}{12\left(16\pi^2\right)^3} \frac{\Gamma\left(-\left(L + 2\right)\epsilon\right)\Gamma\left(1 - \left(L + 1\right)\epsilon\right)\Gamma^3\left(1 - \epsilon\right)}{\Gamma\left(3 - \left(L + 3\right)\epsilon\right) \Gamma\left(2 -  \left(L + 4\right)\epsilon\right)\Gamma\left(2 - 2\epsilon\right)} \left(\frac{4\pi}{s}\right)^{\left(L + 3\right)\epsilon}.
\end{equation}

For $\Sigma_2$, the 2-particle phase spaces do not decouple immediately due to the residual $\ell_2$ and $\ell_3$ dependence. However, upon integrating over the first phase space $\dd{\Phi}_2\left(q_1; \ell_2 + \ell_1\right)$ using \autoref{eq:2ParticlePhaseSpaceStartingPoint}, we find that the expression can then be expressed in terms of the invariant mass $m_1^2$. The resulting mass integral is of the same structure as was found for $\Sigma_1$. Carrying through the integration, we find
\begin{equation}
\Phi_4\left(\Sigma_2\right) = \frac{\Gamma\left(1 - \left(L_2 + 1\right)\epsilon\right)\Gamma\left(2 - 2\epsilon\right)}{\Gamma\left(2 - \left(L_2 + 2\right)\epsilon\right)\Gamma\left(1 - \epsilon\right)} \Phi_4\left(\Sigma_1\right),
\end{equation}
with the understanding that in $\Sigma_1$, $L = L_1 + L_2$. Notice that in the $\epsilon \rightarrow 0$ limit, the leading divergence is the same for both
\begin{equation}
\label{eq:EhnancedLeadingDiv}
\lim_{\epsilon \rightarrow 0} \Phi_4\left(\Sigma_1\right) = \lim_{\epsilon\rightarrow 0} \Phi_4\left(\Sigma_2\right) = -\frac{\pi}{24\left(L + 2\right)\left(16\pi^2\right)^3 \epsilon}. 
\end{equation}

An equivalent procedure can be used to study the following two integrands:
\begin{subequations}
\begin{align}
\Sigma_3 &= \frac{1}{\left(\ell_4 - p\right)^4} \log^{L\epsilon}_{\left(\ell_1 + \ell_2\right)^2}, \\
\Sigma_4 &= \frac{1}{\left(\ell_4 - p\right)^4} \log^{L_1\epsilon}_{\left(\ell_1 + \ell_2\right)^2} \log^{L_2\epsilon}_{\left(\ell_2 + \ell_3\right)^2}.
\end{align}
\end{subequations}
In this case, we will work in exactly four dimensions, $\epsilon = 0$. The mass integral will be the same, with the substitution of $\Sigma_3$ instead of $\Sigma_1$ or $\Sigma_4$ instead of $\Sigma_2$. In this case, however, we are principally interested in the highest power of logarithm, which will arise from the upper bound of the $m_2^2$ integral when it is close to $s$. All other contributions will be subleading. Isolating this contribution, we find 
\begin{equation}
\label{eq:4ParticleLogEnhancement}
\Phi_4\left(\Sigma_3\right) = \Phi_4\left(\Sigma_4\right) = -\frac{\pi}{24\left(16\pi^2\right)\left(L + 1\right)} \log_s^{L + 1} + \dots\;, 
\end{equation}
with the understanding that for $\Sigma_4$, $L = L_1 + L_2$ and that the ellipsis contains further subleading logarithms. 

Note that the 4-particle phase spaces we have considered so far have enhanced divergences or logarithmic power. For example, the logarithmic power of $\Sigma_3$ is just $L$, but the leading contribution from the result of the phase space integral had logarithmic power $L + 1$. This was necessary to constrain the subleading coefficients and divergences in the main text. We argue that the other possible 4-particle phase space integrals do not possess this property. To demonstrate this explicitly, consider a different combination of propagators: 
\begin{equation}
\Sigma_5 = \frac{1}{\left(\ell_4 - p\right)^2 - i\varepsilon} \frac{1}{\left(\ell_3 - p\right)^2 + i\varepsilon}.
\end{equation}
Employing the same reduction to 2-particle phase space and mass integrals, we find that the integral reduces to
\begin{equation}
\Phi_4\left(\Sigma_5\right) = \frac{-\Gamma\left(1 - \epsilon\right)}{192\pi^3 \Gamma\left(2 - 2\epsilon\right)} \left(\frac{4\pi}{s}\right)^{\epsilon} \int \frac{\dd{\Phi}_2\left(p; \ell_4 + q_2\right)\dd{\Phi}_2\left(q_2; \ell_3 + q_1\right) \dd{m_1^2} \dd{m_2^2}}{\left[m_1^2 + i\varepsilon\right]\left[\left(\ell_3 - p\right)^2 + i\varepsilon\right]}. 
\end{equation}
The remaining nested 2-particle phase space integrals can be evaluated and yields a hypergeometric function, 
\begin{equation}
\begin{aligned}
\Phi_4\left(\Sigma_5\right) &= \frac{\pi \Gamma^3\left(1 - \epsilon\right)}{12\left(16\pi^2\right)^3\Gamma^3\left(2 - 2\epsilon\right)} \left(\frac{4\pi}{s}\right)^{3\epsilon} \int_{0}^{1} \dd{x} \int_{0}^{x} \dd{y} \frac{\left(1 - x\right)\left(x - y\right)}{\left(x + i\varepsilon\right)\left(y - i\varepsilon\right)} \\
& \hspace{0.5cm} \times \left(\frac{x}{y\left(x - y\right)^2 \left(1 - x\right)^2}\right)^{\epsilon} 
{}_2F_1\left(1, 1 - \epsilon, 2 - 2\epsilon; \frac{\left(1 - x\right)\left(y - x\right)}{y - i\varepsilon}\right).
\end{aligned}
\end{equation}
To obtain this form, we let $m_2^2 = s x$ and $m_1^2 = sy$. Reduced to this form, we can check that the integral is finite as $\epsilon \rightarrow 0$. In fact, 
\begin{equation}
\Phi_4\left(\Sigma_5\right) = \frac{\pi \Gamma^3\left(1 - \epsilon\right)}{12\left(16\pi^2\right)^3\Gamma^3\left(2 - 2\epsilon\right)} \left(\frac{4\pi}{s}\right)^{3\epsilon}\left(\zeta_2 - 1 + \mathcal{O}\left(\epsilon\right)\right). 
\end{equation}
In other words, this does not produce an additional $\epsilon$ pole as the previous integrals did. We verify this result by matching to explicit computations of diagrams involving this cut, such as the fifth diagram in \autoref{tab:4Particle3LoopDiagrams}. 

\subsection{Extracting Leading Phase Space Divergences}
\label{a:4ParticlePhaseSpaceEnhancement}

In the main text we found, perhaps surprisingly, that it was important to take into account 4-cuts when working at subleading logarithmic or subleading divergence order. Naively, simple power counting arguments suggest that 4-cuts only begin to contribute at sub-subleading order, one order below the considerations of this paper. However, the explicit computation of the type of 4-particle phase space integral in \autoref{eq:enhancedDiv1} and \autoref{eq:enhancedDiv2} reveals that the phase space integral itself contributes an enhancement in the divergence, as we demonstrated in \autoref{eq:EhnancedLeadingDiv}. For intuition, in a Feynman diagram perturbative expansion, such contributions would come from 4-cuts of diagrams involving propagator corrections, such as the last diagram in \autoref{tab:4Particle3LoopDiagrams}. While for the cases of interest of this paper the enhanced phase space integrals could be computed analytically, if one wants to push this analysis beyond subleading order then higher-multiplicity cuts will be needed, which promise to be increasingly difficult to compute analytically. To this end, we use this subsection to present a simple approach, in the spirit of the method of regions \cite{Beneke:1997zp,Jantzen:2012mw}, to extract the leading divergence and derive the phase space enhancement.

In order to demonstrate the approach to extracting the leading divergence we consider \autoref{eq:enhancedDiv1}, whose reduction into mass integrals is given in \autoref{eq:Sigma1integral}. We first note that the parameter ranges are $0\leq m_{1}^{2}\leq m_{2}^{2}$ and $0\leq m_{2}^{2}\leq s$. For simplicity, we now define
\begin{equation}
x= sm_{2}^{2},\qquad y= sm_{1}^{2},
\end{equation}
along with 
\begin{equation}
A=\epsilon-3,\qquad B=-\epsilon\left(L+1\right),\qquad C=1-2\epsilon,\qquad D=1-2\epsilon,
\end{equation}
such that the phase space integral takes the cleaner form 
\begin{equation}
\label{eq:SimplifiedMassIntegral}
\Phi_4\left(\Sigma_1\right) = \frac{\pi \Gamma^3\left(1 - \epsilon\right)}{12\left(16\pi^2\right)^3\Gamma^3\left(2 - 2\epsilon\right)} \left(\frac{4\pi}{s}\right)^{\left(L + 3\right)\epsilon} \int_{0}^{1} \dd{x} \int_{0}^{x} \dd{y} x^A y^B\left(1 - x\right)^C \left(x - y\right)^D.
\end{equation}
The new parameter ranges are $0\leq y\leq x$
and $0\leq x\leq 1$. Since we are interested in the leading divergence, we immediately take the $\epsilon\rightarrow0$ limit of the integral prefactor:
\begin{equation}
\lim_{\epsilon \rightarrow 0} \frac{\pi \Gamma^3\left(1 - \epsilon\right)}{12\left(16\pi^2\right)^3\Gamma^3\left(2 - 2\epsilon\right)} \left(\frac{4\pi}{s}\right)^{\left(L + 3\right)\epsilon} = \frac{\pi}{12\left(16\pi^2\right)^3}.
\end{equation}

\begin{figure}
\begin{center}
\begin{tikzpicture}[scale=0.8]
\begin{scope}
    \path[clip] (0, 0) -- (2, 0) -- (2, 2) -- (0, 0);
    \shade[shading=radial, inner color=Maroon, outer color=white] (0,0) circle (1.5);
\end{scope}
\draw[line width=0.4mm, -Stealth] (0, 0) -- (3, 0);
\draw[line width=0.4mm, -Stealth] (0, 0) -- (0, 3);
\draw[line width=0.3mm, Maroon] (0, 0) -- (2, 0) -- (2, 2) -- (0, 0);
\draw[line width=0.3mm, dashed, Maroon] (0, 2) -- (2, 2);
\node at (3, -0.5) {$x$};
\node at (-0.5, 3) {$y$};
\node[RoyalBlue] at (-0.4, 1) {$A$};
\node[RoyalBlue] at (1, -0.4) {$B$};
\node[RoyalBlue] at (2.4, 1) {$C$};
\node[RoyalBlue] at (0.8, 1.2) {$D$};
\node[Maroon] at (-0.4, 2) {$1$};
\node[Maroon] at (2, -0.4) {$1$};
\end{tikzpicture} \qquad \qquad 
\begin{tikzpicture}[scale=0.8]
\begin{scope}
    \path[clip] (0, 0) -- (2, 0) -- (2, 2) -- (0, 2) -- (0, 0);
    \shade[left color=Maroon, right color=white] (0, 0) rectangle (1, 2);
\end{scope}
\draw[line width=0.4mm, -Stealth] (0, 0) -- (3, 0);
\draw[line width=0.4mm, -Stealth] (0, 0) -- (0, 3);
\draw[line width=0.3mm, Maroon] (0, 0) -- (2, 0) -- (2, 2) -- (0, 2) -- (0, 0);
\node at (3, -0.5) {$x$};
\node at (-0.5, 3) {$z$};
\node[RoyalBlue] at (-0.4, 1) {$A$};
\node[RoyalBlue] at (1, -0.4) {$B$};
\node[RoyalBlue] at (2.4, 1) {$C$};
\node[RoyalBlue] at (1, 2.4) {$D$};
\node[Maroon] at (-0.4, 2) {$1$};
\node[Maroon] at (2, -0.4) {$1$};
\end{tikzpicture}
\end{center}
\caption{Singularity structure of the $\Phi_4\left(\Sigma_1\right)$ mass integral. In the left figure, the integrand becomes singular as $x\rightarrow 0$, which is also the location where the different facets converge. In order to analyze the $x \rightarrow 0$ singularity, we changed variables from $y$ to $z = y/x$, which yields the diagram on the right. The letters on the figure correspond to the exponents associated with each line. For example, in the left figure, the diagonal line $x - y = 0$ is associated to the exponent $D$ in \autoref{eq:SimplifiedMassIntegral}.}
\label{fig:MethodOfRegions}
\end{figure}
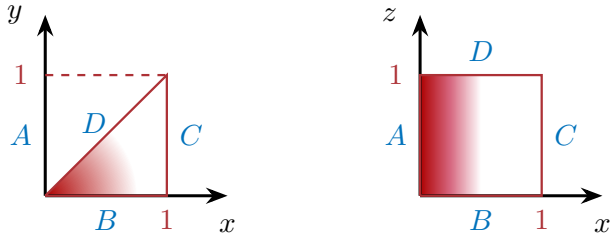

We draw the corresponding integration domain in \autoref{fig:MethodOfRegions}. In the $\epsilon \rightarrow 0$ limit, only the exponent $A$ leads to a singularity, however, as can be seen from \autoref{fig:MethodOfRegions} the facets for singularities corresponding to $A$, $B$ and $D$ all converge at the point $x=y=0$. In order to correctly analyze the genuine singularity from $x \rightarrow 0$, we perform a further variable redefinition: 
\begin{equation}
y=xz.
\end{equation}
$z$ has the range $0\leq z\leq1$ (since the maximum value of $y$ is $x$), and the transformation $\left(x,y\right)\rightarrow\left(x,z\right)$
carries a Jacobian factor $\mathrm{d}x\mathrm{d}y=x\;\mathrm{d}x\mathrm{d}z$,
such that the phase space integral becomes 
\begin{equation}
\label{eq:LeadingDivWithZ}
\eval{\Phi_4\left(\Sigma_1\right)}_{\text{lead. div.}} = \frac{\pi}{12\left(16\pi^2\right)^3} \int_{0}^{1} \dd{x} x^{A + B + D + 1} \left(1 - x\right)^C \int_{0}^{1} \dd{z} z^B \left(1 - z\right)^D. 
\end{equation}
With this judicious choice of variable change, the singularities of the integration region (see the right plot in \autoref{fig:MethodOfRegions}) are now nicely separated and the integral factorizes. The $z$ integral is immediately convergent:
\begin{equation}
\lim_{\epsilon \rightarrow 0} \int_{0}^{1} z^{-\epsilon\left(L + 1\right)} \left(1 - z\right)^{1 - 2\epsilon} \dd{z} = \frac{1}{2}.
\end{equation}
On the other hand, the $x$ integral is divergent near $x = 0$. Focusing on a region near $x = 0$ by imposing a cutoff $\delta \ll 1$, we find
\begin{equation}
\lim_{\epsilon \rightarrow 0} \int_{0}^{\delta} x^{-1 - \left(L + 2\right)\epsilon} \dd{x} = -\frac{\delta^{-\left(L + 2\right)\epsilon}}{\left(L + 2\right)\epsilon}.
\end{equation}
Combining these two integrals with \autoref{eq:LeadingDivWithZ}, we conclude that the leading divergence is 
\begin{equation}
\eval{\Phi_4\left(\Sigma_1\right)}_{\text{lead. div.}} = - \frac{\pi}{24\left(L + 2\right)\left(16\pi^2\right)^3 \epsilon},
\end{equation}
which corresponds with the result obtained by expanding the fully integrated result, \autoref{eq:EhnancedLeadingDiv}.

\newpage


\addcontentsline{toc}{section}{References}
\bibliographystyle{JHEP}
\bibliography{biblio.bib}

\end{document}